\documentclass[12pt]{article}

\usepackage{hyperref}

\usepackage{url}
\usepackage[utf8]{inputenc}
\usepackage[T1]{fontenc} 
\usepackage{lmodern}
\usepackage{multirow}
\usepackage{physics}
\usepackage{dsfont}
\usepackage{slashed}
\usepackage{xifthen}
\usepackage{subfig}
\usepackage{amsmath}
\usepackage{caption}
\usepackage{listings}
\usepackage{pdflscape}
\usepackage{graphics}
\usepackage{placeins}
\usepackage{xfrac}
\usepackage[dvipsnames]{xcolor}
\usepackage{standalone}
\usepackage{enumitem}
\usepackage{amsfonts}

\usepackage{tikz}
\usepackage[compat=1.1.0]{tikz-feynman}
\tikzfeynmanset{warn luatex=false}
\tikzfeynmanset{/tikzfeynman/warn luatex = false}

\def\id{\mathds{1}}

\DeclareMathOperator{\diag}{diag}

\newcommand*{\halfway}{0.5*\pgfdecoratedpathlength+.5*5pt}
\tikzset{->-/.style={decoration={markings, mark=at position #1 with {\arrow{latex}}},postaction={decorate}},
	->-/.default=\halfway}
	
\makeatletter
\newcommand{\subalign}[1]{%
  \vcenter{%
    \Let@ \restore@math@cr \default@tag
    \baselineskip\fontdimen10 \scriptfont\tw@
    \advance\baselineskip\fontdimen12 \scriptfont\tw@
    \lineskip\thr@@\fontdimen8 \scriptfont\thr@@
    \lineskiplimit\lineskip
    \ialign{\hfil$\m@th\scriptstyle##$&$\m@th\scriptstyle{}##$\hfil\crcr
      #1\crcr
    }%
  }%
}
\makeatother
\allowdisplaybreaks

\newcommand{\diags}[3][\relax]{%
  \ifx#1\relax 
 	  \includegraphics[scale=0.75]{img/#2_#3.eps}
   \else 
	  \includegraphics[#1]{img/#2_#3.eps}
  \fi%
}
\newcommand{\gathereddiags}[3][\relax]{%
  \begin{gathered}
	\diags[#1]{#2}{#3}
  \end{gathered}
}

\usepackage{epsfig}
\usepackage{color}
\usepackage{xcolor}
\usepackage{subfig}
\usepackage{a4}
\usepackage{latexsym}
\usepackage{cite}
\usepackage{slashed}

\usepackage[dvipsnames]{xcolor}
\usepackage{color,colordvi}
\def\colour4colour#1{\RubineRed{#1}}

\newcommand{\hspn}{{\hspace{-4mm}}}
\newcommand{\hspp}{{\hspace{4mm}}}
\newcommand{\hspl}{{\hspace{6mm}}}
\newcommand{\beq}{\begin{equation}}
\newcommand{\eeq}{\end{equation}}
\newcommand{\bea}{\begin{eqnarray}}
\newcommand{\eea}{\end{eqnarray}}
\newcommand{\nn}{\nonumber}
\newcommand{\MSb}{$\overline{\mbox{MS}}$}
\newcommand{\ra}{\rightarrow}

\newcommand{\als}{\alpha_{\sf s}}
\newcommand{\ars}{a_{\sf s}}

\newcommand{\ep}{\varepsilon}

\newcommand{\wP}{{\widetilde P}}
\newcommand{\wK}{{\widetilde K}}
\newcommand{\wc}{{\tilde c}}
\newcommand{\wg}{{\widetilde \gamma}}

\newcommand{\wN}{{\widetilde N}}

\begin{document}

\setlength{\parskip}{0.25cm}
\setlength{\baselineskip}{0.55cm}

\def\z#1{{\zeta_{\:\!#1}^{}}}
\def\zss{\zeta_2^{\,2}}
\def\zts{\zeta_3^{\,2}}

\def\nc{{n_c}}
\def\ncs{{n_{c}^{\,2}}}
\def\nct{{n_{c}^{\,3}}}

\def\ca{{C^{}_{\!A}}}
\def\cas{{C^{\,2}_{\!A}}}
\def\cat{{C^{\,3}_{\!A}}}
\def\caf{{C^{\,4}_{\!A}}}

\def\cf{{C^{}_{\:\!\!F}}}
\def\cfs{{C^{\, 2}_{\:\!\!F}}}
\def\cft{{C^{\, 3}_{\:\!\!F}}}
\def\cff{{C^{\, 4}_{\:\!\!F}}}

\def\nf{{n^{}_{\! f}}}
\def\nfz{{n^{\:\!0}_{\! f}}}
\def\nfo{{n^{\:\!1}_{\! f}}}
\def\nfs{{n^{\:\!2}_{\! f}}}
\def\nft{{n^{\:\!3}_{\! f}}}
\def\nff{{n^{\:\!4}_{\! f}}}

\def\tf{{T^{}_{\!F}}}
\def\tfs{{T^{\,2}_{\!F}}}
\def\tft{{T^{\,3}_{\!F}}}
\def\tff{{T^{\,4}_{\!F}}}

\def\dabctnc{{d^{\:\!abc\!}d_{abc}/n_c}}
\def\dabctnr{{ {d^{abc}d_{abc}}\over{n_c} }}

\def\muRs{\mu_{\sf r}^{\:\!2}}
\def\muFs{\mu_{\sf f}^{\:\!2}}

\def\as(#1){{\alpha_{\sf s}^{\:\!#1}}}
\def\ar(#1){{a_{\sf s}^{\:\!#1}}}

\def\xm{x_{m}}
\def\xp{x_{p}}

\def\LntO{\ln(1\!-\!x)}
\def\Lnt(#1){\ln^{\,#1}(1\!-\!x)}

\def\Qs{{Q^{\:\! 2}}}

\def\col{\color{RubineRed}}

\def\S(#1){{\col \mathbf S_{#1}}}
\def\Ss(#1,#2){{\col{ \mathbf S}_{#1,#2}}}
\def\Sss(#1,#2,#3){{\col{ \mathbf S}_{#1,#2,#3}}}
\def\Ssss(#1,#2,#3,#4){{\col{ \mathbf S}_{#1,#2,#3,#4}}}
\def\Sssss(#1,#2,#3,#4,#5){{\col{ \mathbf S}_{#1,#2,#3,#4,#5}}}
\def\Ssssss(#1,#2,#3,#4,#5,#6){{\col{ \mathbf S}_{#1,#2,#3,#4,#5,#6}}}

\def\H(#1){{\col{\rm{H}}_{#1}}}
\def\Hh(#1,#2){{\col{\rm{H}}_{#1,#2}}}
\def\Hhh(#1,#2,#3){{\col{\rm{H}}_{#1,#2,#3}}}
\def\Hhhh(#1,#2,#3,#4){{\col{\rm{H}}_{#1,#2,#3,#4}}}
\def\Hhhhh(#1,#2,#3,#4,#5){{\col{\rm{H}}_{#1,#2,#3,#4,#5}}}\def\dots{..}

\def\ddelta{{\col\delta(1-x)}}
\def\Dplus(#1){\mathcal{D}_{#1}}
\def\D(#1){{ D_{#1}}}
\def\Dd(#1,#2){{  D_{#1}^{\,#2}}}

\def\dots{..}

\def\D(#1){D_{#1}}
\def\Dd(#1,#2){D_{#1}^{\,#2}}

\def\frct#1#2{\mbox{\Large{$\frac{#1}{#2}$}}}

\def\Zfaa{Z_{5,1}}
\def\Zfab{Z_{5,2}}
\def\Zfac{Z_{5,3a}}
\def\Zfad{Z_{5,3b}}
\def\Zfae{Z_{5,3c}}

\def\Lz{L_0}
\def\pqq{p_{qq}}


\begin{titlepage}
\noindent
Nikhef 2022-021 \hfill November 2022\\
LTH 1324 \\
\vspace{1.0cm}
\begin{center}
{\LARGE \bf Four-loop large-n$_{\:\!\bf f}$ contributions to the\\
\vspace{0.3cm}
non-singlet structure functions F$_{\bf 2}$ and F$_{\bf L}$}\\
\vspace{2.0cm}
{\large A. Basdew-Sharma, A. Pelloni}\\
\vspace{0.3cm}
{\it Nikhef Theory Group \\
\vspace{0.5mm}
Science Park 105, 1098 XG Amsterdam, The Netherlands} \\
\vspace{0.6cm}
{\large F. Herzog}\\
\vspace{0.3cm}
{\it Higgs Centre for Theoretical Physics, School of Physics and 
Astronomy,\\
\vspace{0.5mm}
University of Edinburgh, Edinburgh EH9 3FD, Scotland, UK}\\
\vspace{0.6cm}
{\large  A. Vogt}\\
\vspace{0.3cm}
{\it Department of Mathematical Sciences, University of Liverpool\\
\vspace{0.5mm}
Liverpool L69 3BX, United Kingdom}\\
\vspace{2.0cm}
{\large \bf Abstract}
\vspace{-0.2cm}
\end{center}
We have calculated the $\nfs$ and $\nft$ contributions to the flavour 
non-singlet structure functions $F_2$ and $F_L$ in inclusive deep-inelastic 
scattering at the fourth order in the strong coupling~$\als\:\!$.
The coefficient functions have been obtained by computing a very large 
number of Mellin-$N$ moments using the method of differential equations, 
and then  determining the analytic~forms in $N$ and Bjorken-$x$ from these.
Our new $\nfs$ terms are numerically much larger than the $\nft$ leading 
large-$\nf$ parts which were already known; they agree with predictions of 
the threshold and high-energy resummations. Furthermore our calculation 
confirms the earlier determination of the four-loop $\nfs$ part of the 
corresponding anomalous dimension.
Via the no-$\pi^2$ conjecture$/$theorem for Euclidean physical quantities,
we predict the $\z4\:\!\nft$ part of the fifth-order anomalous 
dimension for the evolution of non-singlet quark distributions.
\vspace*{0.5cm}
\end{titlepage}
\newpage

%
\section{Introduction}

Inclusive deep-inelastic lepton nucleon scattering (DIS) via the exchange
of an electro-weak gauge boson is an experimental and theoretical benchmark
process of perturbative QCD.

\vspace{-1mm}
Data on its main structure functions provide a rather direct determination
of (linear combinations of) the quark momentum distributions of the nucleon.
The structure function $F_2(x,\Qs)$, in particular, has been determined 
in the past decades over a wide range of the scale $\Qs$ ($\, = - q^2$, 
where $q$ is the momentum of the exchanged boson) and the Bjorken variable 
$x$ ($\,= \Qs/(2\,p\cdot q)$, where $p$ is the momentum of the nucleon)
in fixed-target experiments and at the electron-proton collider HERA, 
see ref.~\cite{Workman:2022ynf} and references therein.
Further measurements of inclusive DIS are planned for future 
facilities, in particular the Electron Ion Collider (EIC) at Brookhaven 
National Lab \cite{Accardi:2012qut,AbdulKhalek:2021gbh}
and the Large Hadron Electron Collider (LHeC)
\cite{LHeCStudyGroup:2012zhm,LHeC:2020van}.

\vspace{-1mm}
Precise determinations of the quark momentum distributions 
$q_{\:\! \sf i}^{}(\xi,\Qs)$
(with $\xi = x$ at the leading-order of perturbative QCD) as well as, less
directly, of the gluon distribution $g(\xi,\Qs)$ and the strong coupling
$\als$ from structure-function data require higher-order calculations of
the corresponding coefficient functions (partonic structure functions).
These coefficient functions are of relevance also beyond the cross sections
for inclusive DIS, see, e.g., refs.~\cite{Bolzoni:2010xr,Dreyer:2016oyx}
on Higgs production in vector-boson fusion and ref.~\cite{Currie:2018fgr}
on jet production in DIS.

\vspace{-1mm}
For the quantities under consideration in this article, the flavour
non-singlet contributions to $F_2$ and the longitudinal structure function
$F_L$, the second-order corrections have been calculated and verified long
ago~\cite{SanchezGuillen:1990iq,vanNeerven:1991nn,Zijlstra:1992qd,%
Moch:1999eb}. The corresponding three-loop expressions were obtained in 
ref.~\cite{Vermaseren:2005qc} and recently re-calculated in 
ref.~\cite{Blumlein:2022gpp}.
At the fourth order, only the lowest five Mellin-$N$ moments have been 
computed so far \cite{Ruijl:2016pkm,Moch:2022frw,MRUVV-Fns} using the
{\sc Forcer} program \cite{Ruijl:2017cxj}, in addition to the leading 
terms in the limit of a large number of flavours $\nf$ 
\cite{Gracey:1995aj,Mankiewicz:1997gz}. 

\vspace{-1mm}
In the present article, we take the next step towards the determination of 
the fourth-order non-singlet coefficient functions $c_{2,\rm ns}^{\,(4)}(x)$ 
and $c_{L,\rm ns}^{\,(4)}(x)$ and compute their doubly fermionic $\nfs$ 
contributions. 
These results are obtained by a new method which allows the determination
of their moments up to very high (even) values of $N$, beyond $N = 1000$, 
from which the analytic dependence on $N$, and hence on $x$, can be 
re-constructed in terms of harmonic sums \cite{Vermaseren:1998uu} 
and harmonic polylogarithms (HPLs) \cite{Remiddi:1999ew}, respectively.
As a by-product, we have checked the $\nfs$-contributions to the four-loop
non-singlet splitting function $P_{\rm ns}^{\,(3)+}(x)$ of 
ref.~\cite{Davies:2016jie}.

The remainder of this article is organized as follows:
In section 2 we briefly recall the theoretical framework for the 
coefficient functions in inclusive DIS and their determination to the
fourth order in $\als$.
In section 3 we describe our method of the calculation based on 
iteratively solving a system of recurrence relations which is derived 
via the method of differential equations 
\cite{Kotikov:1990kg,Kotikov:1991pm,Remiddi:1997ny,Gehrmann:1999as}.
The analytic results for the coefficient functions in $N$-space are 
presented in section 4, which also includes a resulting partial 
prediction for the five-loop non-singlet splitting function 
$P_{\rm ns}^{\,(4)}(N)$.
The corresponding $x$-space coefficient functions and their threshold 
and high-energy limits are written down and discussed in section 5.
We~summarize our method and results and give a brief outlook in section 6. 

%
\section{Theoretical framework and notations}
\setcounter{equation}{0}

The subject of our computations is unpolarized inclusive lepton-nucleon DIS 
\beq
\label{inclDIS}
  \mbox{lepton}(k) \:+\: \mbox{nucleon}(p) 
     \:\:\ra\:\: 
  \mbox{lepton}(k^{\:\!\prime}) \:+\:  X
\eeq
at the lowest order of QED (i.e., via the exchange of one photon with 
momentum $q = k - k^{\,\prime}$). $X$ stands for all hadronic states 
allowed by quantum number conservation. 
The double-differential cross section 
in $\Qs = - q^2$ and $x = \Qs/(2\,p\cdot q)$ 
for this process can be expressed as the product of a calculable and 
well-known leptonic tensor and the hadronic tensor 
\beq
\label{Wmunu}
  W_{\:\!\!\mu\nu}(p,q) \:\:= \:\:  
     \bigg( \frac{q_{\mu} q_{\nu}}{q^2} - g_{\mu \nu} \bigg)
     \, F_{1}(x,Q^2)
   - \big( q_{\mu} + 2x p_{\mu} \big) \big( q_{\nu} + 2x p_{\nu} \big)
     \, \frac{1}{2xq^2}\, F_{2}(x,Q^2) \:\: .
\eeq

Neglecting contributions that are suppressed at large scales by powers of 
$1/\Qs$, the structure functions 
$\,{\cal F}_2= 1/x\, F_2$ and $\,{\cal F}_{\!L}= 1/x\, (F_2 - 2x\:\! F_1)$ 
can be expressed in terms of the universal but perturbatively incalculable 
quark and gluon parton distribution functions (PDFs), 
$q_{\:\! \sf i}^{}(\xi,\Qs)$ and $g(\xi,\Qs)$,
and the perturbative coefficient functions 
${\cal C}_{a,\rm p}(x,\Qs)$. In the present article, we are specifically 
interested in non-singlet (combinations of) structure functions
$F_{a,\rm ns\,}$, such as $F_a^{\,\rm proton} - F_a^{\,\rm neutron}$, which
decouple from the gluon distribution, viz
\beq
\label{Fa-Cq}
  {\cal F}_{a,\rm ns}(x,\Qs) \:\: = \:\: 
  \big[ \, {\cal C}_{a,\rm ns} \otimes q_{\:\!\rm ns} \big] (x,\Qs) 
\eeq
where $\otimes$ abbreviates the Mellin convolution. The non-singlet
combinations $q_{\:\!\rm ns}$ of quark distributions are normalized such 
that the expansion of the coefficient functions in powers of 
$\ars \equiv \als(\Qs) /(4\:\!\pi)$ is given by
\beq
\label{Cexp4}
  {\cal C}_{a,\rm ns}(x,\Qs) \:\:=\:\: (1-\delta_{aL}) \, \delta (1\!-\!x) 
  \:+\: \sum_{n=1} \ar(n\,) c_{a,\rm ns}^{\,(n)}(x) \:\: .
\eeq
Here and below we identify the \MSb\ renormalization and mass-factorization 
scales $\muRs$ and $\muFs$, at which the strong coupling and the PDFs
are evaluated, with $\Qs$. The dependence on $\muRs$ and $\muFs$ can be 
readily reconstructed a posteriori, see, e.g., eqs.~(2.17) and (2.18) 
of ref.~\cite{vanNeerven:2000uj}.

In terms of the general framework, our determination of the coefficient 
functions uses the method set out (and applied to the lowest 
moments at the third order) in 
refs.~\cite{Larin:1993vu,Larin:1996wd}, 
see also refs.~\cite{Moch:1999eb,Vermaseren:2005qc,Blumlein:2022gpp}:
The cross section for inclusive DIS for quark external states, projected
onto the structure functions ${\cal F}_{a}$, is related by the optical
theorem to the imaginary parts of the corresponding amplitudes for 
photon-quark forward scattering. Via a dispersion relation the 
coefficients of $[(2\,p\cdot q)/\Qs]^N = 1/x^N$ lead to the even-integer
(see also ref.~\cite{Moch:2007gx}) Mellin-$N$ moments
\beq
\label{MelTrf}
  {\widetilde{\cal F}}_{a,\rm ns}(N,\Qs) \:\: = \:\:
  \int_0^1 \! dx\: x^{\,N-1} \, {\widetilde{\cal F}}_{a,\rm ns}(x,\Qs) 
  \:\: ,
\eeq
of the bare partonic structure functions. These are computed from Feynman
diagrams in dimensional regularization with $D = 4 -2\:\!\ep$ dimensions.

After the \MSb\ renormalization of the coupling constant to the fourth order,
\beq
  \ar(\mbox{\scriptsize bare}) \,\: = \:\: \ars 
  \bigg( 1 
  \,-\, \frac{\beta_0^{}}{\ep} \,\ars
  \,+\, \bigg( \frac{\beta_0^2}{\ep^2} 
             - \frac{\beta_1^{}}{2\ep}\bigg) \,\ar(2)
  \,-\, \bigg( \frac{\beta_0^3}{\ep^3} 
             - \frac{7\beta_1^{}\beta_0^{}}{6\ep^2} 
             + \frac{\beta_2^{}}{3\ep} \bigg) \,\ar(3) + \:\ldots
  \bigg)
\eeq
with $\,\beta_0 = 11 - 2/3\,\nf\,$ etc in QCD 
\cite{Tarasov:1980au,Larin:1993tp},
the left-hand-side of eq.~(\ref{MelTrf}) can be written as
\beq
\label{Fb-CZ}
  {\widetilde{\cal F}}_{a,\rm ns}(N,\Qs) \:\: = \:\:
  {\widetilde{\cal C}}_{a,\rm ns}(N,\ars) \, Z_{\rm ns}(N,\ars)
  \:\: .
\eeq
The $D$-dimensional coefficient function $\widetilde{\cal C}_{a,\rm ns}$
includes additional terms with positive powers of $\ep$ on top of 
the Mellin transform of eq.~(\ref{Cexp4}), i.e., 
\beq
\label{CexpD}
  \tilde{c}_{a,\rm ns}^{\,(n)}(N) \:\: = \:\: 
    c_{a,\rm ns}^{\,(n)}(N)
    \,+\, \ep \:\! a_{a,\rm ns}^{(n)}(N)
    \,+\, \ep^2 \:\! b_{a,\rm ns}^{\,(n)}(N)
    \,+\, \ep^3 \:\! d_{a,\rm ns}^{\,(n)}(N)
    \,+\: \ldots
  \:\: .
\eeq
The quantity $Z_{\rm ns}$ which renormalizes the non-singlet quark 
distributions is given by
\bea
\label{Zns4}
  Z_{\rm ns} &\! = &
   1
   + \ars\, \frac{1}{\ep}\,\gamma_{\rm ns}^{(0)}
   + \ar(2) \left[ \:\!
     \frac{1}{2\:\!\ep^2} \left\{ \left(\gamma_{\rm ns}^{(0)} 
     - \beta_0 \right) \gamma_{\rm ns}^{(0)} \right\}
     + \frac{1}{2\:\!\ep}\, \gamma_{\rm ns}^{(1)} \right]
\nn\\[1mm] & & \mbox{}
   + \ar(3) \,\left[ \:\! \frac{1}{6\:\!\ep^3}\,
     \left\{ \left( \gamma_{\rm ns}^{(0)} - 2 \beta_0 \right) 
     \left( \gamma_{\rm ns}^{(0)} - \beta_0 \right)
     \gamma_{\rm ns}^{(0)} \right\}
 \right.  \nn\\ 
 & & \mbox{} \left. \mbox{} \quad\quad
   + \frac{1}{6\:\!\ep^2}
     \left\{ \left( 3 \gamma_{\rm ns}^{(0)} - 2 \beta_0 \right) 
     \gamma_{\rm ns}^{(1)}
     - 2 \beta_1 \gamma_{\rm ns}^{(0)} \right\}
    + \frac{1}{3\:\!\ep}\, \gamma_{\rm ns}^{(2)} \right]
\nn\\[1mm] & & \mbox{}
  + \ar(4) \,\left[ \:\!
    \frac{1}{24\:\!\ep^4}
    \left\{ \left( \gamma_{\rm ns}^{(0)} - 3 \beta_0 \right) 
    \left( \gamma_{\rm ns}^{(0)} - 2 \beta_0 \right)
    \left( \gamma_{\rm ns}^{(0)} - \beta_0 \right)
    \gamma_{\rm ns}^{(0)} \right\}
  \right. \\ 
  & & \left. \mbox{} \quad\quad
  + \frac{1}{12\:\!\ep^3}
    \left\{ \left( \left( 3 \gamma_{\rm ns}^{(0)} - 7 \beta_0\right) 
    \gamma_{\rm ns}^{(0)} + 3 \beta_0^2 \right) \gamma_{\rm ns}^{(1)}
    - 2 \left( 2 \gamma_{\rm ns}^{(0)} - 3 \beta_0 \right) 
    \beta_1 \gamma_{\rm ns}^{(0)} \right\}
  \right.  \nn\\ 
  & & \left. \mbox{} \quad\quad
  + \frac{1}{24\:\!\ep^2}
    \left\{ 2 \left( 4 \gamma_{\rm ns}^{(0)} - 3 \beta_0\right) 
    \gamma_{\rm ns}^{(2)}
    + 3 \left( \gamma_{\rm ns}^{(1)} - 2 \beta_1\right) \gamma_{\rm ns}^{(1)}
    - 6 \beta_2 \gamma_{\rm ns}^{(0)} \right\}
    + \frac{1}{4\:\!\ep}\, \gamma_{\rm ns}^{(3)}
  \right]
\nn
\eea
to the fourth order. Here $\gamma_{\rm ns}^{(n)}(N)$ -- the arguments $N$
have been suppressed in eq.~(\ref{Zns4}) for brevity -- are the N$_{}^{n}$LO
non-singlet anomalous dimensions related by
\beq
\label{gns}
  \gamma_{\rm ns}^{(n)}(N) \:\: = - \int_0^1 \! dx \, x^{N-1}
  P_{\rm ns}^{+(n)}(x)
\eeq
to the expansion coefficients of the \MSb-scheme splitting function for 
the evolution of flavour differences of the sums (hence `+') of quark and 
antiquark PDFs,
\beq
\label{Pplus}
  P_{\rm ns}^{+}(x,\ars) \:\: = \:\: 
  \sum_{n=0} \ar(n+1) P_{\rm ns}^{+(n)}(x) \:\: .
\eeq

Inserting the expansions (\ref{Cexp4}), (\ref{CexpD}) and (\ref{Zns4})
into eq.~(\ref{Fb-CZ}), the anomalous dimension and ($D$-dimensional)
coefficient functions can be extracted order by order from the results
of the diagram calculations. In order to obtain the fourth-order
coefficient functions $c_{a,\rm ns}^{\,(4)}$, the lower-order calculations
need to include terms up to $\ep^{\:\!4-n}$ at order $\as(n)$. In particular,
the determination of the $\nfs$ contributions to $c_{a,\rm ns}^{\,(4)}$
requires the $\nf$ parts of $a_{a,\rm ns}^{(3)}$ which were beyond the
scope of ref.~\cite{Vermaseren:2005qc} -- at the time only the integrals 
required for one simpler Lorentz projection of $W_{\:\!\!\mu\nu}$ were 
extended to this accuracy for ref.~\cite{Moch:2005tm}.


\section{Method and  computations}
\setcounter{equation}{0}

In terms of the diagram sets and the treatment to the point at which 
the Feynman integrals are evaluated, our computation is closely related 
to that of third-order fermionic ($\nf$ and~$\nfs$) contributions
in ref.~\cite{Moch:2002sn} and the non-singlet part of 
ref.~\cite{Davies:2016jie}. 
Our evaluation of the Feynman integrals is entirely different, though, 
from both. 
In ref.~\cite{Moch:2002sn} the analytic $N$-dependence was determined 
by setting up and solving, in a far from fully automated manner, 
complicated systems of difference equations. 
In ref.~\cite{Davies:2016jie} the even moments were computed to 
$N \!=\! 22$ for the $\cf \ca\, \nfs$ terms and to $N \!=\! 42$ for the 
$\cfs\, \nfs$ terms using {\sc Forcer} \cite{Ruijl:2017cxj}. From these 
it was possible, just, to reconstruct the analytic $N$-dependence of 
the four-loop anomalous dimensions using  all available physics 
constraints and systems of Diophantine equations.

Below we give details on the methods used in the present computation.
We believe that some of the techniques we have used here have been 
employed for the first time in a multi-loop calculation, and that these
should be useful not only for tackling DIS at four loops but also for 
other multi-loop calculations. 
Using these techniques we have been able to generate a very large number 
of Mellin moments, up to $N = 1500$, by evaluating the series 
expansion of the forward scattering amplitude around 
$1/x \equiv \omega = 0$, recall the discussion above eq.~(\ref{MelTrf}).
With that many moments, it is possible to reconstruct the analytic 
$N$-dependence of the fourth-order coefficient functions by a direct
(and over-constrained) Gaussian elimination for a sufficiently 
general ansatz in terms of harmonic sums.

Our basic setup relies on \textsc{Qgraf} \cite{Nogueira:1991ex} and 
\textsc{Form} \cite{Vermaseren:2000nd,Kuipers:2012rf,Ruijl:2017dtg},
employing the program \textsc{Minos} \cite{minos} as a diagram database 
tool.
Many of the \textsc{Form} libraries we use have been employed in a 
substantial number of earlier calculations, e.g., in 
refs.~\cite{Moch:1999eb,Vermaseren:2005qc,Ruijl:2016pkm,Davies:2016jie,%
Moch:2002sn} and have been highly optimized for multi-loop perturbative 
QCD calculations.
In a particular, as in refs.~\cite{Ruijl:2016pkm,Davies:2016jie}, the 
database combines diagrams whose underlying graph topology is equivalent 
and whose colour factors are the same.  Such sets of diagrams lead to 
faster evaluation times as they allow to realize algebraic cancellations 
between the individual diagrams.

\begin{figure}[htb]
\vspace*{-1mm}
\centerline{\epsfig{file=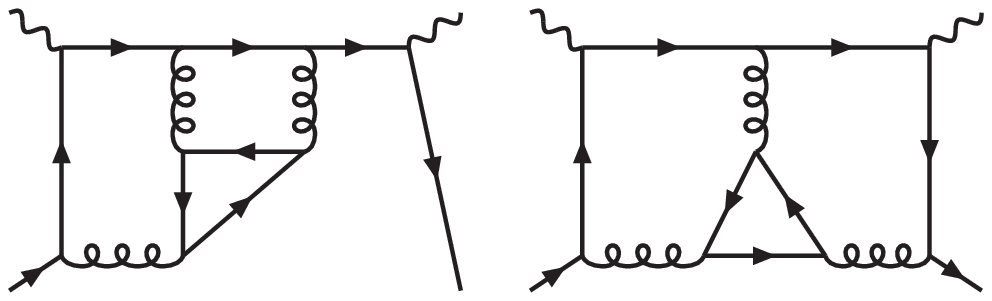,width=7.5cm,angle=0}%
\epsfig{file=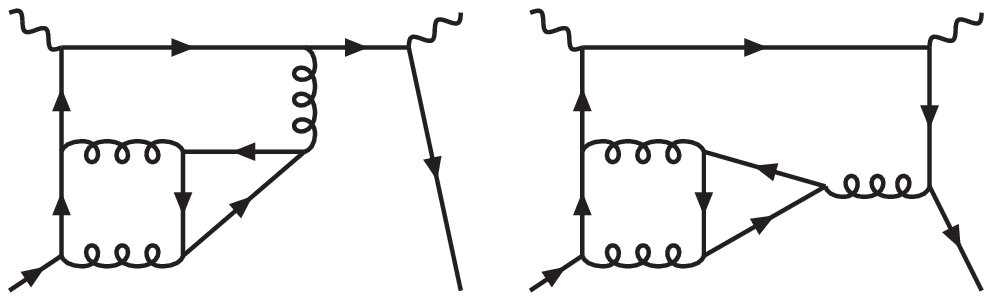,width=7.5cm,angle=0}}
\vspace*{-2mm}
\caption{ \label{CFig1} \small The hardest diagrams contributing to the
 $\nf$ contributions to the third-order coefficient functions for 
 $F_{2,\rm ns}$ and $F_{L,\rm ns}$. In the notation of \textsc{mincer}
 \cite{Gorishnii:1989gt,Larin:1991fz}, the two on the left are of the BE 
 topology and the two on the right are O4 cases. All four diagrams have
 the colour factor $C_{\:\!\!F}\, C_{\!A}\, \nf$.}
\vspace{-2mm}
\end{figure}

As the $\nf$ contributions at three loops, the present $\nfs$ contributions
at four loops are special since they do not yet involve the more difficult
topologies in their respective orders. In fact, the most difficult 
four-loop cases derive from the hardest three-loop diagrams shown in
fig.~\ref{CFig1} by simply inserting an additional quark loop into one of 
the gluon propagators, i.e., no `genuine' (non-insertion) four-loop 
self-energy topologies are required \cite{Ruijl:2016pkm}.

The best route to determine the all-$N$ expressions for the
$\nfs$ four-loop contributions is via the large-$\nc$ limits and the
$\cfs \nfs$ terms. The former do not involve alternating harmonic sums,
as even and odd $N$-values must lead to the same $x$-space function.
For the latter one expects a somewhat simpler form than for the 
$\cf\ca \nfs$ terms, since only two-loop diagrams with two one-loop or
one two-loop self-energy insertion(s) contribute.
In practice, the non-$\zeta$ part of $c_{2,\rm ns}^{\,(4)}$ was the most
difficult case, which we solved in \textsc{Form} using an ansatz with a 
little less than 600 even-$N$ coefficients. 
All other cases required fewer than 400 coefficients.

\vspace{-2mm}
\subsection*{Topologies for the non-singlet $\nfs$ structure functions}
\vspace*{-2mm}

As discussed at the end of section 2, we also need the lower-order 
corrections to a sufficient power in $\ep$; in particular we have to
compute the $\ep^{\,1}$ terms for the $\nf$ parts at three loops.

The diagrams contributing to the final result are organized into 
topologies, and described in terms of a linear independent set of 
propagators and scalar products ready to be reduced to a smaller set 
of master integrals \cite{Laporta:2000dsw}.
In order to obtain a more efficient reduction, especially at four loops, 
we build our topologies by keeping the number of propagators within a 
topology as low as possible.
In practice, a four-loop DIS topology has 18 linearly independent scalar 
products; instead of describing them in terms of as many linearly 
independent propagators, we opt for at most 12 propagators and 6 scalar 
products that will appear only in the numerator. At three-loop, this results 
in a topology described by 10 propagators and 2 scalar products.

We further simplify the expressions of the diagrams by rewriting 
multi-loop self-energy corrections in terms of chained bubbles which has 
the additional benefit of keeping all the propagator powers as whole 
numbers more suited for the Laporta reductions,
\beq
    q\gathereddiags{SelfEnergy}{1}q
    = 
       \frac{ \Pi_L(q^2)}{[\mathrm{Bub}(q^2)]^L}
      \quad q \gathereddiags{SelfEnergy}{2}q \; .
\eeq
 
\begin{figure}[hbt]
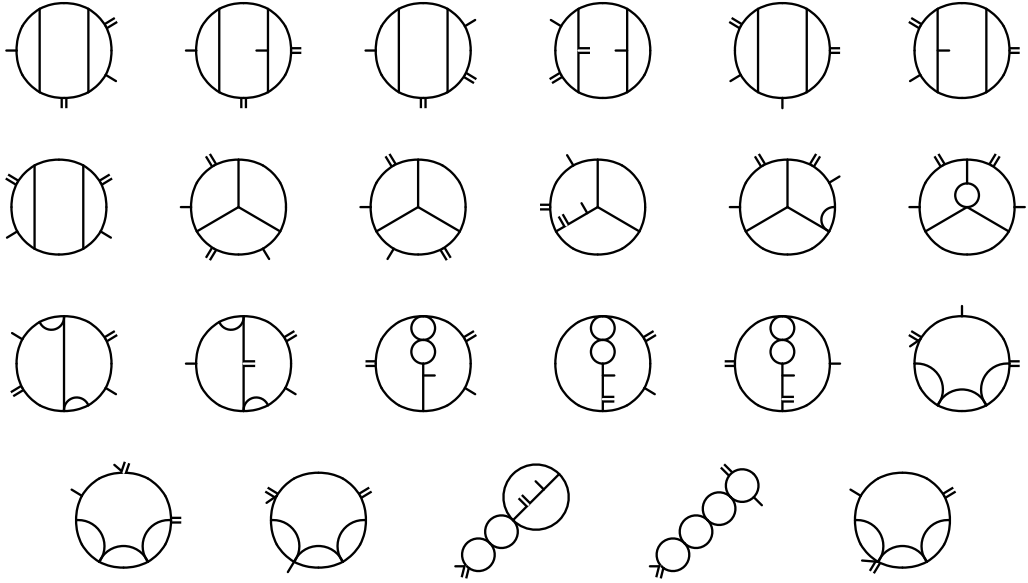

    \centering
\foreach \topoN in {1,...,6}{\subfloat{\diags[width=0.1\textwidth]{Topologies3L}{\topoN}}\hspl}\\
\foreach \topoN in {7,16,18,22}{\subfloat{\diags[width=0.1\textwidth]{Topologies3L}{\topoN}}\hspl}
\foreach \topoN in {1,...,2}{\subfloat{\diags[width=0.1\textwidth]{TopologiesNf2}{\topoN}}\hspl}\\
\foreach \topoN in {3,...,8}{\subfloat{\diags[width=0.1\textwidth]{TopologiesNf2}{\topoN}}\hspl}\\
\foreach \topoN in {9,...,13}{\subfloat{\diags[width=0.1\textwidth]{TopologiesNf2}{\topoN}}\hspp\hspp}\\
\caption{ \small
Three- and four-loop propagator topologies used for the reduction to master
integrals and the expansion in $\omega$ using differential equations. 
The external double line represent the off-shell photon while the simple 
lines the parton. 
Each diagram allows for two independent flows of the external momenta 
resulting in the same topology up to a sign inversion $\omega \leftrightarrow 
-\omega$.}
\label{fig:topologies}
\end{figure}

In fig.~\ref{fig:topologies} we show the range of topologies that are 
required for the  non-singlet $\nfs$ correction at the fourth order, 
with diagrams ranging from 12 to 9 distinct scalar propagators, one 
propagator short from the most complicated case one can encounter 
at four loops. 
 
The number of scalar propagators in the definition of each topology is not 
the only factor to take into account when considering the problem of reducing 
our integrals to master integrals (very important is also, for example, 
the dimensionality of the numerator). 
However, it remains a key aspect since it drastically reduces the degrees of
freedom one has to consider during the reductions by effectively vetoing some 
of the sectors. To perform the reductions to master integrals we have used 
both the \textsc{fire} \cite{Smirnov:2019qkx} and \textsc{kira} 
\cite{Maierhofer:2017gsa} programs.

\vspace{-2mm}
\subsection*{Series expansion and differential equations}
\vspace*{-2mm}

Once all the diagrams appearing in the process are reduced to master 
integrals, we build a system of differential equations by acting on them 
with the differential operator $\partial / \partial\:\! \omega$
and further reduce the resulting integrals. 
This step may also involve extending the initial set of master integrals to 
achieve a closed system. 
The resulting differential equations can be written as
\beq
\label{eq:DE}
\pdv{}{\omega}\vec{M}\,(\omega,\ep) \;=\; 
   A(\omega,\ep)\cdot\vec{M}(\omega,\ep)\,,
\eeq
where $A\in\mathbb{Q}^{\:\!n\times n}(\omega,\ep)$ is a square matrix whose 
elements are rational functions  in $\omega$ and the dimensional regulator 
$\ep$ over the field $\mathbb{Q}$, with $n$ being the number of master 
integrals. The general ansatz for this system of differential equations can 
be written as a linear combination of meromorphic functions,
\beq
\label{eq:masterAnsatz}
	\vec{M}(\omega,\ep) \;=\; \,\omega^{\:\! \vec{\alpha}}
    \sum_{s\,\subset\,\mathbb{Z}}
    \:\sum_{k\,=\,0}^\infty \:\vec{m}^{(s)}_k(\omega,\ep) 
    \,\omega^{\:\!k+s\cdot\ep} \;,
	\quad \vec{\alpha}\in\mathbb{Z}^n \;,
\eeq
where $\vec{\alpha}$ defines the leading power of each master and the sum over 
$s$ goes over a finite set of integers which multiply $\epsilon$ in the power 
of $\omega$, and represent independent (not necessarily physical) solutions to 
the differential equations. The space of solutions is reduced by noticing that 
the boundary conditions for the master integrals in the limit $\omega=0$ are 
finite ($\alpha \geq 0$) and support only the regular sector $s=0$. 
This may in fact not be true for individual integrals, but it is well known 
that the forward amplitudes for DIS are regular at $p=0$ and hence $w=0$. 
We can thus safely ignore solutions with $s \neq 0$. 
The general problem of obtaining series expansions from differential 
equations is of course not new in the context of Feynman integrals and 
indeed has been employed successfully in, e.g., refs.~%
\cite{Laporta:2000dsw,Boughezal:2007ny,Blumlein:2017dxp,%
Lee:2017qql,Liu:2017jxz,Mistlberger:2018etf,Moriello:2019yhu,%
Dubovyk:2022frj,Fael:2022miw,Blumlein:2022gpp},
and public implementations exist 
\cite{Hidding:2020ytt,Liu:2022chg,Armadillo:2022ugh}.

Here we present a new method which takes advantage of the specific analytic 
properties of the problem and which is well suited for obtaining many 
coefficients in the expansion, while at the same time 
being simple enough to be implemented purely in \textsc{Form}. We describe 
this procedure in the following.

We make use of the information coming from the boundary conditions to simplify 
the ansatz for the master integral expansion, and at the same time we consider 
the differential matrix to contain at most simple poles at $\omega=0$, viz
\beq\label{eq:ansatz-simple-poles}
	M_i \:=\: \omega^{\:\!\alpha_i}\,\sum_{k=0}^\infty m_{ik} \,\omega^k 
\;, \quad
	A \:=\: \frac{A_{-1}}{w} + \sum_{k=0}^\infty A_{k} \,\omega^k \;.
\eeq
In general, the matrix A can contain higher order poles around 
$\omega \!=\! 0$, depending on the choice of master integrals. 
While it is known in the Mathematics literature since the 1950s that a basis 
transformation to remove higher poles exists \cite{Moser:1959}, finding the 
transformation to bring the differential equations into a so-called canonical 
form is non-trivial in practice. 
By now there exist nevertheless several strategies/algorithms to solve this 
problem \cite{Henn:2013pwa,Lee:2014ioa}, and some implementations are now 
publicly available \cite{Lee2021,Gituliar2017,Prausa2017}. 
The feasibility to run these algorithms on systems containing hundreds master 
integrals is however questionable. 
For our purpose, we do not actually require the complete reduction of the 
system in to the canonical form -- as it is sufficient to simply remove higher 
order poles at $\omega=0$. 

We achieve this by applying a rescaling transformation $T$ to the master 
integrals,
\beq
 M \,\rightarrow\: T\cdot M \; ,\quad 
 A \:\rightarrow\: \pdv{T}{\omega}\:T^{-1} + T \cdot A \cdot T^{\,-1},
\eeq
with the transformation matrix taking the form
\beq
  T \;=\; \diag\left(\omega^{\beta\,}\right) \;, \;\; \beta\in\mathbb{N}_0^n
 \quad\Rightarrow\quad 
  A_{ij} \:\rightarrow\: \delta_{ij}\,\frac{\beta_i}{\omega} 
  + \omega^{\:\!\beta_i-\beta_j} A_{ij} \; .
\eeq
In general we found that the form of $T$ required to remove all poles is not 
unique. 
We explored several algorithms allowing to construct a suitable $T$; 
the probably simplest was based on constructing $T$ by iteratively removing 
a pole of order $k$ in the $j^{\:\!\rm th}$ row by setting 
$\beta_j=\omega^{\:\!k-1}$. 
While this particular pole would be removed, new poles could be created in 
other rows in the transformed matrix $A$. 
In all cases we found that iterating this procedure allowed to eventually 
remove all non-simple poles from $A$.
We actually have no prove for this simple procedure to terminate and it 
remains to be seen whether it will also work for even more complicated cases. 
The advantage of the procedure in comparison to the much more involved 
algorithms to bring $A$ into a Fuchsian form is that it is computationally 
far simpler and can be applied with 
ease also to comparably large systems of sizes around the 100s or even 1000s. 
It is not clear to us that the same holds for the algorithms mentioned~above.

By using the definitions in eq.~\eqref{eq:ansatz-simple-poles} it is possible 
to rewrite eq.~\eqref{eq:DE} into a recursive expression that allows for an 
efficient extraction of the expansion coefficients of the master integrals,
\beq
\label{eq:exStep_general}
\underbrace{\left((k+1)\id- A_{-1}\right)}_{:=\,B_k}\cdot\,\vec{m}_{k+1}^{}
 \:=\: \sum_{j=0}^k A_{j}\,\vec{m}_{k-j}^{} 
\eeq
where $\id$ is the identity matrix.
Note that we can have at most $n$ cases where $\det(B_k)=0$, corresponding 
to positive integer values of the eigenvalue of $A_{-1}$. 
In these cases the system cannot be inverted and is solved by means of 
Gaussian elimination. 
For high enough $k$ we have $\det(B_k)\neq 0$ and we can write
\beq
\label{eq:exStep_invertible}
	\vec{m}_{k+1}^{} \:=\: B_k^{\:\!-1}\cdot\left(\,
    \sum_{j=0}^k A_{j} \,\vec{m}_{k-j}^{}\right) \; .
\eeq
The inversion of the matrix $B$ is performed for generic values of $k$, 
avoiding the expensive procedure of performing a new inversion for every 
step of the expansion, especially for high values of $k$.

We have implemented the Gaussian elimination for the arbitrary steps of 
eq.~\eqref{eq:exStep_general} in C{}\verb!++! while the case of non-singular 
$B_k$ matrices in eq.~\eqref{eq:exStep_invertible} has been implemented in 
\textsc{Form}. 
The required boundary conditions have been computed using \textsc{Forcer} 
in the limit of $\omega=0$ for all the master integrals, 
where the external parton is taken to be soft. 
The performances of eq.~\eqref{eq:exStep_invertible} can be further improved 
by truncating the expansion in the dimensional regulator to the required order
to obtain an amplitude known up to $\ep^{4-n}$ at the $n$-th order.

One can benefit from the combined expansion only when the two limits 
$\ep,\omega=0$ of the differential matrix $A$ commute. 
For example, the series expansion in $\omega$ of a denominator of the form 
$(\omega-\ep)$ has a convergence radius that goes to zero as $\ep$ vanishes.
These kind of poles are unphysical and will lead to arbitrary high poles in 
$\ep$ when expanding the differential matrix and will cancel in the final 
expression of the expansion coefficients $\vec{m}_{k+1}$.

To circumvent this problem one can transform to a basis of master integrals
in which the reduction coefficients -- and therefore also the coefficients
of the differential equation matrix $A$ -- have the property that their
dependence on $\epsilon$ factorizes from their dependence on $\omega$.
The existence of such a basis was proven in ref.~\cite{BSMF_1992__120_3_371_0},
and two independent implementations which construct the corresponding basis
of master integrals have been published \cite{Smirnov:2020quc,Usovitsch:2020jrk}
as complementary codes to both the \textsc{fire} and \textsc{kira} reduction
programs respectively. We used the former one in our
reductions with \textsc{fire}.

With this implementation, and after obtaining a factorized form for the 
differential equations, we were able to push the recursive algorithm in 
eq.~\eqref{eq:exStep_invertible} up to $\order{\omega^{\:\!1500}}$ within 
no more than a few days for all of the masters. 
Such a large number of coefficients was necessary to be certain to fully 
constrain the expression of the coefficient functions in $N$-space from an 
ansatz in terms of harmonic sums and rational coefficients.

\vspace{-2mm}
\subsection*{Four-loop rescaling example}
\vspace*{-2mm}

Here we expand on the method used to bring the differential system into 
the form with at most simple poles in $\omega$. 
We will illustrate this method with one of the four-loop topologies we 
encountered in our computation.
\newcommand{\Itop}[1]{\text{I}_{\,[#1]}}
\begin{subequations}\label{t14l}
\begin{equation}
\begin{gathered}
\includegraphics[width=.1\textwidth]{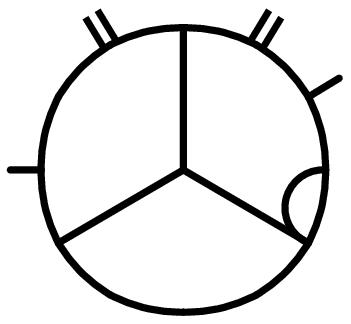}
\end{gathered}:\qquad
\Itop{\nu_1^{},\nu_2^{},\nu_3^{},\dots,\nu_{18}^{}} \,:=\,
\int\left(\prod_{j=1}^4\frac{\dd^d k_j}{(2\pi)^d}\right)
\frac{1}{\prod_{n=1}^{\:\!18}D_n^{\:\!\nu_n}}
\end{equation}
\begin{align}
\nonumber
&D_1=k_1^2,\quad
	&&D_7=\left(k_2-k_3\right){}^2,\quad
	&&D_{13}=2 k_1\cdot k_4,\quad
		\\&
	D_2=k_2^2,\quad
	&&D_8=\left(k_3+p-q\right){}^2,\quad
	&&D_{14}=2 k_2\cdot k_4\,,\quad
\nonumber
		\\&
	D_3=k_3^2,\quad
	&&D_9=\left(k_2-k_3+q\right){}^2,\quad
	&&D_{15}=2 k_1\cdot p\,,\quad
\nonumber
		\\&
	D_4=k_4^2,\quad
	&&D_{10}=\left(k_1-k_3-p+q\right){}^2,\quad
	&&D_{16}=2 k_4\cdot p\,,\quad
\nonumber
		\\&
	D_5=\left(k_3-q\right){}^2,\quad
	&&D_{11}=\left(k_2-k_3-p+q\right){}^2,\quad
	&&D_{17}=2 k_1\cdot q\,,\quad
\nonumber
		\\&
	D_6=\left(k_1-k_2\right){}^2,\quad
	&&D_{12}=\left(k_3+k_4+p-q \right){}^2,\quad
	&&D_{18}=2 k_4\cdot q\,.\quad
\end{align}
\end{subequations}
The reduction of this topology produced a set of $55$ distinct 
master integrals:
\begin{align*}
\{\;
&\Itop{0,0,1,1,0,1,1,0,0,1,0,1,0,0,0,0,0,0},\;
\Itop{0,0,1,1,0,1,1,0,1,1,0,1,0,0,0,0,0,0},\;
\Itop{0,0,1,1,1,1,1,0,1,1,0,1,0,0,0,0,0,0},\;
\\&
\Itop{0,1,0,1,0,1,1,0,0,1,0,1,0,0,0,0,0,0},\;
\Itop{0,1,0,1,0,1,1,0,1,1,0,1,0,0,0,0,0,0},\;
\Itop{0,1,0,1,1,1,1,0,0,1,0,1,0,0,0,0,0,0},\;
\\&
\Itop{0,1,0,1,1,1,1,0,0,2,0,1,0,0,0,0,0,0},\;
\Itop{0,1,1,1,0,1,1,0,1,1,0,1,0,0,0,0,0,0},\;
\Itop{1,0,0,1,0,1,1,0,0,0,0,1,0,0,0,0,0,0},\;
\\&
\Itop{1,0,0,1,0,1,1,0,1,0,0,1,0,0,0,0,0,0},\;
\Itop{1,0,0,1,1,1,1,0,0,0,0,1,0,0,0,0,0,0},\;
\Itop{1,0,0,1,1,1,1,0,0,0,1,1,0,0,0,0,0,0},\;
\\&
\Itop{1,0,0,1,1,1,1,0,0,1,0,1,0,0,0,0,0,0},\;
\Itop{1,0,0,1,1,1,1,0,1,1,0,1,0,0,0,0,0,0},\;
\Itop{1,0,1,1,0,0,1,0,0,1,1,1,0,0,0,0,0,0},\;
\\&
\Itop{1,0,1,1,0,0,1,0,1,1,0,1,0,0,0,0,0,0},\;
\Itop{1,0,1,1,0,1,0,0,1,0,0,1,0,0,0,0,0,0},\;
\Itop{1,0,1,1,0,1,0,0,1,0,0,2,0,0,0,0,0,0},\;
\\&
\Itop{1,0,1,1,0,1,0,0,1,1,0,1,0,0,0,0,0,0},\;
\Itop{1,0,1,1,0,1,1,0,0,0,1,1,0,0,0,0,0,0},\;
\Itop{1,0,1,1,0,1,1,0,0,1,0,1,0,0,0,0,0,0},\;
\\&
\Itop{1,0,1,1,0,1,1,0,1,0,0,1,0,0,0,0,0,0},\;
\Itop{1,0,1,1,0,1,1,0,1,1,0,1,0,0,0,0,0,0},\;
\Itop{1,0,1,1,0,1,1,0,1,1,0,2,0,0,0,0,0,0},\;
\\&
\Itop{1,0,1,1,0,1,1,0,1,2,0,1,0,0,0,0,0,0},\;
\Itop{1,0,1,1,0,1,2,0,1,0,0,1,0,0,0,0,0,0},\;
\Itop{1,0,1,1,0,1,2,0,1,1,0,1,0,0,0,0,0,0},\;
\\&
\Itop{1,0,1,1,1,0,1,0,0,1,1,1,0,0,0,0,0,0},\;
\Itop{1,0,1,1,1,0,1,0,1,1,0,1,0,0,0,0,0,0},\;
\Itop{1,0,1,1,1,1,0,0,1,1,0,1,0,0,0,0,0,0},\;
\\&
\Itop{1,0,1,1,1,1,1,0,0,0,1,1,0,0,0,0,0,0},\;
\Itop{1,0,1,1,1,1,1,0,0,0,1,2,0,0,0,0,0,0},\;
\Itop{1,0,1,1,1,1,1,0,0,1,0,1,0,0,0,0,0,0},\;
\\&
\Itop{1,0,1,1,1,1,1,0,0,1,0,2,0,0,0,0,0,0},\;
\Itop{1,0,1,1,1,1,1,0,0,1,1,1,0,0,0,0,0,0},\;
\Itop{1,0,1,1,1,1,1,0,0,1,1,2,0,0,0,0,0,0},\;
\\&
\Itop{1,0,1,1,1,1,1,0,1,0,0,1,0,0,0,0,0,0},\;
\Itop{1,0,1,1,1,1,1,0,1,1,0,1,0,0,0,0,0,0},\;
\Itop{1,0,1,1,1,1,1,0,1,1,0,2,0,0,0,0,0,0},\;
\\&
\Itop{1,0,1,1,1,1,1,0,1,2,0,1,0,0,0,0,0,0},\;
\Itop{1,0,1,1,1,1,1,0,2,1,0,1,0,0,0,0,0,0},\;
\Itop{1,1,0,1,0,0,1,0,0,1,0,1,0,0,0,0,0,0},\;
\\&
   \Itop{1,1,0,1,0,0,1,0,1,1,0,1,0,0,0,0,0,0},\;
   \Itop{1,1,0,1,1,0,1,0,0,1,0,1,0,0,0,0,0,0},\;
   \Itop{1,1,0,1,1,0,1,0,0,1,1,1,0,0,0,0,0,0},\;
   \\&
   \Itop{1,1,0,1,1,1,1,0,0,1,0,1,0,0,0,0,0,0},\;
   \Itop{1,1,0,1,1,1,1,0,0,1,0,2,0,0,0,0,0,0},\;
   \Itop{1,1,1,1,0,0,0,0,0,1,1,1,0,0,0,0,0,0},\;
   \\&
   \Itop{1,1,1,1,0,0,0,0,1,1,0,1,0,0,0,0,0,0},\;
   \Itop{1,1,1,1,0,0,0,0,1,1,0,2,0,0,0,0,0,0},\;
   \Itop{1,1,1,1,0,0,1,0,1,1,0,1,0,0,0,0,0,0},\;
   \\&
   \Itop{1,1,1,1,0,1,0,0,1,1,0,1,0,0,0,0,0,0},\;
   \Itop{1,1,1,1,0,1,0,0,1,1,0,2,0,0,0,0,0,0},\;
   \Itop{1,1,1,1,1,0,0,0,0,1,1,1,0,0,0,0,0,0},\;
   \\&
   \Itop{1,1,1,1,1,0,1,0,0,1,1,1,0,0,0,0,0,0}\;\}.
\end{align*}

For the purpose of the rescaling, the only information we are concerned about 
is the leading behaviour in $\omega$ of each of the coefficients of the 
matrix $A$ appearing in the differential euqations.
We represent the leading behaviour for the topology of eq.~\eqref{t14l} using 
the following colour coding
\begin{equation*}
w^{\{-4,-3,-2,-1,0,1,2\}}
=\{
\begin{gathered}\tikz{\draw[fill=yellow!0!red ,draw=none]  rectangle ++(0.5em,0.5em);}\end{gathered},
\begin{gathered}\tikz{\draw[fill=yellow!25!red,draw=none]  rectangle ++(0.5em,0.5em);}\end{gathered},
\begin{gathered}\tikz{\draw[fill=yellow!50!red,draw=none]  rectangle ++(0.5em,0.5em);}\end{gathered},
\begin{gathered}\tikz{\draw[fill=yellow!75!red,draw=none]  rectangle ++(0.5em,0.5em);}\end{gathered},
\begin{gathered}\tikz{\draw[fill=green!90!blue,draw=none]  rectangle ++(0.5em,0.5em);}\end{gathered},
\begin{gathered}\tikz{\draw[fill=green!60!blue,draw=none]  rectangle ++(0.5em,0.5em);}\end{gathered},
\begin{gathered}\tikz{\draw[fill=green!30!blue,draw=none]  rectangle ++(0.5em,0.5em);}\end{gathered}
\} \; .
\end{equation*}
We also use a circle, instead of a square, to highlight the position of 
the deepest poles at each step of the transformation.
The elimination is performed row-wise, where we rescale each master whose 
row contains at least one of the deepest poles by a factor of $\omega$.  
In the current example the iterative rescaling procedure leads to the following sequence of transformations:
\def\matrixplotwidth{0.45\textwidth}
\begin{align*}
&\phantom{ \xrightarrow{\;T_1\;\;\;} }
\left(\begin{gathered}
\includegraphics[width=\matrixplotwidth]{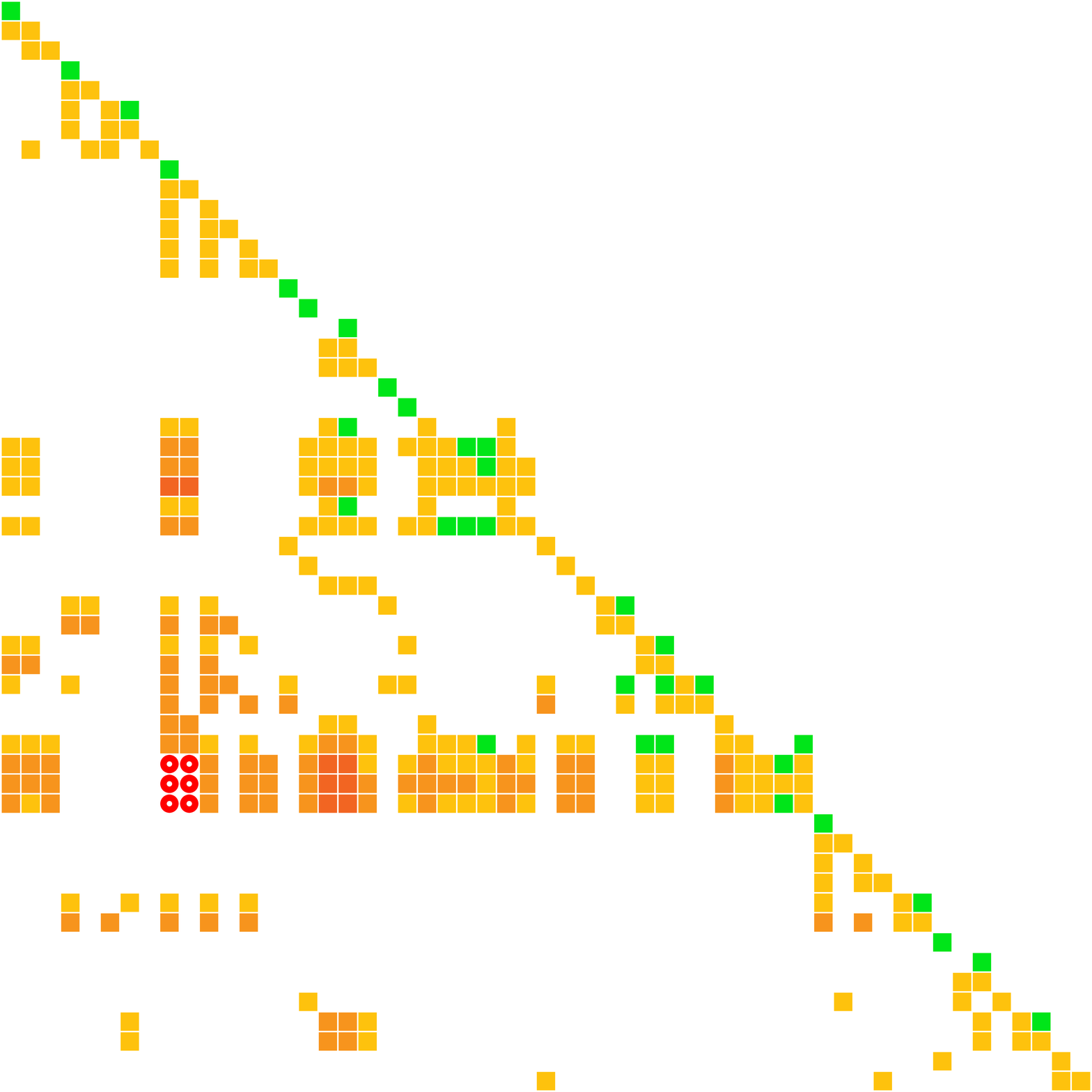}
\end{gathered}\right)
\\
&\xrightarrow{\;T_1\;\;}
\left(\begin{gathered}
\includegraphics[width=\matrixplotwidth]{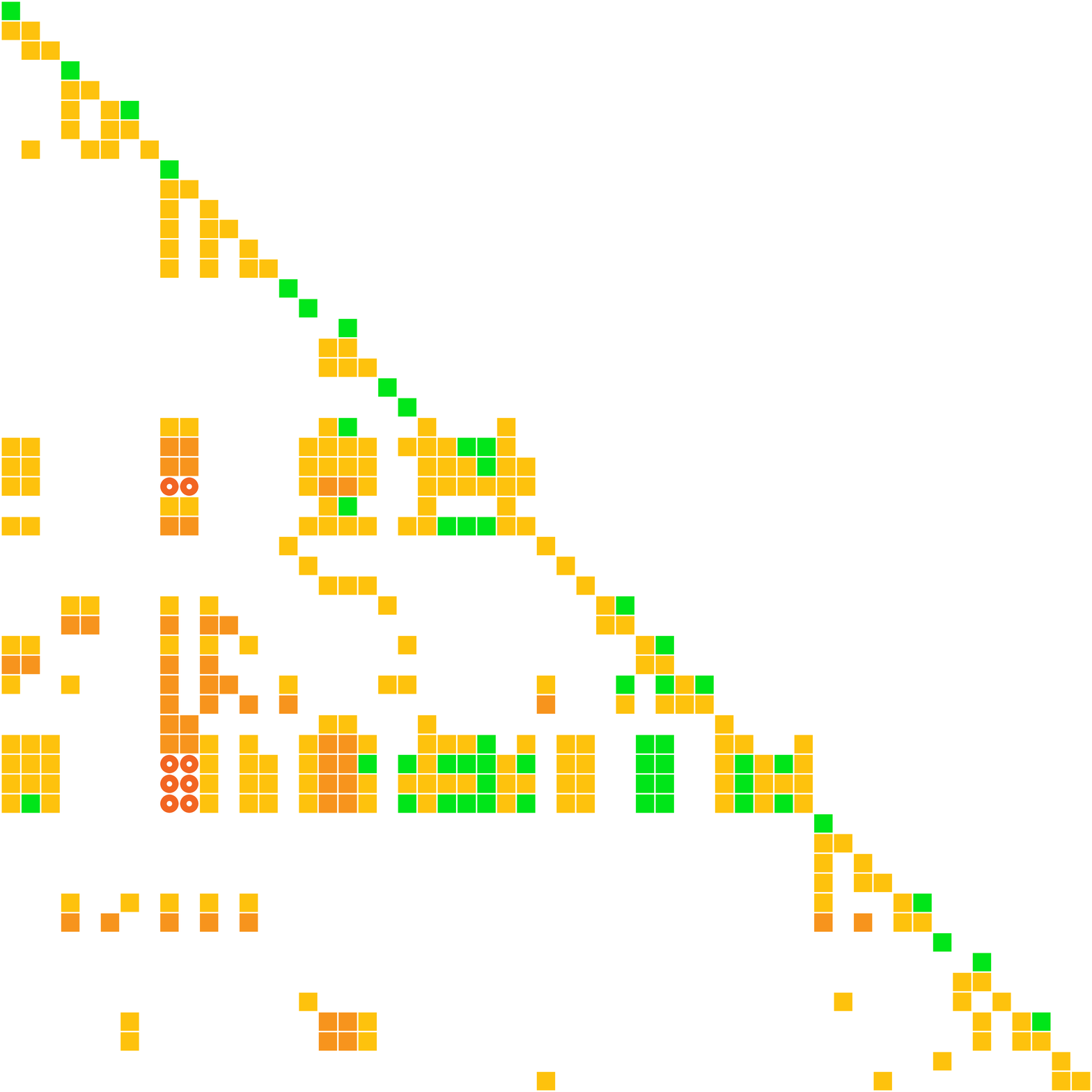}
\end{gathered}\right)
\\
&\xrightarrow{\;T_2\;\;}
\left(\begin{gathered}
\includegraphics[width=\matrixplotwidth]{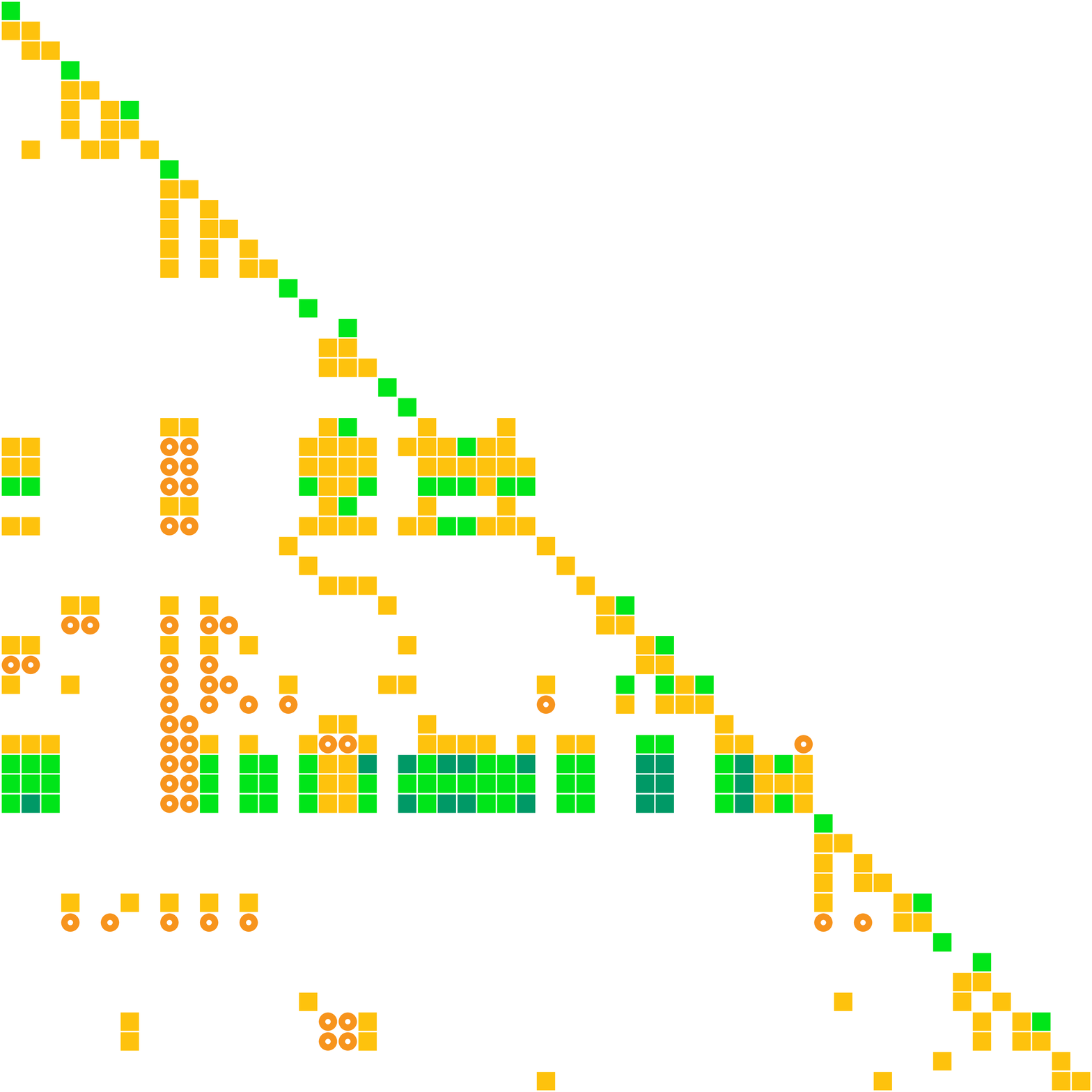}
\end{gathered}\right)
\\
&\xrightarrow{T_3\;\;}
\left(\begin{gathered}
\includegraphics[width=\matrixplotwidth]{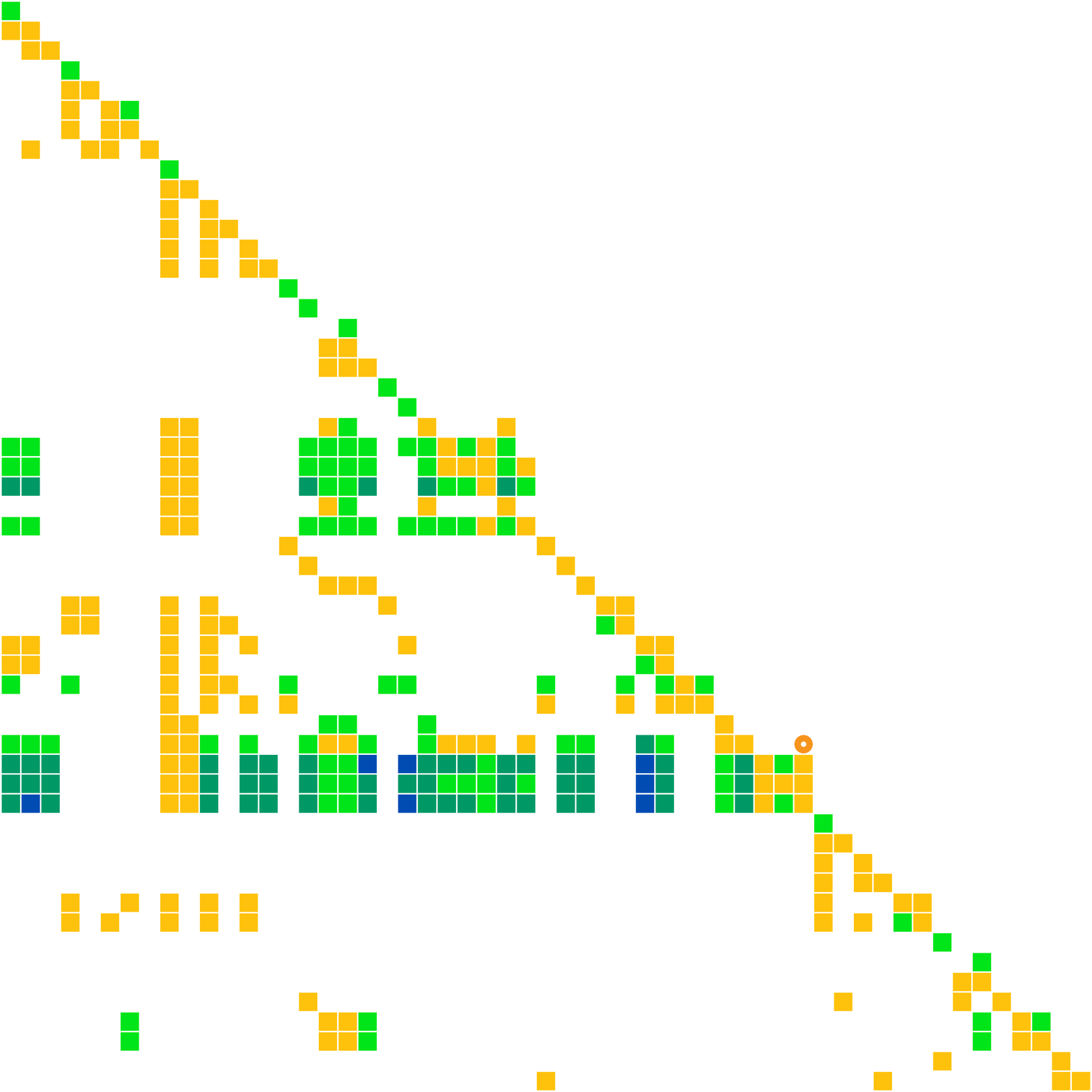}
\end{gathered}\right)
\\
& \xrightarrow{T_4\;\;}
\left(\begin{gathered}
\includegraphics[width=\matrixplotwidth]{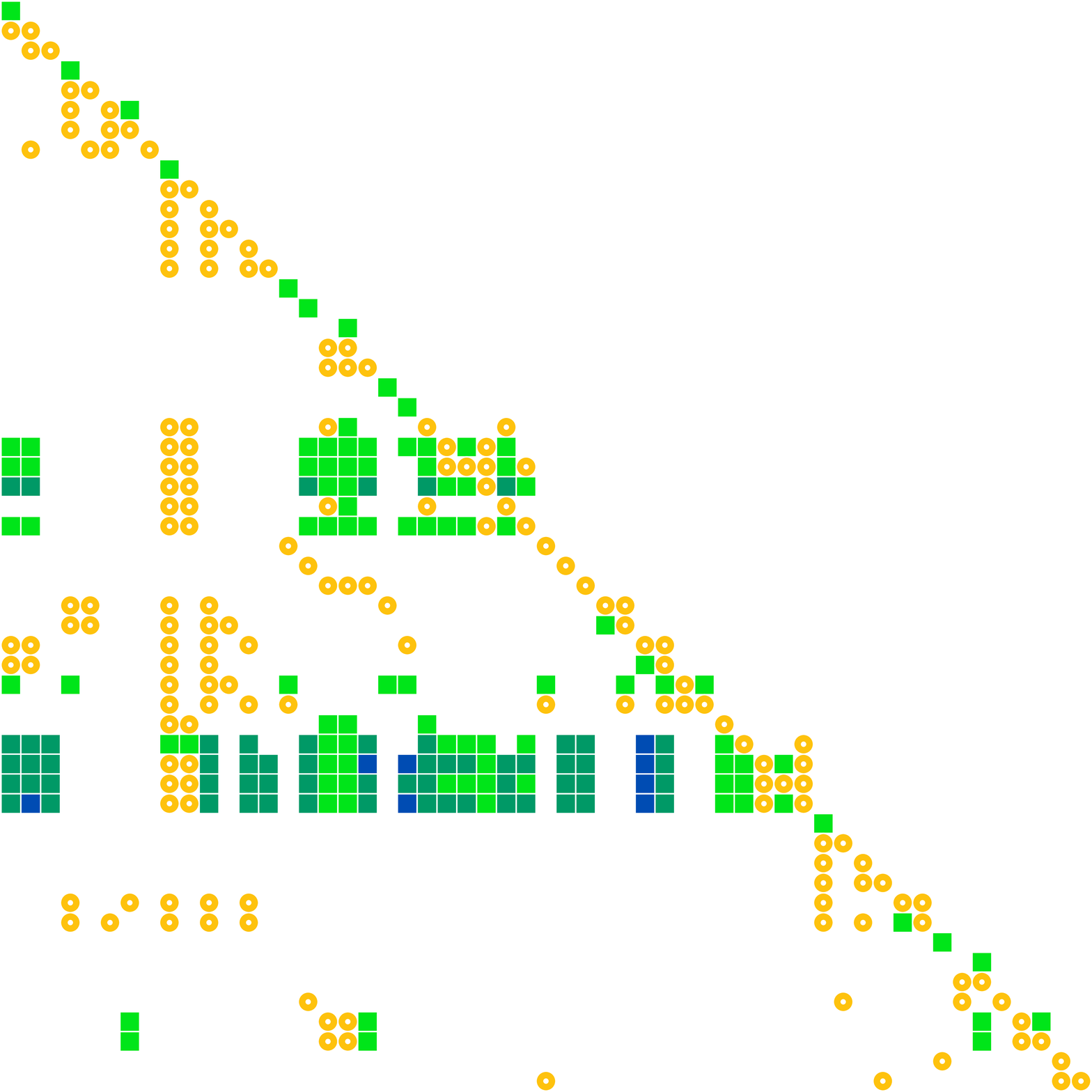}
\end{gathered}\right)
\end{align*}
where the transformation map is defined as
\begin{equation}
  A \;\xrightarrow{\;T_a\;}\; T_a\cdot\left(A \cdot T_a^{\:\!-1} 
  - \pdv{\:\! T_a^{-1}}{\:\!\omega}\right)
\end{equation}
with
\begin{align}
\left(T_{a}\right)_{ij} \;=\;\delta_{ij}\times
\begin{cases}
\;\omega,\quad a=1,\;i\in\{39,40,41\}\\
\;\omega,\quad a=2,\;i\in\{25,39,40,41\}\\
\;\omega,\quad a=3,\;i\in\{23,24,25,27,32,34,\dots,41,47,52,53\}\\
\;\omega,\quad a=4,\;i\in\{38\}\\
\;1 \; .
\end{cases}
\end{align}
Here we use $\dots$ as a short-hand for sequence of consecutive integers.
The complete transformation for the matrix $A$ can then be build by combining 
the intermediate steps
\begin{align}
\nonumber
T= T_4\cdot T_3\cdot T_2\cdot T_1= \text{diag}\Big(&1, 1, 1, 1, 1, 1, 1, 1, 
1, 1, 1, 1, 1, 1, 1, 1, 1, 1, 1, 1, 1, 1, \omega, \omega,
\\& 
\nonumber
\omega^2, 1, \omega, 1, 1, 1, 1, \omega, 1, \omega, \omega, \omega, \omega, 
\omega^2, \omega^3, \omega^3, \omega^3, 1, 1, 1,
\\& 
1, 1, \omega, 1, 1, 1, 1, \omega, \omega, 1, 1\Big) \; .
\end{align}

Alternatively, one could decide remove the deepest pole by rescaling 
column-wise, where each master, whose column contains one of the deepest 
poles, is rescaled by a factor of $\omega^{\:\!-1}$ or, equivalently, 
all the others are rescaled by $\omega$.
We use the latter because it does not affect the ability to express each 
master integral as a Taylor expansion.

If we were to start with the same matrix but cancel all the poles column-wise, 
the resulting transformation matrix would have been
\begin{align}
\left(T_{a}\right)_{ij} \;=\; \delta_{ij}\times
\begin{cases}
\;\omega,\quad a=1,\;i\notin\{9,10\}\\
\;\omega,\quad a=2,\;i\notin\{39,40,41\}\\
\;\omega,\quad a=3,\;i\notin\{1,\dots,6,9,\dots,19,21,22,23,24,26,\dots,
30,37,42,44\}\\
\;1
\end{cases}
\end{align}
and $T = T_3\cdot T_2 \cdot T_1$ with
\begin{align}
\nonumber
T = \text{diag}\Big(&\omega^2, \omega^2, \omega^2, \omega^2, \omega^2, 
 \omega^2, \omega^3, \omega^3, 1, 1, \omega^2, \omega^2, \omega^2, \omega^2,
 \omega^2, \omega^2, \omega, \omega, \omega^2,
\\&
\nonumber
 \omega^3, \omega^2, \omega^2, \omega^2, \omega^2, \omega^3, \omega^2, 
 \omega^2, \omega^2, \omega^2, \omega^2, \omega^3, \omega^3, \omega^3, 
 \omega^3, \omega^3, \omega^3, \omega^2,
\\&
 \omega^3, \omega^3, \omega^3, \omega^3, \omega^2, \omega^3, \omega^2, 
 \omega^3, \omega^3, \omega^3, \omega^3, \omega^3, \omega^3, \omega^3, 
 \omega^3, \omega^3, \omega^3, \omega^3\Big) \; .
\end{align}

In general, one could combine the two approaches by alternating between 
column- and row-wise rescaling at any step of the procedure. 
Such a combination will generate different rescaling matrices $T$. 
We have observed that a column-wise approach is generally preferred for our 
computation because it requires the rescaling of fewer integrals, resulting in
the ability to truncate the expansion for a larger set of master integrals.

%
\section{Results in N-space}
\setcounter{equation}{0}

We are now ready to present our new analytic even-$N$ expressions for 
the $\nfs$ contributions to the fourth-order coefficient functions 
$c_{2,\rm ns}^{\,(4)}$ and $c_{L,\rm ns}^{\,(4)}$ in eq.~(\ref{Cexp4}).
The corresponding $\nft$ results were derived long ago 
\cite{Gracey:1995aj,Mankiewicz:1997gz}. We confirm those results and 
include them below for completeness; for the case of ${\cal C}_2^{}$ 
in a more transparent form than given in ref.~\cite{Mankiewicz:1997gz}. 

The quantities under consideration can be expressed in terms of 
harmonic sums. Following the notation of ref.~\cite{Vermaseren:1998uu}, 
these sums are recursively defined by
\beq
\label{Hsum1}
  {\mathbf S}_{\pm m}(M) \;=\; 
  \sum_{i\,=\,1}^{M}\: \frac{(\pm 1)^i}{i^{\:\! m}} 
\eeq
and
\beq
\label{eq:Hsum2}
  {\mathbf S}_{\pm m_1^{},\,m_2^{},\ldots,\,m_k^{}}(M) \;=\; 
  \sum_{i\,=\,1}^{M}\: \frac{(\pm 1)^{i}}{i^{\:\! m_1^{}}}\: 
  S_{m_2^{},\ldots,\,m_k^{}}(i) \:\: .
\eeq
The sum of the absolute values of the indices $m_k$ defines the weight $w$
of the harmonic sum. In the $n$-loop coefficient functions one encounters 
sums up to weights $2n$ for ${\cal C}_2$ and $2n-1$ for ${\cal C}_L$.
The present non-singlet $\nfs$ contributions only include sums op to 
$w = 6$ for ${\cal C}_2$ and $w = 5$ for ${\cal C}_L^{}$, for the $\nft$ 
terms the corresponding maximal weights are lower by 1.
Below all harmonic sums have the argument $N$, which is omitted in the 
formulae for brevity, and we use the short-hand 
\beq
  D_a \:= \: (N+a)^{-1} \:\: .
\eeq

We first present the result for the coefficient function for $F_L$
which we write as
\bea
\label{cLNdec}
  c_{L,\rm ns}^{\,(4)}(N) & \!=\! &
  \nfz \mbox{ and } \nfo \mbox{ contributions }
  \nn \\[1mm] & & \mbox{\hspn}
  + \cf \* \ca \,\* \nfs \*\, \frac{16}{9} \*\:
    c_{L,\rm ns}^{\,(4)\mathrm{L}}(N)
  + \cf\, (\cf-\frct{1}{2}\,\* \ca) \,\* \nfs \*\, \frac{16}{9} \*\:
    c_{L,\rm ns}^{\,(4)\mathrm{N}}(N)
\nn \\[1mm] & & \mbox{\hspn}
  + \cf \* \nft \* \,\frac{16}{27} \*\:
    c_{L,\rm ns}^{\,(4)\mathrm{F}}(N)
\eea
where $\ca = \nc$ and $\cf = (\nc - n_c^{-1})/2$ in SU($\nc$), with 
$\nc = 3$ colours in QCD.
For compactness, we have decomposed the $\nfs$ part into leading (L) 
and non-leading (N) contributions in the large-$\nc$ limit. The factors
16/9 and 16/27 have been put in order to shorten, on average, the lengths
of the fractions in the expressions below. 
The $\nfs$ contributions are given by
\bea
\label{cLns4L}
\lefteqn{ c_{L,\rm ns}^{\,(4)\mathrm{L}}(N) \;=\;  } \nn \\[0.5mm] & & \mbox{}
%
%
%
%
%
%
+\Ss(1,4)\,(-80\,\D(-2)
+40\,\D(-1)
-40\,\D(1)
+80\,\D(2)
-120\,\D(3))
+\Ss(2,3)\,(-32\,\D(-2)
+16\,\D(-1)
\nn \\[0.5mm] & & \mbox{}
-16\,\D(1)
+32\,\D(2)
-48\,\D(3))
+\Ss(3,2)\,(32\,\D(-2)
-16\,\D(-1)
+16\,\D(1)
-32\,\D(2)
+48\,\D(3))
\nn \\[0.5mm] & & \mbox{}
+\Ss(4,1)\,(80\,\D(-2)
-40\,\D(-1)
+40\,\D(1)
-80\,\D(2)
+120\,\D(3))
+\Sss(1,1,3)\,(40\,\D(-2)
-20\,\D(-1)
\nn \\[0.5mm] & & \mbox{}
+20\,\D(1)
-40\,\D(2)
+60\,\D(3))
+\Sss(1,2,2)\,(8\,\D(-2)
-4\,\D(-1)
+4\,\D(1)
-8\,\D(2)
+12\,\D(3))
\nn \\[0.5mm] & & \mbox{}
+\Sss(1,3,1)\,(-8\,\D(-2)
+4\,\D(-1)
-4\,\D(1)
+8\,\D(2)
-12\,\D(3))
+\Sss(2,1,2)\,(8\,\D(-2)
-4\,\D(-1)
\nn \\[0.5mm] & & \mbox{}
+4\,\D(1)
-8\,\D(2)
+12\,\D(3))
+\Sss(2,2,1)\,(-8\,\D(-2)
+4\,\D(-1)
-4\,\D(1)
+8\,\D(2)
-12\,\D(3))
\nn \\[0.5mm] & & \mbox{}
+\Sss(3,1,1)\,(-40\,\D(-2)
+20\,\D(-1)
-20\,\D(1)
+40\,\D(2)
-60\,\D(3))
+\Ssss(1,1,2,1)\,(-18\,\D(1)
+72\,\D(2)
\nn \\[0.5mm] & & \mbox{}
-60\,\D(3))
+\Ssss(1,2,1,1)\,(18\,\D(1)
-72\,\D(2)
+60\,\D(3))
+\S(4)\,(120\,\D(-2)
-40\,\D(-1)
-55/3\,\D(1)
\nn \\[0.5mm] & & \mbox{}
-80\,\D(2)
+180\,\D(3))
+\Ss(1,3)\,(-40\,\D(-2)
+32\,\D(-1)
-16\,\D(0)
+65\,\D(1)
-24\,\D(2)
+12\,\D(3)
\nn \\[0.5mm] & & \mbox{}
-48\,\Dd(-2,2)
+24\,\Dd(-1,2))
+\Ss(2,2)\,(-12\,\D(-2)
+4\,\D(-1)
+38\,\D(1)
+8\,\D(2)
-18\,\D(3))
\nn \\[0.5mm] & & \mbox{}
+\Ss(3,1)\,(-8\,\D(-2)
-16\,\D(-1)
+16\,\D(0)
+9\,\D(1)
+56\,\D(2)
-84\,\D(3)
+48\,\Dd(-2,2)
-24\,\Dd(-1,2))
\nn \\[0.5mm] & & \mbox{}
+\Sss(1,1,2)\,(-34\,\D(1)
+16\,\D(2)
-8\,\D(3))
+\Sss(1,2,1)\,(-38\,\D(1)
-72\,\D(2)
+90\,\D(3))
\nn \\[0.5mm] & & \mbox{}
+\Sss(2,1,1)\,(-46\,\D(1)
+56\,\D(2)
-82\,\D(3))
+36\,\D(1)\,\*\Ssss(1,1,1,1)
+\S(3)\,(-30\,\D(-2)
-4\,\D(-1)
\nn \\[0.5mm] & & \mbox{}
-26\,\D(0)
+1097/9\,\D(1)
+56\,\D(2)
-108\,\D(3)
+72\,\Dd(-2,2)
-24\,\Dd(-1,2)
-18\,\Dd(1,2))
\nn \\[0.5mm] & & \mbox{}
+\Ss(1,2)\,(-10\,\D(-2)
+4\,\D(-1)
+65/3\,\D(0)
-895/6\,\D(1)
+56\,\D(2)
-116/3\,\D(3)
+29\,\Dd(1,2))
\nn \\[0.5mm] & & \mbox{}
+\Ss(2,1)\,(10\,\D(-2)
-4\,\D(-1)
+91/3\,\D(0)
-913/6\,\D(1)
-163/3\,\D(3)
+25\,\Dd(1,2))
\nn \\[0.5mm] & & \mbox{}
+\Sss(1,1,1)\,(-25\,\D(0)
+293/2\,\D(1)
-25\,\Dd(1,2))
+\S(2)\,(-30\,\D(-2)
+16\,\D(-1)
+410/3\,\D(0)
\nn \\[0.5mm] & & \mbox{}
-1493/4\,\D(1)
-16\,\D(2)
-11/2\,\D(3)
-35\,\Dd(0,2)
+263/3\,\Dd(1,2)
-27\,\Dd(1,3))
+\Ss(1,1)\,(45\,\D(-2)
\nn \\[0.5mm] & & \mbox{}
-20\,\D(-1)
-140\,\D(0)
+14351/36\,\D(1)
+16\,\D(2)
-5\,\D(3)
+39\,\Dd(0,2)
-99\,\Dd(1,2)
+25\,\Dd(1,3))
\nn \\[0.5mm] & & \mbox{}
+\,\S(1)\,(135/2\,\D(-2)
-20\,\D(-1)
-4045/12\,\D(0)
+1030465/1296\,\D(1)
-152\,\D(2)
\nn \\[0.5mm] & & \mbox{}
+2431/12\,\D(3)
-54\,\Dd(-2,2)
+24\,\Dd(-1,2)
+911/6\,\Dd(0,2)
-75/2\,\Dd(0,3)
-10565/36\,\Dd(1,2)
\nn \\[0.5mm] & & \mbox{}
+245/3\,\Dd(1,3)
-37/2\,\Dd(1,4))
-75/2\,\D(-2)
-714425/1296\,\D(0)
+3332269/2592\,\D(1)
\nn \\[0.5mm] & & \mbox{}
-935/12\,\D(3)
+18\,\Dd(-2,2)
+3355/12\,\Dd(0,2)
-3661/36\,\Dd(0,3)
+115/6\,\Dd(0,4)
\nn \\[0.5mm] & & \mbox{}
-785749/1296\,\Dd(1,2)
+3491/18\,\Dd(1,3)
-125/36\,\Dd(1,4)
-115/6\,\Dd(1,5)
\nn \\[0.5mm] & & \mbox{\hspn}
     + \z3 \,\* \Big[
%
%
%
%
%
%
 \S(2)\,(112\,\D(-2)
-56\,\D(-1)
+56\,\D(1)
-112\,\D(2)
+168\,\D(3))
+\Ss(1,1)\,(-80\,\D(-2)
+40\,\D(-1)
\nn \\[0.5mm] & & \mbox{}
+32\,\D(1)
-208\,\D(2)
+120\,\D(3))
+\,\S(1)\,(80\,\D(-2)
-64\,\D(-1)
+32\,\D(0)
-254/3\,\D(1)
\nn \\[0.5mm] & & \mbox{}
+368\,\D(2)
-400\,\D(3)
+96\,\Dd(-2,2)
-48\,\Dd(-1,2))
+32\,\D(-1)
-14/3\,\D(0)
-253/3\,\D(1)
\nn \\[0.5mm] & & \mbox{}
+356\,\D(3)
-144\,\Dd(-2,2)
+48\,\Dd(-1,2)
+50/3\,\Dd(1,2)
\Big]
\nn \\[0.5mm] & & \mbox{\hspn}
     + \z5 \,\* \Big[
%
%
%
%
%
%
-240\,\D(-2)
+120\,\D(-1)
+240\,\D(1)
-1200\,\D(2)
+840\,\D(3)
\Big]

\eea
and
\bea
\label{cLns4N}
\lefteqn{ c_{L,\rm ns}^{\,(4)\mathrm{N}}(N) \;=\;  } \nn \\[0.5mm] & & \mbox{}
%
%
%
%
%
%
+288\,\D(1)\,\*\S(-5)
-288\,\D(1)\,\*\S(5)
-256\,\D(1)\,\*\Ss(-3,-2)
-288\,\D(1)\,\*\Ss(-2,-3)
-576\,\D(1)\,\*\Ss(1,-4)
\nn \\[0.5mm] & & \mbox{}
-560\,\D(1)\,\*\Ss(2,-3)
-72\,\D(1)\,\*\Ss(2,3)
-368\,\D(1)\,\*\Ss(3,-2)
+112\,\D(1)\,\*\Ss(3,2)
+344\,\D(1)\,\*\Ss(4,1)
\nn \\[0.5mm] & & \mbox{}
+192\,\D(1)\,\*\Sss(-2,-2,1)
+128\,\D(1)\,\*\Sss(1,-2,-2)
+544\,\D(1)\,\*\Sss(1,1,-3)
+88\,\D(1)\,\*\Sss(1,1,3)
\nn \\[0.5mm] & & \mbox{}
+352\,\D(1)\,\*\Sss(1,2,-2)
+64\,\D(1)\,\*\Sss(1,2,2)
-184\,\D(1)\,\*\Sss(1,3,1)
+352\,\D(1)\,\*\Sss(2,1,-2)
+56\,\D(1)\,\*\Sss(2,1,2)
\nn \\[0.5mm] & & \mbox{}
-16\,\D(1)\,\*\Sss(2,2,1)
-168\,\D(1)\,\*\Sss(3,1,1)
-320\,\D(1)\,\*\Ssss(1,1,1,-2)
-72\,\D(1)\,\*\Ssss(1,1,1,2)
\nn \\[0.5mm] & & \mbox{}
+32\,\D(1)\,\*\Ssss(1,1,2,1)
+40\,\D(1)\,\*\Ssss(1,2,1,1)
+\S(-4)\,(576/5\,\D(-2)
+576\,\D(0)
-664\,\D(1)
\nn \\[0.5mm] & & \mbox{}
+864/5\,\D(3)
-432\,\Dd(1,2))
+\S(4)\,(1138/3\,\D(1)
-144\,\Dd(1,2))
+\Ss(-3,1)\,(-224/5\,\D(-2)
\nn \\[0.5mm] & & \mbox{}
+112\,\D(-1)
-224\,\D(0)
+224\,\D(2)
-336/5\,\D(3)
+224\,\Dd(1,2))
+\Ss(-2,-2)\,(-64/5\,\D(-2)
\nn \\[0.5mm] & & \mbox{}
-32\,\D(-1)
-64\,\D(0)
+336\,\D(1)
-64\,\D(2)
-96/5\,\D(3)
-32\,\Dd(1,2))
+\Ss(-2,2)\,(-144/5\,\D(-2)
\nn \\[0.5mm] & & \mbox{}
+72\,\D(-1)
-144\,\D(0)
+144\,\D(2)
-216/5\,\D(3)
+144\,\Dd(1,2))
+\Ss(1,-3)\,(-272/5\,\D(-2)
\nn \\[0.5mm] & & \mbox{}
-136\,\D(-1)
-272\,\D(0)
+1040\,\D(1)
-272\,\D(2)
-408/5\,\D(3)
-16\,\Dd(1,2))
+\Ss(1,3)\,(-16/5\,\D(-2)
\nn \\[0.5mm] & & \mbox{}
-8\,\D(-1)
-16\,\D(0)
+126\,\D(1)
-72\,\D(2)
-724/5\,\D(3)
+72\,\Dd(1,2))
+\Ss(2,-2)\,(-176/5\,\D(-2)
\nn \\[0.5mm] & & \mbox{}
-88\,\D(-1)
-176\,\D(0)
+688\,\D(1)
-176\,\D(2)
-264/5\,\D(3)
-16\,\Dd(1,2))
+\Ss(2,2)\,(-32/5\,\D(-2)
\nn \\[0.5mm] & & \mbox{}
-16\,\D(-1)
-32\,\D(0)
+176\,\D(1)
-32\,\D(2)
-48/5\,\D(3)
+8\,\Dd(1,2))
+\Ss(3,1)\,(64/5\,\D(-2)
\nn \\[0.5mm] & & \mbox{}
+32\,\D(-1)
+64\,\D(0)
-370\,\D(1)
+120\,\D(2)
+796/5\,\D(3))
+\Sss(-2,1,1)\,(128/5\,\D(-2)
-64\,\D(-1)
\nn \\[0.5mm] & & \mbox{}
+128\,\D(0)
-128\,\D(2)
+192/5\,\D(3)
-128\,\Dd(1,2))
+\Sss(1,1,-2)\,(32\,\D(-2)
+80\,\D(-1)
+160\,\D(0)
\nn \\[0.5mm] & & \mbox{}
-672\,\D(1)
+160\,\D(2)
+48\,\D(3)
+32\,\Dd(1,2))
+\Sss(1,1,2)\,(16/5\,\D(-2)
+8\,\D(-1)
+16\,\D(0)
\nn \\[0.5mm] & & \mbox{}
-188\,\D(1)
+48\,\D(2)
+544/5\,\D(3)
-32\,\Dd(1,2))
+\Sss(1,2,1)\,(-16/5\,\D(-2)
-8\,\D(-1)
-16\,\D(0)
\nn \\[0.5mm] & & \mbox{}
+8\,\D(1)
-16\,\D(2)
-24/5\,\D(3)
-8\,\Dd(1,2))
+\Sss(2,1,1)\,(-56\,\D(1)
-32\,\D(2)
-104\,\D(3)
+40\,\Dd(1,2))
\nn \\[0.5mm] & & \mbox{}
+72\,\D(1)\,\*\Ssss(1,1,1,1)
+\S(-3)\,(-1576/25\,\D(-2)
+136\,\D(-1)
-1384\,\D(0)
+2492/3\,\D(1)
\nn \\[0.5mm] & & \mbox{}
+272\,\D(2)
-2724/25\,\D(3)
+576/5\,\Dd(-2,2)
+576\,\Dd(0,2)
+824\,\Dd(1,2)
-240\,\Dd(1,3))
\nn \\[0.5mm] & & \mbox{}
+\S(3)\,(24/5\,\D(-2)
+8\,\D(-1)
-48\,\D(0)
-2672/9\,\D(1)
+72\,\D(2)
+1086/5\,\D(3)
+160\,\Dd(1,2)
\nn \\[0.5mm] & & \mbox{}
-48\,\Dd(1,3))
+\Ss(-2,1)\,(1232/25\,\D(-2)
-160\,\D(-1)
+448\,\D(0)
-48\,\D(1)
-368\,\D(2)
\nn \\[0.5mm] & & \mbox{}
+1968/25\,\D(3)
-192/5\,\Dd(-2,2)
+96\,\Dd(-1,2)
-192\,\Dd(0,2)
-352\,\Dd(1,2)
+96\,\Dd(1,3))
\nn \\[0.5mm] & & \mbox{}
+\Ss(1,-2)\,(32/25\,\D(-2)
+80\,\D(-1)
+432\,\D(0)
-2200/3\,\D(1)
+208\,\D(2)
+168/25\,\D(3)
\nn \\[0.5mm] & & \mbox{}
-192/5\,\Dd(-2,2)
-96\,\Dd(-1,2)
-192\,\Dd(0,2)
+16\,\Dd(1,2))
+\Ss(1,2)\,(-24/5\,\D(-2)
-8\,\D(-1)
\nn \\[0.5mm] & & \mbox{}
+178/3\,\D(0)
-319/3\,\D(1)
+100\,\D(2)
-8/15\,\D(3)
-34\,\Dd(1,2)
+24\,\Dd(1,3))
+\Ss(2,1)\,(24/5\,\D(-2)
\nn \\[0.5mm] & & \mbox{}
+8\,\D(-1)
+134/3\,\D(0)
+5/3\,\D(1)
-132\,\D(2)
-2332/15\,\D(3)
+106\,\Dd(1,2)
-24\,\Dd(1,3))
\nn \\[0.5mm] & & \mbox{}
+\Sss(1,1,1)\,(-50\,\D(0)
+73\,\D(1)
-50\,\Dd(1,2))
+\S(-2)\,(5432/375\,\D(-2)
-136\,\D(-1)
\nn \\[0.5mm] & & \mbox{}
+3856/3\,\D(0)
-2194/3\,\D(1)
-320\,\D(2)
+3606/125\,\D(3)
-944/25\,\Dd(-2,2)
+384/5\,\Dd(-2,3)
\nn \\[0.5mm] & & \mbox{}
+96\,\Dd(-1,2)
-928\,\Dd(0,2)
+384\,\Dd(0,3)
-1828/3\,\Dd(1,2)
+344\,\Dd(1,3)
-96\,\Dd(1,4))
\nn \\[0.5mm] & & \mbox{}
+\S(2)\,(-12\,\D(-2)
-56\,\D(-1)
+244/3\,\D(0)
+801/2\,\D(1)
-276\,\D(2)
-98\,\D(3)
-30\,\Dd(0,2)
\nn \\[0.5mm] & & \mbox{}
-320/3\,\Dd(1,2)
+6\,\Dd(1,3))
+\Ss(1,1)\,(96/5\,\D(-2)
+64\,\D(-1)
-82\,\D(0)
-3721/18\,\D(1)
+96\,\D(2)
\nn \\[0.5mm] & & \mbox{}
-506/5\,\D(3)
+34\,\Dd(0,2)
-34\,\Dd(1,2)
+42\,\Dd(1,3))
+\,\S(1)\,(924/25\,\D(-2)
+208\,\D(-1)
\nn \\[0.5mm] & & \mbox{}
-499/6\,\D(0)
-418193/648\,\D(1)
+244\,\D(2)
-21269/150\,\D(3)
-144/5\,\Dd(-2,2)
-96\,\Dd(-1,2)
\nn \\[0.5mm] & & \mbox{}
+65/3\,\Dd(0,2)
-15\,\Dd(0,3)
+4699/18\,\Dd(1,2)
+46/3\,\Dd(1,3)
+3\,\Dd(1,4))
+10663/125\,\D(-2)
\nn \\[0.5mm] & & \mbox{}
-1031/648\,\D(0)
-421403/1296\,\D(1)
+153647/750\,\D(3)
-2388/25\,\Dd(-2,2)
+288/5\,\Dd(-2,3)
\nn \\[0.5mm] & & \mbox{}
+307/6\,\Dd(0,2)
-1285/18\,\Dd(0,3)
+115/3\,\Dd(0,4)
+256829/648\,\Dd(1,2)
-2269/9\,\Dd(1,3)
\nn \\[0.5mm] & & \mbox{}
+2251/18\,\Dd(1,4)
-115/3\,\Dd(1,5)
\nn \\[0.5mm] & & \mbox{\hspn}
     + \z3 \,\* \Big[
%
%
%
%
%
%
-192\,\D(1)\,\*\S(-2)
-432\,\D(1)\,\*\S(2)
+288\,\D(1)\,\*\Ss(1,1)
+\,\S(1)\,(-48\,\D(-2)
-120\,\D(-1)
\nn \\[0.5mm] & & \mbox{}
-240\,\D(0)
+1964/3\,\D(1)
-64\,\D(2)
+416\,\D(3)
-192\,\Dd(1,2))
-1776/25\,\D(-2)
\nn \\[0.5mm] & & \mbox{}
-40\,\D(-1)
-3916/3\,\D(0)
+3196/3\,\D(1)
+584\,\D(2)
+3076/25\,\D(3)
+576/5\,\Dd(-2,2)
\nn \\[0.5mm] & & \mbox{}
+96\,\Dd(-1,2)
+576\,\Dd(0,2)
+172/3\,\Dd(1,2)
\Big]
    \:+\: \z5 \,\* \Big[
%
%
%
%
%
%
180\,\D(1)
\Big]

\; .
\eea
  
The $\cfs$ contribution (\ref{cLns4N}) includes, as the corresponding
lower-order quantities, only the denominator $D_1 = 1/(N\!+\!1)$ at the
maximal overall weight $w = 5$ of the sums and Riemann $\zeta$-values. 
The $\cf\ca$ part, and hence the large-$\nc$ coefficient (\ref{cLns4L}), 
does not have this expected property, but instead involves the linear 
combinations
$$
  2 D_{-2} - D_{-1} + D_1 - 2 D_2 + 3 D_3
 \quad \mbox{ and } \quad
  3 D_1 - 12 D_2 + 10 D_3
\:\: .
$$
The presence of terms with $D_{-2} = 1/(N\!-2)$ and $D_{-1} = 1/(N\!-1)$
does not imply poles at $N=2$ or $N=1$, as the corresponding numerators also 
vanish at this point. This feature already occurred in the second-order 
coefficient functions of 
refs.~\cite{SanchezGuillen:1990iq,vanNeerven:1991nn,Zijlstra:1992qd,%
Moch:1999eb}. At $N\!=\!2$ these functions are given by
\bea
\label{cLns4N2}
 c_{L,\rm ns}^{\,(4)\mathrm{L}}(N\!=\!2) & \!=\! & 
    \frac{1058755}{2916}
  - \frac{6713}{135} \:\*\z3
  - 32\,\*\z5
  + 24\,\*\zts
\:\: ,
\\[1.5mm]
 c_{L,\rm ns}^{\,(4)\mathrm{N}}(N\!=\!2) & \!=\! &
   \frac{1720051}{29160}
 + \frac{247}{270}\:\z3
 + 30\,\*\z5
\:\: .
\eea
Note the $\zts$ term which does not occur at higher $N$.
The $\nft$ contribution to eq.~(\ref{cLNdec}) reads
\bea
\label{cLns4F}
\lefteqn{c_{L,\rm ns}^{\,(4)\mathrm{F}}(N) \;=\;  } \nn \\[0.5mm] & & \mbox{}
%
%
%
%
%
%
-12\,\D(1)\,\*\S(3)
+12\,\D(1)\,\*\Ss(1,2)
+12\,\D(1)\,\*\Ss(2,1)
-12\,\D(1)\,\*\Sss(1,1,1)
+\S(2)\,(-12\,\D(0)
+50\,\D(1)
\nn \\[0.5mm] & & \mbox{}
-12\,\Dd(1,2))
+\Ss(1,1)\,(12\,\D(0)
-50\,\D(1)
+12\,\Dd(1,2))
+\,\S(1)\,(38\,\D(0)
-317/3\,\D(1)
-12\,\Dd(0,2)
\nn \\[0.5mm] & & \mbox{}
+50\,\Dd(1,2)
-12\,\Dd(1,3))
+203/3\,\D(0)
-8609/54\,\D(1)
-38\,\Dd(0,2)
+12\,\Dd(0,3)
+317/3\,\Dd(1,2)
\nn \\[0.5mm] & & \mbox{}
-50\,\Dd(1,3)
+12\,\Dd(1,4)

\:\: .
\eea

The corresponding result for ${\cal C}_2$ can be decomposed as
\bea
\label{c2Ndec}
  c_{2,\rm ns}^{\,(4)}(N) & \!=\! &
  \nfz \mbox{ and } \nfo \mbox{ contributions }
  \nn \\[1mm] & & \mbox{}
  + \cf \* \ca \,\* \nfs \*\, \frac{16}{9} \*\: 
    c_{2,\rm ns}^{\,(4)\mathrm{L}}(N)
  + \cf\, (\cf-\frct{1}{2}\,\* \ca) \,\* \nfs \*\, \frac{16}{9} \*\:
    c_{2,\rm ns}^{\,(4)\mathrm{N}}(N)
\nn \\[1mm] & & \mbox{}
  + \cf \,\* (\cf-\ca) \,\* \nfs \*\, \frac{1}{3} \*\, \z4 \*\, 
    c_{2,\rm ns}^{\,(4)\mathrm{Z}}(N)
  + \cf \* \nft \* \,\frac{16}{27} \*\: 
    c_{2,\rm ns}^{\,(4)\mathrm{F}}(N)
 \:\: .
\eea
In addition to the structures present in eq.~(\ref{cLNdec}), the $\nfs$
coefficient function for $F_2$ includes a $\z4$ contribution,
which is proportional to $(\cf\! - \ca)$ and hence vanishes vanish for 
$ \ca = \cf$  which is part of the choice of the colour factors that leads 
to a ${\cal N}\! =\! 1$ supersymmetric theory. 
The three $\nfs$ coefficients in eq.~(\ref{c2Ndec}) read 
\bea
\label{c2ns4L}
\lefteqn{ c_{2,\rm ns}^{\,(4)\mathrm{L}}(N)\;=\;  } \nn \\[0.5mm] & & \mbox{}
%
%
%
%
%
%
-1951/12\,\*\S(6)
+671/6\,\*\Ss(1,5)
+352/3\,\*\Ss(2,4)
+335/2\,\*\Ss(3,3)
+643/3\,\*\Ss(4,2)
+1445/6\,\*\Ss(5,1)
\nn \\[0.5mm] & & \mbox{}
-265/3\,\*\Sss(1,1,4)
-89\,\*\Sss(1,2,3)
-117\,\*\Sss(1,3,2)
-412/3\,\*\Sss(1,4,1)
-119\,\*\Sss(2,1,3)
-125\,\*\Sss(2,2,2)
\nn \\[0.5mm] & & \mbox{}
-137\,\*\Sss(2,3,1)
-169\,\*\Sss(3,1,2)
-188\,\*\Sss(3,2,1)
-688/3\,\*\Sss(4,1,1)
+71\,\*\Ssss(1,1,1,3)
+81\,\*\Ssss(1,1,2,2)
\nn \\[0.5mm] & & \mbox{}
+98\,\*\Ssss(1,1,3,1)
+103\,\*\Ssss(1,2,1,2)
+108\,\*\Ssss(1,2,2,1)
+135\,\*\Ssss(1,3,1,1)
+109\,\*\Ssss(2,1,1,2)
+122\,\*\Ssss(2,1,2,1)
\nn \\[0.5mm] & & \mbox{}
+139\,\*\Ssss(2,2,1,1)
+176\,\*\Ssss(3,1,1,1)
-70\,\*\Sssss(1,1,1,1,2)
-82\,\*\Sssss(1,1,1,2,1)
-94\,\*\Sssss(1,1,2,1,1)
-114\,\*\Sssss(1,2,1,1,1)
\nn \\[0.5mm] & & \mbox{}
-115\,\*\Sssss(2,1,1,1,1)
+80\,\*\Ssssss(1,1,1,1,1,1)
+\S(5)\,(20567/36
-671/12\,\D(0)
+671/12\,\D(1))
\nn \\[0.5mm] & & \mbox{}
+\Ss(1,4)\,(-9995/36
-20\,\D(-2)
+265/6\,\D(0)
-505/6\,\D(1)
+120\,\D(2)
-180\,\D(3))
\nn \\[0.5mm] & & \mbox{}
+\Ss(2,3)\,(-13373/36
-8\,\D(-2)
+89/2\,\D(0)
-121/2\,\D(1)
+48\,\D(2)
-72\,\D(3))
\nn \\[0.5mm] & & \mbox{}
+\Ss(3,2)\,(-9817/18
+8\,\D(-2)
+117/2\,\D(0)
-85/2\,\D(1)
-48\,\D(2)
+72\,\D(3))
\nn \\[0.5mm] & & \mbox{}
+\Ss(4,1)\,(-12343/18
+20\,\D(-2)
+206/3\,\D(0)
-86/3\,\D(1)
-120\,\D(2)
+180\,\D(3))
\nn \\[0.5mm] & & \mbox{}
+\Sss(1,1,3)\,(8495/36
+10\,\D(-2)
-71/2\,\D(0)
+111/2\,\D(1)
-60\,\D(2)
+90\,\D(3))
+\Sss(1,2,2)\,(803/3
\nn \\[0.5mm] & & \mbox{}
+2\,\D(-2)
-81/2\,\D(0)
+89/2\,\D(1)
-12\,\D(2)
+18\,\D(3))
+\Sss(1,3,1)\,(12263/36
-2\,\D(-2)
-49\,\D(0)
\nn \\[0.5mm] & & \mbox{}
+45\,\D(1)
+12\,\D(2)
-18\,\D(3))
+\Sss(2,1,2)\,(4283/12
+2\,\D(-2)
-103/2\,\D(0)
+111/2\,\D(1)
\nn \\[0.5mm] & & \mbox{}
-12\,\D(2)
+18\,\D(3))
+\Sss(2,2,1)\,(4835/12
-2\,\D(-2)
-54\,\D(0)
+50\,\D(1)
+12\,\D(2)
-18\,\D(3))
\nn \\[0.5mm] & & \mbox{}
+\Sss(3,1,1)\,(19967/36
-10\,\D(-2)
-135/2\,\D(0)
+95/2\,\D(1)
+60\,\D(2)
-90\,\D(3))
\nn \\[0.5mm] & & \mbox{}
+\Ssss(1,1,1,2)\,(-1279/6
+35\,\D(0)
-35\,\D(1))
+\Ssss(1,1,2,1)\,(-1511/6
+41\,\D(0)
-71\,\D(1)
+108\,\D(2)
\nn \\[0.5mm] & & \mbox{}
-90\,\D(3))
+\Ssss(1,2,1,1)\,(-605/2
+47\,\D(0)
-17\,\D(1)
-108\,\D(2)
+90\,\D(3))
+\Ssss(2,1,1,1)\,(-2135/6
\nn \\[0.5mm] & & \mbox{}
+57\,\D(0)
-57\,\D(1))
+\Sssss(1,1,1,1,1)\,(226
-40\,\D(0)
+40\,\D(1))
+\S(4)\,(-152383/144
+30\,\D(-2)
\nn \\[0.5mm] & & \mbox{}
+5929/36\,\D(0)
-1906/9\,\D(1)
-120\,\D(2)
+270\,\D(3)
-319/6\,\Dd(0,2)
+20\,\Dd(1,2))
\nn \\[0.5mm] & & \mbox{}
+\Ss(1,3)\,(3800/9
-13\,\D(-2)
+6\,\D(-1)
-4369/36\,\D(0)
+7735/36\,\D(1)
-36\,\D(2)
+18\,\D(3)
\nn \\[0.5mm] & & \mbox{}
-12\,\Dd(-2,2)
+54\,\Dd(0,2)
-71/2\,\Dd(1,2))
+\Ss(2,2)\,(24305/36
-3\,\D(-2)
-827/6\,\D(0)
\nn \\[0.5mm] & & \mbox{}
+1235/6\,\D(1)
+12\,\D(2)
-27\,\D(3)
+57\,\Dd(0,2)
-55/2\,\Dd(1,2))
+\Ss(3,1)\,(17689/18
+\D(-2)
\nn \\[0.5mm] & & \mbox{}
-6\,\D(-1)
-6559/36\,\D(0)
+8071/36\,\D(1)
+84\,\D(2)
-126\,\D(3)
+12\,\Dd(-2,2)
+63\,\Dd(0,2)
\nn \\[0.5mm] & & \mbox{}
-19\,\Dd(1,2))
+\Sss(1,1,2)\,(-27307/72
+1237/12\,\D(0)
-4073/24\,\D(1)
+24\,\D(2)
-12\,\D(3)
\nn \\[0.5mm] & & \mbox{}
-49\,\Dd(0,2)
+59/2\,\Dd(1,2))
+\Sss(1,2,1)\,(-10459/24
+805/6\,\D(0)
-593/3\,\D(1)
-108\,\D(2)
\nn \\[0.5mm] & & \mbox{}
+135\,\D(3)
-111/2\,\Dd(0,2)
+51/2\,\Dd(1,2))
+\Sss(2,1,1)\,(-23191/36
+613/4\,\D(0)
-1837/8\,\D(1)
\nn \\[0.5mm] & & \mbox{}
+84\,\D(2)
-123\,\D(3)
-64\,\Dd(0,2)
+47/2\,\Dd(1,2))
+\Ssss(1,1,1,1)\,(25979/72
-231/2\,\D(0)
\nn \\[0.5mm] & & \mbox{}
+357/2\,\D(1)
+52\,\Dd(0,2)
-23\,\Dd(1,2))
+\S(3)\,(842039/648
-3\,\D(-2)
-6\,\D(-1)
-18863/72\,\D(0)
\nn \\[0.5mm] & & \mbox{}
+11233/24\,\D(1)
+84\,\D(2)
-162\,\D(3)
+18\,\Dd(-2,2)
+3107/18\,\Dd(0,2)
-283/4\,\Dd(0,3)
\nn \\[0.5mm] & & \mbox{}
-421/4\,\Dd(1,2)
+91/4\,\Dd(1,3))
+\Ss(1,2)\,(-147071/324
-5/2\,\D(-2)
+2005/9\,\D(0)
\nn \\[0.5mm] & & \mbox{}
-34157/72\,\D(1)
+84\,\D(2)
-58\,\D(3)
-1969/12\,\Dd(0,2)
+143/2\,\Dd(0,3)
+134\,\Dd(1,2)
-24\,\Dd(1,3))
\nn \\[0.5mm] & & \mbox{}
+\Ss(2,1)\,(-1178369/1296
+5/2\,\D(-2)
+557/2\,\D(0)
-11939/24\,\D(1)
-163/2\,\D(3)
\nn \\[0.5mm] & & \mbox{}
-1127/6\,\Dd(0,2)
+81\,\Dd(0,3)
+323/3\,\Dd(1,2)
-49/2\,\Dd(1,3))
+\Sss(1,1,1)\,(46483/108
\nn \\[0.5mm] & & \mbox{}
-8321/36\,\D(0)
+32905/72\,\D(1)
+2009/12\,\Dd(0,2)
-75\,\Dd(0,3)
-1331/12\,\Dd(1,2)
+21\,\Dd(1,3))
\nn \\[0.5mm] & & \mbox{}
+\S(2)\,(-5764837/5184
-15/2\,\D(-2)
+1006649/2592\,\D(0)
-2225699/2592\,\D(1)
-24\,\D(2)
\nn \\[0.5mm] & & \mbox{}
-33/4\,\D(3)
-12665/36\,\Dd(0,2)
+16325/72\,\Dd(0,3)
-1025/12\,\Dd(0,4)
+875/3\,\Dd(1,2)
\nn \\[0.5mm] & & \mbox{}
-7073/72\,\Dd(1,3)
+263/12\,\Dd(1,4))
+\Ss(1,1)\,(2134163/5184
+45/4\,\D(-2)
-356983/864\,\D(0)
\nn \\[0.5mm] & & \mbox{}
+780301/864\,\D(1)
+24\,\D(2)
-15/2\,\D(3)
+1090/3\,\Dd(0,2)
-17063/72\,\Dd(0,3)
+1115/12\,\Dd(0,4)
\nn \\[0.5mm] & & \mbox{}
-7363/24\,\Dd(1,2)
+6887/72\,\Dd(1,3)
-239/12\,\Dd(1,4))
+\S(1)\,(33182/81
+27/2\,\D(-2)
+6\,\D(-1)
\nn \\[0.5mm] & & \mbox{}
-1115063/1728\,\D(0)
+243035/162\,\D(1)
-228\,\D(2)
+2431/8\,\D(3)
-27/2\,\Dd(-2,2)
\nn \\[0.5mm] & & \mbox{}
+275219/432\,\Dd(0,2)
-22763/48\,\Dd(0,3)
+19123/72\,\Dd(0,4)
-1069/12\,\Dd(0,5)
\nn \\[0.5mm] & & \mbox{}
-1669825/2592\,\Dd(1,2)
+34919/144\,\Dd(1,3)
-604/9\,\Dd(1,4)
+32/3\,\Dd(1,5))
-18199451/27648
\nn \\[0.5mm] & & \mbox{}
-33/4\,\D(-2)
-2362801/2304\,\D(0)
+5233867/2304\,\D(1)
-935/8\,\D(3)
+9/2\,\Dd(-2,2)
\nn \\[0.5mm] & & \mbox{}
+9889087/10368\,\Dd(0,2)
-1874987/2592\,\Dd(0,3)
+63839/144\,\Dd(0,4)
-14161/72\,\Dd(0,5)
\nn \\[0.5mm] & & \mbox{}
+1951/48\,\Dd(0,6)
-3790549/3456\,\Dd(1,2)
+367649/864\,\Dd(1,3)
-12191/144\,\Dd(1,4)
\nn \\[0.5mm] & & \mbox{}
-41/3\,\Dd(1,5)
+149/16\,\Dd(1,6)
\nn \\[0.5mm] & & \mbox{\hspn}
     + \z3 \,\* \Big[ 
%
%
%
%
%
%
-331/6\,\*\S(3)
+121/3\,\*\Ss(1,2)
+166/3\,\*\Ss(2,1)
-14\,\*\Sss(1,1,1)
+\S(2)\,(215/2
+28\,\D(-2)
\nn \\[0.5mm] & & \mbox{}
-121/6\,\D(0)
+457/6\,\D(1)
-168\,\D(2)
+252\,\D(3))
+\Ss(1,1)\,(-111/2
-20\,\D(-2)
+7\,\D(0)
\nn \\[0.5mm] & & \mbox{}
-312\,\D(2)
+180\,\D(3))
+\S(1)\,(-274/3
+26\,\D(-2)
-12\,\D(-1)
-29/3\,\D(0)
-497/4\,\D(1)
\nn \\[0.5mm] & & \mbox{}
+73\,\D(1)
+552\,\D(2)
-600\,\D(3)
+24\,\Dd(-2,2)
-25\,\Dd(0,2)
+133/3\,\Dd(1,2))
+2083/32
-9\,\D(-2)
\nn \\[0.5mm] & & \mbox{}
+12\,\D(-1)
-101/24\,\D(0)
-6115/24\,\D(1)
+534\,\D(3)
-36\,\Dd(-2,2)
-122/3\,\Dd(0,2)
\nn \\[0.5mm] & & \mbox{}
+257/12\,\Dd(0,3)
+401/3\,\Dd(1,2)
-33/4\,\Dd(1,3)
\Big]
\\[0.5mm] & & \mbox{\hspn}
     + \z5 \,\* \Big[ 
%
%
%
%
%
%
-191/2\,\*\S(1)
+693/8
-60\,\D(-2)
+191/4\,\D(0)
+1729/4\,\D(1)
-1800\,\D(2)
+1260\,\D(3)
\Big]
\nn

\:\: , \\[3mm]
\label{c2ns4N}
\lefteqn{c_{2,\rm ns}^{\,(4)\mathrm{N}}(N) \;=\;  } \nn \\[0.5mm] & & \mbox{}
%
%
%
%
%
%
-150\,\*\S(-6)
-1051/6\,\*\S(6)
+6\,\*\Ss(-5,1)
+408\,\*\Ss(-4,-2)
+510\,\*\Ss(-3,-3)
+352\,\*\Ss(-2,-4)
\nn \\[0.5mm] & & \mbox{}
+450\,\*\Ss(1,-5)
+446/3\,\*\Ss(1,5)
+538\,\*\Ss(2,-4)
+548/3\,\*\Ss(2,4)
+586\,\*\Ss(3,-3)
+273\,\*\Ss(3,3)
\nn \\[0.5mm] & & \mbox{}
+434\,\*\Ss(4,-2)
+968/3\,\*\Ss(4,2)
+932/3\,\*\Ss(5,1)
+8\,\*\Sss(-4,1,1)
-264\,\*\Sss(-3,-2,1)
+4\,\*\Sss(-3,1,-2)
\nn \\[0.5mm] & & \mbox{}
-216\,\*\Sss(-2,-3,1)
-136\,\*\Sss(-2,-2,2)
-4\,\*\Sss(1,-4,1)
-220\,\*\Sss(1,-3,-2)
-252\,\*\Sss(1,-2,-3)
-520\,\*\Sss(1,1,-4)
\nn \\[0.5mm] & & \mbox{}
-770/3\,\*\Sss(1,1,4)
-496\,\*\Sss(1,2,-3)
-302\,\*\Sss(1,2,3)
-336\,\*\Sss(1,3,-2)
-160\,\*\Sss(1,3,2)
+4/3\,\*\Sss(1,4,1)
\nn \\[0.5mm] & & \mbox{}
-4\,\*\Sss(2,-3,1)
-140\,\*\Sss(2,-2,-2)
-496\,\*\Sss(2,1,-3)
-326\,\*\Sss(2,1,3)
-320\,\*\Sss(2,2,-2)
-314\,\*\Sss(2,2,2)
\nn \\[0.5mm] & & \mbox{}
-64\,\*\Sss(2,3,1)
-4\,\*\Sss(3,-2,1)
-344\,\*\Sss(3,1,-2)
-344\,\*\Sss(3,1,2)
-276\,\*\Sss(3,2,1)
-1016/3\,\*\Sss(4,1,1)
\nn \\[0.5mm] & & \mbox{}
+8\,\*\Ssss(-3,1,1,1)
+112\,\*\Ssss(-2,-2,1,1)
-8\,\*\Ssss(1,-3,1,1)
+144\,\*\Ssss(1,-2,-2,1)
-8\,\*\Ssss(1,-2,1,-2)
+112\,\*\Ssss(1,1,-2,-2)
\qquad \nn \\[0.5mm] & & \mbox{}
+448\,\*\Ssss(1,1,1,-3)
+274\,\*\Ssss(1,1,1,3)
+288\,\*\Ssss(1,1,2,-2)
+252\,\*\Ssss(1,1,2,2)
+58\,\*\Ssss(1,1,3,1)
+288\,\*\Ssss(1,2,1,-2)
\nn \\[0.5mm] & & \mbox{}
+254\,\*\Ssss(1,2,1,2)
+188\,\*\Ssss(1,2,2,1)
+88\,\*\Ssss(1,3,1,1)
-8\,\*\Ssss(2,-2,1,1)
+288\,\*\Ssss(2,1,1,-2)
+314\,\*\Ssss(2,1,1,2)
\nn \\[0.5mm] & & \mbox{}
+210\,\*\Ssss(2,1,2,1)
+216\,\*\Ssss(2,2,1,1)
+302\,\*\Ssss(3,1,1,1)
-16\,\*\Sssss(1,-2,1,1,1)
-256\,\*\Sssss(1,1,1,1,-2)
-244\,\*\Sssss(1,1,1,1,2)
\nn \\[0.5mm] & & \mbox{}
-148\,\*\Sssss(1,1,1,2,1)
-144\,\*\Sssss(1,1,2,1,1)
-184\,\*\Sssss(1,2,1,1,1)
-230\,\*\Sssss(2,1,1,1,1)
+160\,\*\Ssssss(1,1,1,1,1,1)
\nn \\[0.5mm] & & \mbox{}
+\S(-5)\,(310/3
-297\,\D(0)
+585\,\D(1))
+\S(5)\,(3463/9
-7/3\,\D(0)
-857/3\,\D(1))
\nn \\[0.5mm] & & \mbox{}
+\Ss(-4,1)\,(-40/3
+2\,\D(0)
-2\,\D(1))
+\Ss(-3,-2)\,(-1192/3
+174\,\D(0)
-430\,\D(1))
\nn \\[0.5mm] & & \mbox{}
+\Ss(-2,-3)\,(-392
+198\,\D(0)
-486\,\D(1))
+\Ss(1,-4)\,(-448
+404\,\D(0)
-980\,\D(1))
\nn \\[0.5mm] & & \mbox{}
+\Ss(1,4)\,(-6395/18
+385/3\,\D(0)
-385/3\,\D(1))
+\Ss(2,-3)\,(-460
+388\,\D(0)
-948\,\D(1))
\nn \\[0.5mm] & & \mbox{}
+\Ss(2,3)\,(-6125/18
+169\,\D(0)
-241\,\D(1))
+\Ss(3,-2)\,(-376
+260\,\D(0)
-628\,\D(1))
\nn \\[0.5mm] & & \mbox{}
+\Ss(3,2)\,(-5683/9
+52\,\D(0)
+60\,\D(1))
+\Ss(4,1)\,(-7129/9
-260/3\,\D(0)
+1292/3\,\D(1))
\nn \\[0.5mm] & & \mbox{}
+\Sss(-3,1,1)\,(-40/3
+4\,\D(0)
-4\,\D(1))
+\Sss(-2,-2,1)\,(160
-120\,\D(0)
+312\,\D(1))
\nn \\[0.5mm] & & \mbox{}
+\Sss(-2,1,-2)\,(4\,\D(0)
-4\,\D(1))
+40/3\,\*\Sss(1,-3,1)
+\Sss(1,-2,-2)\,(512/3
-88\,\D(0)
+216\,\D(1))
\nn \\[0.5mm] & & \mbox{}
+\Sss(1,1,-3)\,(1360/3
-360\,\D(0)
+904\,\D(1))
+\Sss(1,1,3)\,(7679/18
-159\,\D(0)
+247\,\D(1))
\nn \\[0.5mm] & & \mbox{}
+\Sss(1,2,-2)\,(880/3
-232\,\D(0)
+584\,\D(1))
+\Sss(1,2,2)\,(1417/3
-142\,\D(0)
+206\,\D(1))
\nn \\[0.5mm] & & \mbox{}
+\Sss(1,3,1)\,(5885/18
+17\,\D(0)
-201\,\D(1))
+40/3\,\*\Sss(2,-2,1)
+\Sss(2,1,-2)\,(880/3
-232\,\D(0)
\nn \\[0.5mm] & & \mbox{}
+584\,\D(1))
+\Sss(2,1,2)\,(1027/2
-141\,\D(0)
+197\,\D(1))
+\Sss(2,2,1)\,(1250/3
-90\,\D(0)
+74\,\D(1))
\nn \\[0.5mm] & & \mbox{}
+\Sss(3,1,1)\,(6313/9
-2\,\D(0)
-166\,\D(1))
+\Ssss(-2,1,1,1)\,(8\,\D(0)
-8\,\D(1))
+80/3\,\*\Ssss(1,-2,1,1)
\nn \\[0.5mm] & & \mbox{}
+\Ssss(1,1,1,-2)\,(-800/3
+208\,\D(0)
-528\,\D(1))
+\Ssss(1,1,1,2)\,(-1163/3
+140\,\D(0)
-212\,\D(1))
\nn \\[0.5mm] & & \mbox{}
+\Ssss(1,1,2,1)\,(-1015/3
+66\,\D(0)
-34\,\D(1))
+\Ssss(1,2,1,1)\,(-413
+62\,\D(0)
-22\,\D(1))
\nn \\[0.5mm] & & \mbox{}
+\Ssss(2,1,1,1)\,(-1343/3
+92\,\D(0)
-92\,\D(1))
+\Sssss(1,1,1,1,1)\,(320
-80\,\D(0)
+80\,\D(1))
\nn \\[0.5mm] & & \mbox{}
+\S(-4)\,(-1270/9
+144/5\,\D(-2)
+1094\,\D(0)
-1182\,\D(1)
+1296/5\,\D(3)
-625\,\Dd(0,2)
\nn \\[0.5mm] & & \mbox{}
-707\,\Dd(1,2))
+\S(4)\,(-18371/72
+1105/18\,\D(0)
+2344/9\,\D(1)
-133/3\,\Dd(0,2)
-206\,\Dd(1,2))
\nn \\[0.5mm] & & \mbox{}
+\Ss(-3,1)\,(38/9
-56/5\,\D(-2)
-692/3\,\D(0)
+20/3\,\D(1)
+336\,\D(2)
-504/5\,\D(3)
+166\,\Dd(0,2)
\nn \\[0.5mm] & & \mbox{}
+386\,\Dd(1,2))
+\Ss(-2,-2)\,(200
-16/5\,\D(-2)
-796/3\,\D(0)
+1612/3\,\D(1)
-96\,\D(2)
-144/5\,\D(3)
\nn \\[0.5mm] & & \mbox{}
+110\,\Dd(0,2)
-54\,\Dd(1,2))
+\Ss(-2,2)\,(-36/5\,\D(-2)
-140\,\D(0)
-4\,\D(1)
+216\,\D(2)
-324/5\,\D(3)
\nn \\[0.5mm] & & \mbox{}
+104\,\Dd(0,2)
+248\,\Dd(1,2))
+\Ss(1,-3)\,(2338/9
-68/5\,\D(-2)
-2624/3\,\D(0)
+4928/3\,\D(1)
\nn \\[0.5mm] & & \mbox{}
-408\,\D(2)
-612/5\,\D(3)
+388\,\Dd(0,2)
-44\,\Dd(1,2))
+\Ss(1,3)\,(4369/9
-4/5\,\D(-2)
-5005/18\,\D(0)
\nn \\[0.5mm] & & \mbox{}
+3524/9\,\D(1)
-108\,\D(2)
-1086/5\,\D(3)
+156\,\Dd(0,2)
+53\,\Dd(1,2))
+\Ss(2,-2)\,(1522/9
\nn \\[0.5mm] & & \mbox{}
-44/5\,\D(-2)
-1712/3\,\D(0)
+3248/3\,\D(1)
-264\,\D(2)
-396/5\,\D(3)
+252\,\Dd(0,2)
-36\,\Dd(1,2))
\nn \\[0.5mm] & & \mbox{}
+\Ss(2,2)\,(12331/36
-8/5\,\D(-2)
-920/3\,\D(0)
+1541/3\,\D(1)
-48\,\D(2)
-72/5\,\D(3)
+160\,\Dd(0,2)
\nn \\[0.5mm] & & \mbox{}
-32\,\Dd(1,2))
+\Ss(3,1)\,(23429/36
+16/5\,\D(-2)
+1067/18\,\D(0)
-2239/9\,\D(1)
+180\,\D(2)
\nn \\[0.5mm] & & \mbox{}
+1194/5\,\D(3)
-25\,\Dd(0,2)
-14\,\Dd(1,2))
+\Sss(-2,1,1)\,(32/5\,\D(-2)
+296/3\,\D(0)
+88/3\,\D(1)
\nn \\[0.5mm] & & \mbox{}
-192\,\D(2)
+288/5\,\D(3)
-84\,\Dd(0,2)
-220\,\Dd(1,2))
-76/9\,\*\Sss(1,-2,1)
+\Sss(1,1,-2)\,(-496/3
+8\,\D(-2)
\nn \\[0.5mm] & & \mbox{}
+1600/3\,\D(0)
-3136/3\,\D(1)
+240\,\D(2)
+72\,\D(3)
-232\,\Dd(0,2)
+56\,\Dd(1,2))
\nn \\[0.5mm] & & \mbox{}
+\Sss(1,1,2)\,(-15511/36
+4/5\,\D(-2)
+799/3\,\D(0)
-1372/3\,\D(1)
+72\,\D(2)
+816/5\,\D(3)
\nn \\[0.5mm] & & \mbox{}
-156\,\Dd(0,2)
+6\,\Dd(1,2))
+\Sss(1,2,1)\,(-1839/4
-4/5\,\D(-2)
+398/3\,\D(0)
-593/3\,\D(1)
-24\,\D(2)
\nn \\[0.5mm] & & \mbox{}
-36/5\,\D(3)
-84\,\Dd(0,2)
+32\,\Dd(1,2))
+\Sss(2,1,1)\,(-3491/9
+184\,\D(0)
-341\,\D(1)
-48\,\D(2)
\nn \\[0.5mm] & & \mbox{}
-156\,\D(3)
-97\,\Dd(0,2)
+97\,\Dd(1,2))
+\Ssss(1,1,1,1)\,(12119/36
-165\,\D(0)
+291\,\D(1)
+104\,\Dd(0,2)
\nn \\[0.5mm] & & \mbox{}
-46\,\Dd(1,2))
+\S(-3)\,(4346/27
-214/25\,\D(-2)
-16850/9\,\D(0)
+11870/9\,\D(1)
+408\,\D(2)
\nn \\[0.5mm] & & \mbox{}
-4086/25\,\D(3)
+144/5\,\Dd(-2,2)
+1533\,\Dd(0,2)
-776\,\Dd(0,3)
+3839/3\,\Dd(1,2)
-408\,\Dd(1,3))
\nn \\[0.5mm] & & \mbox{}
+\S(3)\,(-103447/1296
+6/5\,\D(-2)
-3599/36\,\D(0)
-1963/12\,\D(1)
+108\,\D(2)
+1629/5\,\D(3)
\nn \\[0.5mm] & & \mbox{}
+2131/18\,\Dd(0,2)
-197/2\,\Dd(0,3)
+530/3\,\Dd(1,2)
-171/2\,\Dd(1,3))
+\Ss(-2,1)\,(248/25\,\D(-2)
\nn \\[0.5mm] & & \mbox{}
+24\,\D(-1)
+4208/9\,\D(0)
-608/9\,\D(1)
-552\,\D(2)
+2952/25\,\D(3)
-48/5\,\Dd(-2,2)
\nn \\[0.5mm] & & \mbox{}
-1220/3\,\Dd(0,2)
+206\,\Dd(0,3)
-1636/3\,\Dd(1,2)
+178\,\Dd(1,3))
+\Ss(1,-2)\,(-2428/27
-52/25\,\D(-2)
\nn \\[0.5mm] & & \mbox{}
-24\,\D(-1)
+2420/3\,\D(0)
-1108\,\D(1)
+312\,\D(2)
+252/25\,\D(3)
-48/5\,\Dd(-2,2)
-1766/3\,\Dd(0,2)
\nn \\[0.5mm] & & \mbox{}
+266\,\Dd(0,3)
+106/3\,\Dd(1,2)
-10\,\Dd(1,3))
+\Ss(1,2)\,(-110893/324
-6/5\,\D(-2)
+4537/18\,\D(0)
\nn \\[0.5mm] & & \mbox{}
-11255/36\,\D(1)
+150\,\D(2)
-4/5\,\D(3)
-237\,\Dd(0,2)
+146\,\Dd(0,3)
+475/6\,\Dd(1,2)
+5\,\Dd(1,3))
\nn \\[0.5mm] & & \mbox{}
+\Ss(2,1)\,(-81845/648
+6/5\,\D(-2)
+146\,\D(0)
-3817/12\,\D(1)
-198\,\D(2)
-1166/5\,\D(3)
\nn \\[0.5mm] & & \mbox{}
-935/6\,\Dd(0,2)
+112\,\Dd(0,3)
+1411/6\,\Dd(1,2)
-58\,\Dd(1,3))
+\Sss(1,1,1)\,(26339/108
-2957/18\,\D(0)
\nn \\[0.5mm] & & \mbox{}
+11089/36\,\D(1)
+1091/6\,\Dd(0,2)
-129\,\Dd(0,3)
-1061/6\,\Dd(1,2)
+43\,\Dd(1,3))
+\S(-2)\,(-232/3
\nn \\[0.5mm] & & \mbox{}
+23/375\,\D(-2)
+24\,\D(-1)
+88627/54\,\D(0)
-58711/54\,\D(1)
-480\,\D(2)
+5409/125\,\D(3)
\nn \\[0.5mm] & & \mbox{}
-116/25\,\Dd(-2,2)
+96/5\,\Dd(-2,3)
-27859/18\,\Dd(0,2)
+3611/3\,\Dd(0,3)
-589\,\Dd(0,4)
\nn \\[0.5mm] & & \mbox{}
-17027/18\,\Dd(1,2)
+1609/3\,\Dd(1,3)
-177\,\Dd(1,4))
+\S(2)\,(162721/1296
-3\,\D(-2)
\nn \\[0.5mm] & & \mbox{}
-66817/1296\,\D(0)
+566587/1296\,\D(1)
-414\,\D(2)
-147\,\D(3)
-4849/36\,\Dd(0,2)
\nn \\[0.5mm] & & \mbox{}
+6167/36\,\Dd(0,3)
-707/6\,\Dd(0,4)
-1307/12\,\Dd(1,2)
-1355/36\,\Dd(1,3)
-61/6\,\Dd(1,4))
\nn \\[0.5mm] & & \mbox{}
+\Ss(1,1)\,(239633/2592
+24/5\,\D(-2)
+26327/432\,\D(0)
-71081/432\,\D(1)
+144\,\D(2)
\nn \\[0.5mm] & & \mbox{}
-759/5\,\D(3)
+385/3\,\Dd(0,2)
-5999/36\,\Dd(0,3)
+731/6\,\Dd(0,4)
-1475/12\,\Dd(1,2)
\nn \\[0.5mm] & & \mbox{}
+5471/36\,\Dd(1,3)
-113/6\,\Dd(1,4))
+\S(1)\,(161929/1728
+186/25\,\D(-2)
-24\,\D(-1)
\nn \\[0.5mm] & & \mbox{}
+33065/96\,\D(0)
-1073755/1296\,\D(1)
+366\,\D(2)
-21269/100\,\D(3)
-36/5\,\Dd(-2,2)
\nn \\[0.5mm] & & \mbox{}
-1819/216\,\Dd(0,2)
-687/8\,\Dd(0,3)
+4237/36\,\Dd(0,4)
-287/3\,\Dd(0,5)
+493793/1296\,\Dd(1,2)
\nn \\[0.5mm] & & \mbox{}
+2093/72\,\Dd(1,3)
-1285/18\,\Dd(1,4)
-43/6\,\Dd(1,5))
-61555/512
+1807/125\,\D(-2)
\nn \\[0.5mm] & & \mbox{}
-269059/1296\,\D(0)
-810827/5184\,\D(1)
+153647/500\,\D(3)
-507/25\,\Dd(-2,2)
\nn \\[0.5mm] & & \mbox{}
+72/5\,\Dd(-2,3)
+460601/2592\,\Dd(0,2)
-271195/2592\,\Dd(0,3)
+3533/72\,\Dd(0,4)
-595/9\,\Dd(0,5)
\nn \\[0.5mm] & & \mbox{}
+1951/24\,\Dd(0,6)
+57113/96\,\Dd(1,2)
-380303/864\,\Dd(1,3)
+17011/72\,\Dd(1,4)
-1087/12\,\Dd(1,5)
\nn \\[0.5mm] & & \mbox{}
+149/8\,\Dd(1,6)
\nn \\[0.5mm] & & \mbox{\hspn}
     + \z3 \,\* \Big[ 
%
%
%
%
%
%
 436\,\*\S(-3)
+1061/3\,\*\S(3)
-200\,\*\Ss(1,-2)
-496/3\,\*\Ss(1,2)
-100/3\,\*\Ss(2,1)
+44\,\*\Sss(1,1,1)
\nn \\[0.5mm] & & \mbox{}
+\S(-2)\,(-388
+148\,\D(0)
-340\,\D(1))
+\S(2)\,(-28
+572/3\,\D(0)
-1868/3\,\D(1))
\nn \\[0.5mm] & & \mbox{}
+\Ss(1,1)\,(35
-94\,\D(0)
+382\,\D(1))
+\S(1)\,(-397/6
-12\,\D(-2)
-1396/3\,\D(0)
+1005\,\D(1)
\nn \\[0.5mm] & & \mbox{}
-96\,\D(2)
+624\,\D(3)
+142\,\Dd(0,2)
-688/3\,\Dd(1,2))
-15883/16
-264/25\,\D(-2)
+24\,\D(-1)
\nn \\[0.5mm] & & \mbox{}
-19655/12\,\D(0)
+22715/12\,\D(1)
+876\,\D(2)
+4614/25\,\D(3)
+144/5\,\Dd(-2,2)
+3947/3\,\Dd(0,2)
\nn \\[0.5mm] & & \mbox{}
-3955/6\,\Dd(0,3)
+964/3\,\Dd(1,2)
-51/2\,\Dd(1,3)
\nn \\[0.5mm] & & \mbox{\hspn}
     + \z5 \,\* \Big[ 
%
%
%
%
%
%
 81\,\*\S(1)
-513/4
-171/2\,\D(0)
+531/2\,\D(1)
\Big]

\eea
and
\bea
\label{c2ns4Z}
 c_{2,\rm ns}^{\,(4)\mathrm{Z}}(N) \;=\;  
%
%
%
%
%
%
 24\,\*\S(2)
-80\,\*\S(1)
+33
+64\,\D(0)
-64\,\D(1)
-12\,\Dd(0,2)
+12\,\Dd(1,2)

\; .
\eea
Unlike the above expressions for ${\cal C}_L$, eqs.~(\ref{c2ns4L}) and
(\ref{c2ns4N}) include harmonic sums (up to $w=6$) without prefactors 
$D_a$; these are the terms that do not vanish in the soft-gluon limit
$N \!\ra \infty$. As in eq.~(\ref{cLns4L}), only non-alternating sums
occur in the large-$\nc$ result (\ref{c2ns4L}), and additional denominators 
(here in addition to $D_0$ and $D_1$) occur with $w=5$ sums in the form
$$
  D_{-2} -  6D_2 + 9 D_3 \:\: .
$$
Again as for ${\cal C}_L$, terms with $D_a$, $a \neq 0,\,1$, are not
present with $w=5$ sums in the $\cfs\, \nfs$ coefficient (\ref{c2ns4N}).
The situation at $N=2$ is the same as that for ${\cal C}_L$ above, with
\bea
\label{c2ns4N2}
 c_{2,\rm ns}^{\,(4)\mathrm{L}}(N\!=\!2) & \!=\! &
    \frac{1163533}{5832}
  - \frac{4613}{540} \:\*\z3
  - \frac{290}{3} \:\*\z5
  + 6\,\*\zts
\:\: ,
\\[1.5mm]
 c_{2,\rm ns}^{\,(4)\mathrm{N}}(N\!=\!2) & \!=\! &
    \frac{1720051}{29160}
  + \frac{247}{270} \:\*\z3
  + 30\,\*\z5
\:\: .
\eea

\noindent 
Finally the $\nft$ coefficient in eq.~(\ref{c2Ndec}) is given by
\bea
\label{c2ns4F}
\lefteqn{c_{2,\rm ns}^{\,(4)\mathrm{F}}(N) \;=\;  } \nn \\[0.5mm] & & \mbox{}
%
%
%
%
%
%
-119/2\,\*\S(5)
+12\,\*\Ss(1,4)
+24\,\*\Ss(2,3)
+36\,\*\Ss(3,2)
+48\,\*\Ss(4,1)
-12\,\*\Sss(1,1,3)
-12\,\*\Sss(1,2,2)
-12\,\*\Sss(1,3,1)
\nn \\[0.5mm] & & \mbox{}
-24\,\*\Sss(2,1,2)
-24\,\*\Sss(2,2,1)
-36\,\*\Sss(3,1,1)
+12\,\*\Ssss(1,1,1,2)
+12\,\*\Ssss(1,1,2,1)
+12\,\*\Ssss(1,2,1,1)
+24\,\*\Ssss(2,1,1,1)
\nn \\[0.5mm] & & \mbox{}
-12\,\*\Sssss(1,1,1,1,1)
+\S(4)\,(853/6
-6\,\D(0)
+6\,\D(1))
+\Ss(1,3)\,(-29
+6\,\D(0)
-6\,\D(1))
\nn \\[0.5mm] & & \mbox{}
+\Ss(2,2)\,(-67
+6\,\D(0)
-6\,\D(1))
+\Ss(3,1)\,(-105
+6\,\D(0)
-6\,\D(1))
+\Sss(1,1,2)\,(29
-6\,\D(0)
+6\,\D(1))
\qquad \nn \\[0.5mm] & & \mbox{}
+\Sss(1,2,1)\,(29
-6\,\D(0)
+6\,\D(1))
+\Sss(2,1,1)\,(67
-6\,\D(0)
+6\,\D(1))
+\Ssss(1,1,1,1)\,(-29
+6\,\D(0)
\nn \\[0.5mm] & & \mbox{}
-6\,\D(1))
+\S(3)\,(-524/3
+13\,\D(0)
-34\,\D(1)
-12\,\Dd(0,2)
+6\,\Dd(1,2))
+\Ss(1,2)\,(235/6
-13\,\D(0)
\nn \\[0.5mm] & & \mbox{}
+34\,\D(1)
+12\,\Dd(0,2)
-6\,\Dd(1,2))
+\Ss(2,1)\,(641/6
-13\,\D(0)
+34\,\D(1)
+12\,\Dd(0,2)
-6\,\Dd(1,2))
\nn \\[0.5mm] & & \mbox{}
+\Sss(1,1,1)\,(-235/6
+13\,\D(0)
-34\,\D(1)
-12\,\Dd(0,2)
+6\,\Dd(1,2))
+\S(2)\,(14321/108
-161/6\,\D(0)
\nn \\[0.5mm] & & \mbox{}
+280/3\,\D(1)
+29\,\Dd(0,2)
-18\,\Dd(0,3)
-34\,\Dd(1,2)
+6\,\Dd(1,3))
+\Ss(1,1)\,(-4429/108
+161/6\,\D(0)
\nn \\[0.5mm] & & \mbox{}
-280/3\,\D(1)
-29\,\Dd(0,2)
+18\,\Dd(0,3)
+34\,\Dd(1,2)
-6\,\Dd(1,3))
+\,\S(1)\,(-25279/648
+5729/108\,\D(0)
\nn \\[0.5mm] & & \mbox{}
-9259/54\,\D(1)
-325/6\,\Dd(0,2)
+45\,\Dd(0,3)
-24\,\Dd(0,4)
+280/3\,\Dd(1,2)
-34\,\Dd(1,3)
+6\,\Dd(1,4))
\nn \\[0.5mm] & & \mbox{}
+281971/3456
+55157/648\,\D(0)
-39803/162\,\D(1)
-9847/108\,\Dd(0,2)
+161/2\,\Dd(0,3)
\nn \\[0.5mm] & & \mbox{}
-721/12\,\Dd(0,4)
+119/4\,\Dd(0,5)
+9241/54\,\Dd(1,2)
-277/3\,\Dd(1,3)
+397/12\,\Dd(1,4)
-23/4\,\Dd(1,5)
\nn \\[0.5mm] & & \mbox{\hspn}
     + \z3 \,\* \Big[ 
%
%
%
%
%
%
\S(2)
-5/3\,\*\,\S(1)
+1/8
+11/6\,\D(0)
-11/6\,\D(1)
-1/2\,\Dd(0,2)
+1/2\,\Dd(1,2)
\Big]
\nn \\[0.5mm] & & \mbox{\hspn}
     + \z4 \,\* \Big[ 
%
%
%
%
%
%
3/2\,\*\,\S(1)
-9/8
-3/4\,\D(0)
+3/4\,\D(1)
\Big]

\:\: .
\eea

\begin{figure}[p]
\vspace*{-1mm}
\centerline{\epsfig{file=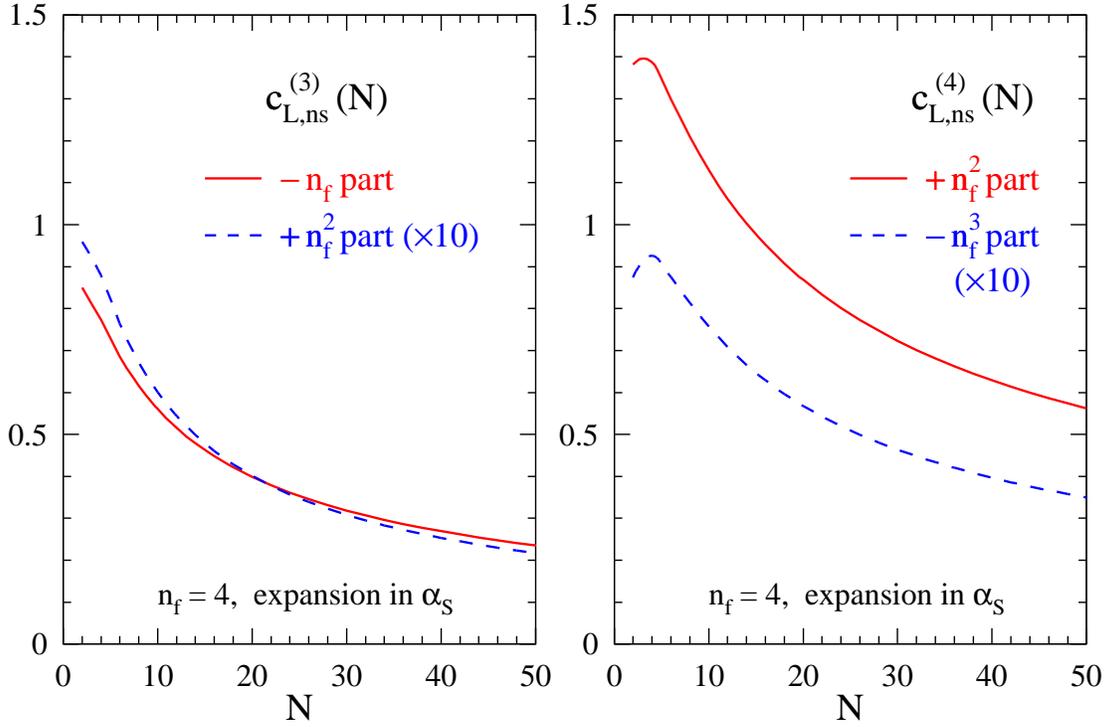,width=15.0cm,angle=0}\quad}
\vspace{-3mm}
\caption{ \label{RFig1} \small
 The leading (dashed) and sub-leading (solid) large-$\nf$ contributions
 to the three-loop (left panel) and four-loop (right panel) coefficient
 functions for the structure function $F_{L,\rm ns}$ for QCD with
 $\nf = 4$ flavours. The results in eqs.~(\ref{cLNdec}) -- (\ref{c2ns4F})
 have been converted to an expansion in $\als$, and the leading
 large-$\nf$ curves have been scaled up by a factor of 10 for better
 visibility. }
\vspace{-2mm}

\end{figure}

\begin{figure}[p]
\vspace*{-2mm}
\centerline{\epsfig{file=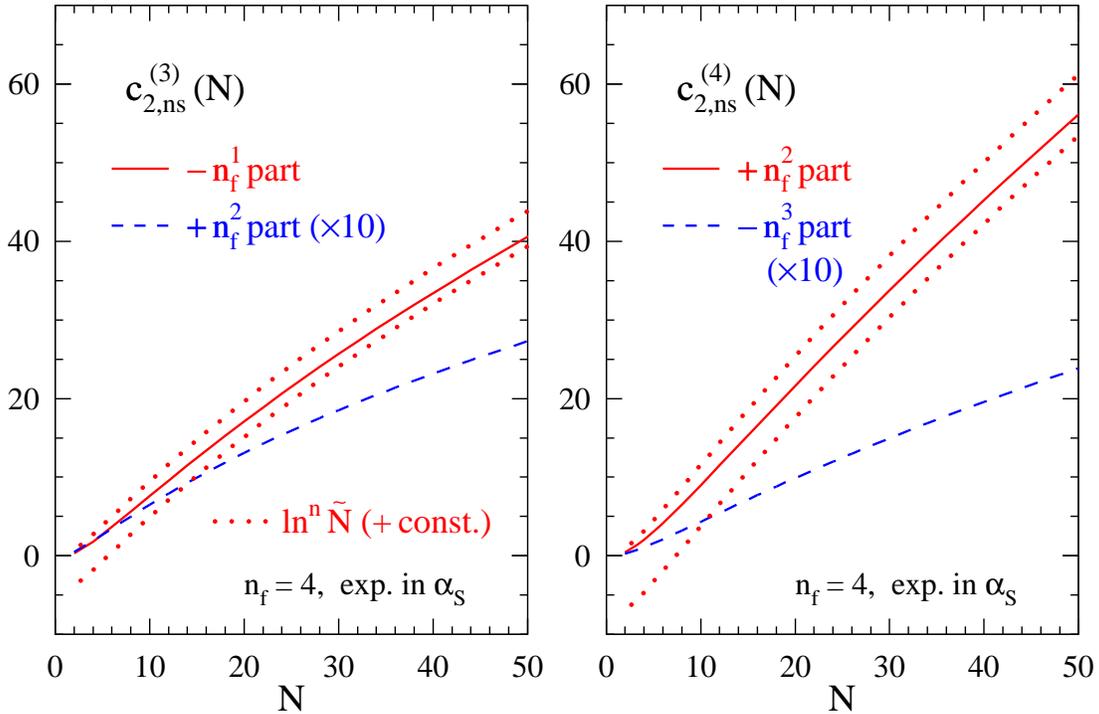,width=15.0cm,angle=0}\quad}
\vspace{-3mm}
\caption{ \label{RFig2} \small
 As fig.~\ref{RFig1}, but for the structure function $F_{2,\rm ns}$.
 In addition the $\ln^\ell \!\wN$ and $\ln^\ell \! \wN \!+\! 
 \mbox{const.}$ threshold contributions are shown here, respectively, 
 by the upper and lower dotted curves. }
\vspace{-2mm}
\end{figure}

The numerical size of the above fourth-order results is shown in 
figs.~\ref{RFig1} and \ref{RFig2} for QCD, i.e., $\ca = 3$ and 
$\cf = 4/3$, with $\nf = 4$ light flavours, together with the 
corresponding third-order contributions for the physically very 
wide range $2 \leq N \leq 50$.
The coefficient functions $c_{L,\rm ns}^{\,(n)}$ vanish for 
$N \!\ra \infty$, yet only slowly: the size of the $\nfs$ parts of
$c_{L,\rm ns}^{\,(4)}$ in the right part of fig.~\ref{RFig1} 
decreases only by a factor of 2 from $N \!=\! 10$ to $N \!=\! 50$.
A further reduction by another factor of 2 and 4 is only reached 
at $N \!=\! 175$ and $N = 540$, respectively.

The shape of the leading large-$\nf$ contributions 
($ \sim n_{\! f}^{\,n-1}$ at order $\as(n)$) in fig.~\ref{RFig1} is 
similar to that of the subleading large-$\nf$ contributions 
($\sim n_{\! f}^{\,n-2}$ at order $\as(n)$) in the $N$-range of the 
figure; its relative size is decreasing for $\nf = 4$ from about 
1/10 at $n=3$ to about 1/15 at $n=4$. 
This pattern of a reduced numerical significance of the leading
large-$\nf$ term at the fourth order is also seen for 
$c_{2,\rm ns}^{\,(n)}$ in fig.~\ref{RFig2}. Here the subleading 
large-$\nf$ contributions are larger by factors between 10 and 15 
at $4 \leq N \leq 50$ for $n=3$; the corresponding range for $n=4$
is 19 to 24.


Also shown in fig.~\ref{RFig2} are the dominant contributions to
${\cal C}_{2,\rm ns}$ in the large-$N$ threshold limit, where our
new contributions (\ref{c2ns4L}) -- (\ref{c2ns4Z}) to 
$c_{2,\rm ns}^{\,(4)}$ lead to the numerical expansion
\bea
\label{c2Ninf}
  c_{2,\rm ns}^{\,(4)}(N) \Big|_{\nfs} &\!=\!& \;
      0.7233196 \,\ln^{\:\! 6} \!\wN 
    + 12.339095 \,\ln^{\:\! 5} \!\wN 
  \nn \\[-2.5mm] & & \mbox{\hspn}
    + 87.721224 \,\ln^{\:\! 4} \!\wN
    + 293.04552 \,\ln^{\:\! 3} \!\wN
  \nn \\[0.5mm] & & \mbox{\hspn}
    + 233.48456 \,\ln^{\:\! 2} \!\wN
    + 65.035706 \,\ln \wN - 12175.412
  \nn \\[1.5mm] & & \mbox{\hspn\hspn\hspn\hspn} \;
  + N^{\,-1} \left(
      2.1069959 \,\ln^{\:\! 5} \!\wN
    + 67.193416 \,\ln^{\:\! 4} \!\wN
  \right.\nn \\[-0.5mm] & & \left. \mbox{\hspn}
    + 691.16782 \,\ln^{\:\! 3} \!\wN
    + 3429.9787 \,\ln^{\:\! 2} \!\wN
  \right.\nn \\[0.5mm] & & \left. \mbox{\hspn}
    + 8755.5832 \,\ln \wN
    + 12282.167 \right)
  + {\cal O} \left( N^{-2} \right) 
\eea
with $\ln \wN = \ln N + \gamma_e$, where $\gamma_e$ is the 
Euler-Mascheroni constant.
Keeping only the $\ln^{\:\!\ell} \!\wN$ terms in the first three lines,
and the corresponding contributions at the third order, one arrives
at the upper dotted curves in the figure. Adding to these the constant-$N$
contributions yields the lower dotted curves. Neither of these results
can quantitatively replace the exact expressions at physically interesting
moderate values of $N$.

Analytic $x$-space expression corresponding to eq.~(\ref{c2Ninf}) and
a further discussion of the threshold and high-energy limits can be found
in section 4 below.

\noindent
{\bf \large A five-loop prediction} 

\noindent
It is possible to predict a small part, the $\z4$ coefficients of $\nft$
and $\nff$, of the $n\!=\!4$ five-loop non-singlet anomalous dimension 
in eq.~(\ref{gns}), from our result (\ref{c2ns4Z}) and eq.~(\ref{c2ns4F})
for $c_{2,\rm ns}^{\,(4)}(N)$. This possibility arises from the 
no$\:\!$-$\:\!\pi^2$ conjecture$/$theorem \cite{Jamin:2017mul,Baikov:2018wgs} 
for Euclidean (space-like) physical quantities in a suitable renormalization
scheme, which was investigated in the context of inclusive DIS in
ref.~\cite{Davies:2017hyl}.
For this one considers the physical evolution kernels which were 
expressed in terms of the splitting functions and coefficient functions
for the non-singlet case to the fifth order in eqs.~(2.7) -- (2.9) of
ref.~\cite{vanNeerven:2001pe}.

Keeping only the terms with $\z4$ simplifies the fifth-order \MSb\ kernel
for $F_{2,\rm ns}$ to
\beq
  \label{K2ns4z4}
  \wK_{2\,\rm ns}^{\,(4)}(N) \:=\: \mbox{}
  - \wg_{\,\rm ns}^{\,(4)}(N) 
  - 3\:\! \beta_1\, \wc_a^{\,(3)}(N) 
  - 4\:\! \beta_0 \left( \wc_{2\,\rm ns}^{\,(4)}(N) 
    - c_{2\,\rm ns}^{\,(1)}(N)\, \wc_{2\,\rm ns}^{\,(3)}(N) \right)
\:\: ,
\eeq
where the tilde indicates the contribution with $\z4\!=\! 1/90\:\pi^{\,4}$.
At this order, a scheme transformation removing the $\z4$ term of the 
five-loop beta function \cite{Baikov:2016tgj,Herzog:2017ohr,Luthe:2017ttg} 
needs to be performed, and the prediction of the no$\:\!$-$\:\!\pi^2$ 
theorem becomes \cite{Davies:2017hyl} 
\beq
\label{K2ns4z4a}
  \beta_0\, \wK_a^{\:(4)} 
  + \frct{1}{3}\: \widetilde{\beta}_{\,4} \, K_a^{\,(0)} \;=\; 0 \; .
\eeq
Since $\beta_0$ includes $\nf$ and $K_{2,\rm ns}^{\,(0)} \sim \cf$, 
only the $\zeta_4 \,\nff$ terms in $\beta_4$ can contribute, but 
there is no such term. Hence already the quantity (\ref{K2ns4z4}) 
has to vanish for the $\nft$ and $\nff$ terms.

The three-loop coefficient function $c_{2,\rm ns}^{\,(3)}(N)$ does not
include a $\z4$ term with $\nfs$. Since the prefactors $\beta_1$ and
$\beta_0 \, c_{2,\rm ns}^{\,(1)}$ include no more than one power in 
$\nf$, the two terms with $c_{2,\rm ns}^{\,(3)}$ do not contribute to 
the $\nft$ and $\nff$ parts of eq.~(\ref{K2ns4z4}).
So we end up with a simple relation for the $\z4\,\nft$ and
$\z4\,\nff$ contributions, 
\beq
  \wP_{\rm ns}^{\,(4)}(N) 
  \;=\; -\:\! \wg_{\rm ns}^{\,(4)}(N) 
  \;=\; 4\:\! \beta_0\, \wc_{2\,\rm ns}^{\,(4)}(N)
\:\: ,
\eeq
which leads to
\bea
\label{Pns4z4}
\lefteqn{P_{\rm ns}^{\,(4)}(N) \Big|_{\:\!\z4} \;=\;  
\nfz,\, \nfo \mbox{ and } \nfs \mbox{ contributions } } 
\nn \\[0.5mm] & & \mbox{}
+\cf\*\ca\,\*\nft\,\* \*\bigg( \:
 \frac{64}{3}\,\*\S(2)
-\frac{1568}{27}\,\*\,\*\S(1)
+\frac{176}{9}
+\frac{1360}{27}\,\*\D(0)
-\frac{1360}{27}\,\*\D(1)
-\frac{32}{3}\,\*\Dd(0,2)
+\frac{32}{3}\,\*\Dd(1,2)\bigg)
\quad \nn \\[0.5mm] & & \mbox{}
+\cfs\,\*\nft\,\*\bigg(
-\frac{64}{3}\,\*\S(2)
+\frac{640}{9}\,\*\,\*\S(1)
-\frac{88}{3}
-\frac{512}{9}\,\*\D(0)
+\frac{512}{9}\,\*\D(1)
+\frac{32}{3}\,\*\Dd(0,2)
-\frac{32}{3}\,\*\Dd(1,2)\bigg)
\nn \\[0.5mm] & & \mbox{}
+\cf\,\*\nff\,\*\bigg(
-\frac{64}{27}\,\*\,\*\S(1)
+ \frac{16}{9}
+\frac{32}{27}\,\*\D(0)
-\frac{32}{27}\,\*\D(1)\bigg)

\; .
\eea
At $N \!=\!2 $ and $N \!=\!3 $ this result agrees with those obtained 
by diagram calculations in ref.~\cite{Herzog:2018kwj} using the program 
of ref.~\cite{Herzog:2017bjx} for the $R^{\:\!\ast}$ operation.
Its last line was derived long ago as part of the complete leading 
large-$\nf$ result~\cite{Gracey:1994nn}.
Due to eqs.~(\ref{c2Ndec}) and (\ref{c2ns4F}), the whole of (\ref{Pns4z4}) 
is proportional to $\beta_0$ and the lowest-order anomalous dimension 
$\gamma_{\rm qq}^{\,(0)}(N)$ for $\cf = \ca$.

%
\section{The x-space coefficient functions}
\setcounter{equation}{0}

The coefficient functions $c_{a,\rm ns}^{\,(n)}(x)$ are obtained from 
the $N$-space results of the previous section by an inverse Mellin
transformation, which expresses these functions in terms of harmonic
polylogarithms (HPLs) $H_{m_1,...,m_w}(x)$, $m_j = 0,\pm 1$. Following
ref.~\cite{Remiddi:1999ew}, to which the reader is referred for a 
detailed discussion, the lowest-weight ($w = 1$) functions $H_m(x)$ are 
given by
\beq
\label{hpol1}
  H_0(x)       \: = \: \ln x \:\: , \quad\quad
  H_{\pm 1}(x) \: = \: \mp \, \ln (1 \mp x) \:\: ,
\eeq
and the higher-weight ($w \geq 2$) functions are recursively defined as
\beq
\label{hpol2}
  H_{m_1,...,m_w}(x) \: = \:
    \left\{ \begin{array}{cl}
    \displaystyle{ \frac{1}{w!}\,\ln^w x \:\: ,}
       & \quad {\rm if} \:\:\: m^{}_1,...,m^{}_w = 0,\ldots ,0 \\[2ex]
    \displaystyle{ \int_0^x \! dz\: f_{m_1}(z) \, H_{m_2,...,m_w}(z)
       \:\: , } & \quad {\rm else}
    \end{array} \right.
\eeq
with
\beq
\label{hpolf}
  f_0(x)       \: = \: \frac{1}{x} \:\: , \quad\quad
  f_{\pm 1}(x) \: = \: \frac{1}{1 \mp x} \:\: .
\eeq
The inverse Mellin transformation exploits an isomorphism between 
the set of harmonic sums for even or odd $N$ and the set of HPLs.
Hence it can be performed by a completely algebraic procedure~%
\cite{Moch:1999eb,Remiddi:1999ew}, based on the fact that harmonic 
sums occur as coefficients of the Taylor expansion of harmonic 
polylogarithms.
A {\sc Fortran} program for the HPLs up to weight $w=4$ has been 
provided in ref.~\cite{Gehrmann:2001pz}, later this was extended 
to $w=5$ and $w=6$.

In our results below, the argument $x$ of the HPLs is suppressed 
for brevity, and we use the abbreviations
\beq
  \xm \; = \; 1-x \quad \mbox{ and} \quad \xp \; = \; 1+x \:\: .
\eeq
Analogous to eq.~(\ref{cLxdec}) above, the $x$-space coefficient 
function for $F_{L,\rm ns}$ is decomposed as
\bea
\label{cLxdec}
  c_{L,\rm ns}^{\,(4)}(x) & \!=\! &
  \nfz \mbox{ and } \nfo \mbox{ contributions }
  \nn \\[1mm] & & \mbox{\hspn\hspn\hspn}
  + \cf \* \ca \,\* \nfs \*\, \frac{16}{9} \*\:
    c_{L,\rm ns}^{\,(4)\mathrm{L}}(x)
  + \cf\, (\cf-\frct{1}{2}\,\* \ca) \,\* \nfs \*\, \frac{16}{9} \*\:
    c_{L,\rm ns}^{\,(4)\mathrm{N}}(x)
  + \cf \* \nft \* \,\frac{16}{27} \*\:
    c_{L,\rm ns}^{\,(4)\mathrm{F}}(x)
 \; . \qquad	
\eea
The two $\nfs$ contributions in the second line are given by
\bea
\label{cLnsX4L}
\lefteqn{c_{L,\rm ns}^{\,(4)\mathrm{L}}(x) \;=\;  } \nn \\[0.5mm] & & \mbox{ \hspn }
%
%
%
%
%
%
+\Hhhhh(0,0,0,0,1)\,(-40\,x
+80\,x^2
-120\,x^3)
+\Hhhhh(0,0,0,1,0)\,(-16\,x
+32\,x^2
-48\,x^3)
\nn \\[0.5mm] & & \mbox{ \hspn }
+\Hhhhh(0,0,0,1,1)\,(-20\,x
+40\,x^2
-60\,x^3)
+\Hhhhh(0,0,1,0,0)\,(16\,x
-32\,x^2
+48\,x^3)
+\Hhhhh(0,0,1,0,1)\,(-4\,x
\nn \\[0.5mm] & & \mbox{ \hspn }
+8\,x^2
-12\,x^3)
+\Hhhhh(0,0,1,1,0)\,(4\,x
-8\,x^2
+12\,x^3)
+\Hhhhh(0,1,0,0,0)\,(40\,x
-80\,x^2
+120\,x^3)
\nn \\[0.5mm] & & \mbox{ \hspn }
+\Hhhhh(0,1,0,0,1)\,(-48\,x^{-2}
+24\,x^{-1}
-4\,x
+8\,x^2
-12\,x^3)
+\Hhhhh(0,1,0,1,0)\,(4\,x
-8\,x^2
+12\,x^3)
\nn \\[0.5mm] & & \mbox{ \hspn }
+\Hhhhh(0,1,0,1,1)\,(-18\,x
+72\,x^2
-60\,x^3)
+\Hhhhh(0,1,1,0,0)\,(48\,x^{-2}
-24\,x^{-1}
+20\,x
-40\,x^2
+60\,x^3)
\nn \\[0.5mm] & & \mbox{ \hspn }
+\Hhhhh(0,1,1,0,1)\,(18\,x
-72\,x^2
+60\,x^3)
+\Hhhhh(1,0,0,0,1)\,(-80\,x^{-2}
+40\,x^{-1}
-40\,x
+80\,x^2
\nn \\[0.5mm] & & \mbox{ \hspn }
-120\,x^3)
+\Hhhhh(1,0,0,1,0)\,(-32\,x^{-2}
+16\,x^{-1}
-16\,x
+32\,x^2
-48\,x^3)
+\Hhhhh(1,0,0,1,1)\,(-40\,x^{-2}
\nn \\[0.5mm] & & \mbox{ \hspn }
+20\,x^{-1}
-20\,x
+40\,x^2
-60\,x^3)
+\Hhhhh(1,0,1,0,0)\,(32\,x^{-2}
-16\,x^{-1}
+16\,x
-32\,x^2
+48\,x^3)
\nn \\[0.5mm] & & \mbox{ \hspn }
+\Hhhhh(1,0,1,0,1)\,(-8\,x^{-2}
+4\,x^{-1}
-4\,x
+8\,x^2
-12\,x^3)
+\Hhhhh(1,0,1,1,0)\,(8\,x^{-2}
-4\,x^{-1}
+4\,x
-8\,x^2
\nn \\[0.5mm] & & \mbox{ \hspn }
+12\,x^3)
+\Hhhhh(1,1,0,0,0)\,(80\,x^{-2}
-40\,x^{-1}
+40\,x
-80\,x^2
+120\,x^3)
+\Hhhhh(1,1,0,0,1)\,(-8\,x^{-2}
\nn \\[0.5mm] & & \mbox{ \hspn }
+4\,x^{-1}
-4\,x
+8\,x^2
-12\,x^3)
+\Hhhhh(1,1,0,1,0)\,(8\,x^{-2}
-4\,x^{-1}
+4\,x
-8\,x^2
+12\,x^3)
\nn \\[0.5mm] & & \mbox{ \hspn }
+\Hhhhh(1,1,0,1,1)\,(-18\,x
+72\,x^2
-60\,x^3)
+\Hhhhh(1,1,1,0,0)\,(40\,x^{-2}
-20\,x^{-1}
+20\,x
-40\,x^2
+60\,x^3)
\nn \\[0.5mm] & & \mbox{ \hspn }
+\Hhhhh(1,1,1,0,1)\,(18\,x
-72\,x^2
+60\,x^3)
+188/3\,x\,\*\Hhhh(0,0,0,0)
+\Hhhh(0,0,0,1)\,(80\,x^{-1}
+115/2\,x
\nn \\[0.5mm] & & \mbox{ \hspn }
+184\,x^2
-102\,x^3)
+\Hhhh(0,0,1,0)\,(32\,x^{-1}
+86\,x
+48\,x^2)
+\Hhhh(0,0,1,1)\,(40\,x^{-1}
+86\,x
+76\,x^2
\nn \\[0.5mm] & & \mbox{ \hspn }
-8\,x^3)
+\Hhhh(0,1,0,0)\,(-32\,x^{-1}
+91\,x
-112\,x^2
+102\,x^3)
+\Hhhh(0,1,0,1)\,(8\,x^{-1}
+61\,x
+12\,x^2)
\nn \\[0.5mm] & & \mbox{ \hspn }
+\Hhhh(0,1,1,0)\,(-8\,x^{-1}
+65\,x
-28\,x^2
+8\,x^3)
+61\,x\,\*\Hhhh(0,1,1,1)
+\Hhhh(1,0,0,0)\,(-80\,x^{-1}
+115/3\,x
\nn \\[0.5mm] & & \mbox{ \hspn }
-120\,x^2)
+\Hhhh(1,0,0,1)\,(16
-20\,x^{-2}
-4\,x^{-1}
+7\,x
+76\,x^2
-102\,x^3)
+\Hhhh(1,0,1,0)\,(-8\,x^{-1}
\nn \\[0.5mm] & & \mbox{ \hspn }
+40\,x
-12\,x^2)
+\Hhhh(1,0,1,1)\,(4\,x
+76\,x^2
-8\,x^3)
+\Hhhh(1,1,0,0)\,(-16
+20\,x^{-2}
-28\,x^{-1}
+75\,x
\nn \\[0.5mm] & & \mbox{ \hspn }
-124\,x^2
+102\,x^3)
+\Hhhh(1,1,0,1)\,(80\,x
-60\,x^2)
+\Hhhh(1,1,1,0)\,(34\,x
-16\,x^2
+8\,x^3)
\nn \\[0.5mm] & & \mbox{ \hspn }
+36\,x\,\*\Hhhh(1,1,1,1)
+\Hhh(0,0,0)\,(125/6
+10969/36\,x
+120\,x^2
+40\,\z2\,x
-80\,\z2\,x^2
+120\,\z2\,x^3)
\nn \\[0.5mm] & & \mbox{ \hspn }
+\Hhh(0,0,1)\,(-23/2
-12\,x^{-1}
+1613/6\,x
+290\,x^2
-197/3\,x^3
+4\,\z2\,x
-8\,\z2\,x^2
+12\,\z2\,x^3)
\nn \\[0.5mm] & & \mbox{ \hspn }
+\Hhh(0,1,0)\,(-23
-32\,x^{-1}
+1307/6\,x
-20\,x^2
+197/3\,x^3
+48\,\z2\,x^{-2}
-24\,\z2\,x^{-1}
+4\,\z2\,x
\nn \\[0.5mm] & & \mbox{ \hspn }
-8\,\z2\,x^2
+12\,\z2\,x^3)
+\Hhh(0,1,1)\,(-29
-40\,x^{-1}
+471/2\,x
+8\,x^2
-18\,\z2\,x
+72\,\z2\,x^2
\nn \\[0.5mm] & & \mbox{ \hspn }
-60\,\z2\,x^3)
+\Hhh(1,0,0)\,(-24
-36\,x^{-1}
+1439/9\,x
-90\,x^2
+80\,\z2\,x^{-2}
-40\,\z2\,x^{-1}
\nn \\[0.5mm] & & \mbox{ \hspn }
+40\,\z2\,x
-80\,\z2\,x^2
+120\,\z2\,x^3)
+\Hhh(1,0,1)\,(-85/3
-8\,x^{-1}
+721/6\,x
+152\,x^2
\nn \\[0.5mm] & & \mbox{ \hspn }
-197/3\,x^3
+8\,\z2\,x^{-2}
-4\,\z2\,x^{-1}
+4\,\z2\,x
-8\,\z2\,x^2
+12\,\z2\,x^3)
+\Hhh(1,1,0)\,(-71/3
\nn \\[0.5mm] & & \mbox{ \hspn }
+8\,x^{-1}
+997/6\,x
-100\,x^2
+197/3\,x^3
+8\,\z2\,x^{-2}
-4\,\z2\,x^{-1}
+4\,\z2\,x
-8\,\z2\,x^2
\nn \\[0.5mm] & & \mbox{ \hspn }
+12\,\z2\,x^3)
+\Hhh(1,1,1)\,(-25
+293/2\,x
-18\,\z2\,x
+72\,\z2\,x^2
-60\,\z2\,x^3)
\nn \\[0.5mm] & & \mbox{ \hspn }
+\Hh(0,0)\,(-3301/36
+2471/3\,x
+210\,x^2
-80\,\z2\,x^{-1}
-115/2\,\z2\,x
-184\,\z2\,x^2
\nn \\[0.5mm] & & \mbox{ \hspn }
+102\,\z2\,x^3
-28\,\z3\,x
+56\,\z3\,x^2
-84\,\z3\,x^3)
+\Hh(0,1)\,(-941/6
-36\,x^{-1}
+10169/18\,x
\nn \\[0.5mm] & & \mbox{ \hspn }
+611/3\,x^2
-8\,\z2\,x^{-1}
-61\,\z2\,x
-12\,\z2\,x^2
-48\,\z3\,x^{-2}
+24\,\z3\,x^{-1}
+62\,\z3\,x
\nn \\[0.5mm] & & \mbox{ \hspn }
-304\,\z3\,x^2
+216\,\z3\,x^3)
+\Hh(1,0)\,(-464/3
+32\,x^{-1}
+4565/12\,x
+43/3\,x^2
+20\,\z2\,x^{-2}
\nn \\[0.5mm] & & \mbox{ \hspn }
+4\,\z2\,x^{-1}
-16\,\z2
-7\,\z2\,x
-76\,\z2\,x^2
+102\,\z2\,x^3
-56\,\z3\,x^{-2}
+28\,\z3\,x^{-1}
-28\,\z3\,x
\nn \\[0.5mm] & & \mbox{ \hspn }
+56\,\z3\,x^2
-84\,\z3\,x^3)
+\Hh(1,1)\,(-155
+40\,x^{-1}
+14459/36\,x
+8\,x^2
-80\,\z2\,x
+60\,\z2\,x^2
\nn \\[0.5mm] & & \mbox{ \hspn }
-56\,\z3\,x^{-2}
+28\,\z3\,x^{-1}
+62\,\z3\,x
-304\,\z3\,x^2
+216\,\z3\,x^3)
+\,\H(0)\,(-3679/12
\nn \\[0.5mm] & & \mbox{ \hspn }
+839905/648\,x
+587/3\,x^2
+12\,\z2\,x^{-1}
+23/2\,\z2
-1613/6\,\z2\,x
-290\,\z2\,x^2
\nn \\[0.5mm] & & \mbox{ \hspn }
+197/3\,\z2\,x^3
+56\,\z3\,x^{-1}
-605/6\,\z3\,x
+164\,\z3\,x^2
-110\,\z3\,x^3
-53\,\z4\,x
\nn \\[0.5mm] & & \mbox{ \hspn }
+106\,\z4\,x^2
-159\,\z4\,x^3)
+\,\H(1)\,(-4483/12
+84\,x^{-1}
+842437/1296\,x
+587/3\,x^2
\nn \\[0.5mm] & & \mbox{ \hspn }
+8\,\z2\,x^{-1}
+85/3\,\z2
-721/6\,\z2\,x
-152\,\z2\,x^2
+197/3\,\z2\,x^3
-20\,\z3\,x^{-2}
+44\,\z3\,x^{-1}
\nn \\[0.5mm] & & \mbox{ \hspn }
+16\,\z3
+397/3\,\z3\,x
-136\,\z3\,x^2
-110\,\z3\,x^3
-106\,\z4\,x^{-2}
+53\,\z4\,x^{-1}
-53\,\z4\,x
\nn \\[0.5mm] & & \mbox{ \hspn }
+106\,\z4\,x^2
-159\,\z4\,x^3)
-763025/1296
+3130309/2592\,x
+36\,\z2\,x^{-1}
+941/6\,\z2
\nn \\[0.5mm] & & \mbox{ \hspn }
-10169/18\,\z2\,x
-611/3\,\z2\,x^2
+12\,\z3\,x^{-1}
-13/6\,\z3
-391\,\z3\,x
+650\,\z3\,x^2
\nn \\[0.5mm] & & \mbox{ \hspn }
-197\,\z3\,x^3
+22\,\z3\,\z2\,x
-80\,\z3\,\z2\,x^2
+72\,\z3\,\z2\,x^3
+106\,\z4\,x^{-1}
+33\,\z4\,x
\nn \\[0.5mm] & & \mbox{ \hspn }
+263\,\z4\,x^2
-192\,\z4\,x^3
+140\,\z5\,x
-640\,\z5\,x^2
+480\,\z5\,x^3

\eea
and
\bea
\label{cLnsX4N}
\lefteqn{c_{L,\rm ns}^{\,(4)\mathrm{N}}(x) \;=\;  } \nn \\[0.5mm] & & \mbox{ \hspn }
%
%
%
%
%
%
-320\,x\,\*\Hhhhh(-1,-1,-1,-1,0)
+544\,x\,\*\Hhhhh(-1,-1,-1,0,0)
+352\,x\,\*\Hhhhh(-1,-1,0,-1,0)
-576\,x\,\*\Hhhhh(-1,-1,0,0,0)
\nn \\[0.5mm] & & \mbox{ \hspn }
+352\,x\,\*\Hhhhh(-1,0,-1,-1,0)
-560\,x\,\*\Hhhhh(-1,0,-1,0,0)
-368\,x\,\*\Hhhhh(-1,0,0,-1,0)
+288\,x\,\*\Hhhhh(-1,0,0,0,0)
\nn \\[0.5mm] & & \mbox{ \hspn }
+352\,x\,\*\Hhhhh(0,-1,-1,-1,0)
-560\,x\,\*\Hhhhh(0,-1,-1,0,0)
-368\,x\,\*\Hhhhh(0,-1,0,-1,0)
+144\,x\,\*\Hhhhh(0,-1,0,0,0)
\nn \\[0.5mm] & & \mbox{ \hspn }
-224\,x\,\*\Hhhhh(0,-1,0,0,1)
-144\,x\,\*\Hhhhh(0,-1,0,1,0)
-128\,x\,\*\Hhhhh(0,-1,0,1,1)
-384\,x\,\*\Hhhhh(0,0,-1,-1,0)
\nn \\[0.5mm] & & \mbox{ \hspn }
+48\,x\,\*\Hhhhh(0,0,-1,0,0)
-288\,x\,\*\Hhhhh(0,0,-1,0,1)
-48\,x\,\*\Hhhhh(0,0,1,0,0)
-160\,x\,\*\Hhhhh(0,1,0,-1,0)
\nn \\[0.5mm] & & \mbox{ \hspn }
-144\,x\,\*\Hhhhh(0,1,0,0,0)
-184\,x\,\*\Hhhhh(0,1,0,0,1)
+56\,x\,\*\Hhhhh(0,1,0,1,0)
+16\,x\,\*\Hhhhh(0,1,1,0,0)
-40\,x\,\*\Hhhhh(0,1,1,0,1)
\nn \\[0.5mm] & & \mbox{ \hspn }
+40\,x\,\*\Hhhhh(0,1,1,1,0)
-288\,x\,\*\Hhhhh(1,0,-1,0,0)
-192\,x\,\*\Hhhhh(1,0,-1,0,1)
-256\,x\,\*\Hhhhh(1,0,0,-1,0)
\nn \\[0.5mm] & & \mbox{ \hspn }
-288\,x\,\*\Hhhhh(1,0,0,0,0)
-344\,x\,\*\Hhhhh(1,0,0,0,1)
-112\,x\,\*\Hhhhh(1,0,0,1,0)
-168\,x\,\*\Hhhhh(1,0,0,1,1)
+72\,x\,\*\Hhhhh(1,0,1,0,0)
\nn \\[0.5mm] & & \mbox{ \hspn }
-16\,x\,\*\Hhhhh(1,0,1,0,1)
+56\,x\,\*\Hhhhh(1,0,1,1,0)
-128\,x\,\*\Hhhhh(1,1,0,-1,0)
-184\,x\,\*\Hhhhh(1,1,0,0,1)
+64\,x\,\*\Hhhhh(1,1,0,1,0)
\nn \\[0.5mm] & & \mbox{ \hspn }
-40\,x\,\*\Hhhhh(1,1,0,1,1)
+88\,x\,\*\Hhhhh(1,1,1,0,0)
-32\,x\,\*\Hhhhh(1,1,1,0,1)
+72\,x\,\*\Hhhhh(1,1,1,1,0)
+\Hhhh(-1,-1,-1,0)\,(160
\nn \\[0.5mm] & & \mbox{ \hspn }
+32\,x^{-2}
-80\,x^{-1}
+672\,x
+160\,x^2
-48\,x^3)
+\Hhhh(-1,-1,0,0)\,(-272
-272/5\,x^{-2}
+136\,x^{-1}
\nn \\[0.5mm] & & \mbox{ \hspn }
-1040\,x
-272\,x^2
+408/5\,x^3)
+\Hhhh(-1,0,-1,0)\,(-176
-176/5\,x^{-2}
+88\,x^{-1}
-688\,x
\nn \\[0.5mm] & & \mbox{ \hspn }
-176\,x^2
+264/5\,x^3)
+\Hhhh(-1,0,0,0)\,(576
+576/5\,x^{-2}
+664\,x
-864/5\,x^3)
+\Hhhh(-1,0,0,1)\,(224
\nn \\[0.5mm] & & \mbox{ \hspn }
+224/5\,x^{-2}
+112\,x^{-1}
-224\,x^2
-336/5\,x^3)
+\Hhhh(-1,0,1,0)\,(144
+144/5\,x^{-2}
+72\,x^{-1}
\nn \\[0.5mm] & & \mbox{ \hspn }
-144\,x^2
-216/5\,x^3)
+\Hhhh(-1,0,1,1)\,(128
+128/5\,x^{-2}
+64\,x^{-1}
-128\,x^2
-192/5\,x^3)
\nn \\[0.5mm] & & \mbox{ \hspn }
+\Hhhh(0,-1,-1,0)\,(-192
-192/5\,x^{-2}
+96\,x^{-1}
-688\,x
-160\,x^2
+48\,x^3)
+\Hhhh(0,-1,0,0)\,(576
\nn \\[0.5mm] & & \mbox{ \hspn }
+576/5\,x^{-2}
+216\,x
+272\,x^2
-408/5\,x^3)
+\Hhhh(0,-1,0,1)\,(192
+192/5\,x^{-2}
+96\,x^{-1}
\nn \\[0.5mm] & & \mbox{ \hspn }
-352\,x)
+\Hhhh(0,0,-1,0)\,(384
+384/5\,x^{-2}
+24\,x
+240\,x^2
-168/5\,x^3)
+\Hhhh(0,0,0,0)\,(-248/3\,x
\nn \\[0.5mm] & & \mbox{ \hspn }
+864/5\,x^3)
+\Hhhh(0,0,0,1)\,(127\,x
+344\,x^2
+1132/5\,x^3)
+\Hhhh(0,0,1,0)\,(136\,x
+112\,x^2
\nn \\[0.5mm] & & \mbox{ \hspn }
+168/5\,x^3)
+\Hhhh(0,0,1,1)\,(148\,x
+160\,x^2
+712/5\,x^3)
+\Hhhh(0,1,0,0)\,(-34\,x
-72\,x^2
\nn \\[0.5mm] & & \mbox{ \hspn }
-724/5\,x^3)
+\Hhhh(0,1,0,1)\,(98\,x
+16\,x^2
+24/5\,x^3)
+\Hhhh(0,1,1,0)\,(154\,x
-48\,x^2
-544/5\,x^3)
\nn \\[0.5mm] & & \mbox{ \hspn }
+122\,x\,\*\Hhhh(0,1,1,1)
+\Hhhh(1,0,-1,0)\,(64
+64/5\,x^{-2}
+32\,x^{-1}
-336\,x
+64\,x^2
+96/5\,x^3)
\nn \\[0.5mm] & & \mbox{ \hspn }
-1138/3\,x\,\*\Hhhh(1,0,0,0)
+\Hhhh(1,0,0,1)\,(64
+64/5\,x^{-2}
+32\,x^{-1}
-370\,x
+120\,x^2
+796/5\,x^3)
\nn \\[0.5mm] & & \mbox{ \hspn }
+\Hhhh(1,0,1,0)\,(-32
-32/5\,x^{-2}
-16\,x^{-1}
+176\,x
-32\,x^2
-48/5\,x^3)
+\Hhhh(1,0,1,1)\,(56\,x
+32\,x^2
\nn \\[0.5mm] & & \mbox{ \hspn }
+104\,x^3)
+\Hhhh(1,1,0,0)\,(-16
-16/5\,x^{-2}
-8\,x^{-1}
+126\,x
-72\,x^2
-724/5\,x^3)
+\Hhhh(1,1,0,1)\,(16
\nn \\[0.5mm] & & \mbox{ \hspn }
+16/5\,x^{-2}
+8\,x^{-1}
-8\,x
+16\,x^2
+24/5\,x^3)
+\Hhhh(1,1,1,0)\,(-16
-16/5\,x^{-2}
-8\,x^{-1}
\nn \\[0.5mm] & & \mbox{ \hspn }
+188\,x
-48\,x^2
-544/5\,x^3)
+72\,x\,\*\Hhhh(1,1,1,1)
-160\,\z2\,x\,\*\Hhh(-1,-1,-1)
+\Hhh(-1,-1,0)\,(-336
\nn \\[0.5mm] & & \mbox{ \hspn }
-1232/25\,x^{-2}
+128\,x^{-1}
-2752/3\,x
-320\,x^2
+1968/25\,x^3
+112\,\z2\,x)
\nn \\[0.5mm] & & \mbox{ \hspn }
+176\,\z2\,x\,\*\Hhh(-1,0,-1)
+\Hhh(-1,0,0)\,(6104/5
+3616/25\,x^{-2}
+272/5\,x^{-1}
+17152/15\,x
\nn \\[0.5mm] & & \mbox{ \hspn }
-408/5\,x^2
-5784/25\,x^3
-56\,\z2\,x)
+\Hhh(-1,0,1)\,(448
+1232/25\,x^{-2}
+160\,x^{-1}
+48\,x
\nn \\[0.5mm] & & \mbox{ \hspn }
-368\,x^2
-1968/25\,x^3)
+176\,\z2\,x\,\*\Hhh(0,-1,-1)
+\Hhh(0,-1,0)\,(784
+2384/25\,x^{-2}
+112/5\,x^{-1}
\nn \\[0.5mm] & & \mbox{ \hspn }
+1264/5\,x
+296\,x^2
-1968/25\,x^3
+120\,\z2\,x)
+96\,\z2\,x\,\*\Hhh(0,0,-1)
+\Hhh(0,0,0)\,(289/15
\nn \\[0.5mm] & & \mbox{ \hspn }
-576/5\,x^{-1}
-73891/90\,x
+864/5\,x^2
+5784/25\,x^3)
+\Hhh(0,0,1)\,(-143
-288/5\,x^{-1}
\nn \\[0.5mm] & & \mbox{ \hspn }
-143/15\,x
+424\,x^2
+18104/75\,x^3
+192\,\z2\,x)
+\Hhh(0,1,0)\,(-342/5
-112/5\,x^{-1}
\nn \\[0.5mm] & & \mbox{ \hspn }
+2383/15\,x
-476/5\,x^2
-488/3\,x^3
+288\,\z2\,x)
+\Hhh(0,1,1)\,( 216\,\z2\,x
-426/5
-128/5\,x^{-1}
\nn \\[0.5mm] & & \mbox{ \hspn }
+659/5\,x
-328/5\,x^2)
+192\,\z2\,x\,\*\Hhh(1,0,-1)
+\Hhh(1,0,0)\,(-192/5
+16/5\,x^{-1}
-6862/45\,x
\nn \\[0.5mm] & & \mbox{ \hspn }
+724/5\,x^2
+400\,\z2\,x)
+\Hhh(1,0,1)\,( 192\,\z2\,x
-814/15
-16/5\,x^{-1}
-301/15\,x
+716/5\,x^2
\nn \\[0.5mm] & & \mbox{ \hspn }
+488/3\,x^3)
+\Hhh(1,1,0)\,(-746/15
+16/5\,x^{-1}
+3131/15\,x
-196/5\,x^2
-488/3\,x^3
\nn \\[0.5mm] & & \mbox{ \hspn }
+296\,\z2\,x)
+\Hhh(1,1,1)\,(-50
+73\,x
+192\,\z2\,x)
+\Hh(-1,-1)\,(16\,\z2\,x^{-2}
-40\,\z2\,x^{-1}
+80\,\z2
\nn \\[0.5mm] & & \mbox{ \hspn }
+336\,\z2\,x
+80\,\z2\,x^2
-24\,\z2\,x^3
+160\,\z3\,x)
+\Hh(-1,0)\,(85732/75
+32252/375\,x^{-2}
\nn \\[0.5mm] & & \mbox{ \hspn }
+1632/25\,x^{-1}
+79732/75\,x
-2568/25\,x^2
-17916/125\,x^3
-56\,\z2\,x^{-2}
-84\,\z2\,x^{-1}
\nn \\[0.5mm] & & \mbox{ \hspn }
-280\,\z2
-176\,\z2\,x
+168\,\z2\,x^2
+84\,\z2\,x^3
-80\,\z3\,x)
+\Hh(0,-1)\,(-288/5\,\z2\,x^{-2}
\nn \\[0.5mm] & & \mbox{ \hspn }
-48\,\z2\,x^{-1}
-288\,\z2
+8\,\z2\,x
-80\,\z2\,x^2
+24\,\z2\,x^3
-112\,\z3\,x)
+\Hh(0,0)\,(-12541/450
\nn \\[0.5mm] & & \mbox{ \hspn }
-3616/25\,x^{-1}
-41282/25\,x
+4444/25\,x^2
+17916/125\,x^3
-103\,\z2\,x
-344\,\z2\,x^2
\nn \\[0.5mm] & & \mbox{ \hspn }
-260\,\z2\,x^3
+96\,\z3\,x)
+\Hh(0,1)\,(-7937/75
-752/25\,x^{-1}
-67831/225\,x
-12836/75\,x^2
\nn \\[0.5mm] & & \mbox{ \hspn }
-96/5\,\z2\,x^{-2}
-48\,\z2\,x^{-1}
-96\,\z2
+246\,\z2\,x
-96\,\z2\,x^2
-144/5\,\z2\,x^3
+160\,\z3\,x)
\nn \\[0.5mm] & & \mbox{ \hspn }
+\Hh(1,0)\,(-536/15
+112/5\,x^{-1}
-799/6\,x
+320/3\,x^2
-24\,\z2\,x^{-2}
-60\,\z2\,x^{-1}
-120\,\z2
\nn \\[0.5mm] & & \mbox{ \hspn }
+546\,\z2\,x
-176\,\z2\,x^2
-176\,\z2\,x^3
+56\,\z3\,x)
+\Hh(1,1)\,(-122/5
+128/5\,x^{-1}
\nn \\[0.5mm] & & \mbox{ \hspn }
-13169/90\,x
-328/5\,x^2
-96/5\,\z2\,x^{-2}
-48\,\z2\,x^{-1}
-96\,\z2
+344\,\z2\,x
-96\,\z2\,x^2
\nn \\[0.5mm] & & \mbox{ \hspn }
-144/5\,\z2\,x^3
+80\,\z3\,x)
+\,\H(-1)\,(-1848/25\,\z2\,x^{-2}
-96\,\z2\,x^{-1}
-616\,\z2
\nn \\[0.5mm] & & \mbox{ \hspn }
-1520/3\,\z2\,x
+208\,\z2\,x^2
+2952/25\,\z2\,x^3
-144/5\,\z3\,x^{-2}
+8\,\z3\,x^{-1}
-144\,\z3
\nn \\[0.5mm] & & \mbox{ \hspn }
-336\,\z3\,x
-16\,\z3\,x^2
+216/5\,\z3\,x^3
-100\,\z4\,x)
+\,\H(0)\,(-16081/250
+628/375\,x^{-1}
\nn \\[0.5mm] & & \mbox{ \hspn }
-39052717/40500\,x
+41888/375\,x^2
+80\,\z2\,x^{-1}
+143\,\z2
+787/3\,\z2\,x
-424\,\z2\,x^2
\nn \\[0.5mm] & & \mbox{ \hspn }
-24008/75\,\z2\,x^3
+115/3\,\z3\,x
+16\,\z3\,x^2
+872/5\,\z3\,x^3
+12\,\z4\,x)
\nn \\[0.5mm] & & \mbox{ \hspn }
+\,\H(1)\,(13997/150
+1712/25\,x^{-1}
-7089221/16200\,x
-7916/75\,x^2
-616/25\,\z2\,x^{-2}
\nn \\[0.5mm] & & \mbox{ \hspn }
-304/5\,\z2\,x^{-1}
-1706/15\,\z2
+7181/15\,\z2\,x
-1516/5\,\z2\,x^2
-15152/75\,\z2\,x^3
\nn \\[0.5mm] & & \mbox{ \hspn }
-88/5\,\z3\,x^{-2}
-44\,\z3\,x^{-1}
-88\,\z3
+362/3\,\z3\,x
+1088/5\,\z3\,x^3
-190\,\z4\,x)
\nn \\[0.5mm] & & \mbox{ \hspn }
-3918163/81000
+49532/375\,x^{-1}
-54681799/162000\,x
+27156/125\,x^2
\nn \\[0.5mm] & & \mbox{ \hspn }
+2384/25\,\z2\,x^{-1}
+7937/75\,\z2
+307027/225\,\z2\,x
+12836/75\,\z2\,x^2
\nn \\[0.5mm] & & \mbox{ \hspn }
-17916/125\,\z2\,x^3
+232/5\,\z3\,x^{-1}
+1591/15\,\z3
+1052/15\,\z3\,x
-4/5\,\z3\,x^2
\nn \\[0.5mm] & & \mbox{ \hspn }
+1456/5\,\z3\,x^3
-136\,\z3\,\z2\,x
+301\,\z4\,x
+302\,\z4\,x^2
+677/5\,\z4\,x^3
+296\,\z5\,x

\:\: .
\eea
The $\nft$ coefficient in the last line of eq.~(\ref{cLxdec}) reads
\bea
\label{cLnsX4F}
\lefteqn{c_{L,\rm ns}^{\,(4)\mathrm{F}}(x) \;=\;  } \nn \\[0.5mm] & & \mbox{ \hspn }
%
%
%
%
%
%
-48\,x\,\*\Hhh(0,0,0)
-36\,x\,\*\Hhh(0,0,1)
-24\,x\,\*\Hhh(0,1,0)
-24\,x\,\*\Hhh(0,1,1)
-12\,x\,\*\Hhh(1,0,0)
-12\,x\,\*\Hhh(1,0,1)
\nn \\[0.5mm] & & \mbox{ \hspn }
-12\,x\,\*\Hhh(1,1,0)
-12\,x\,\*\Hhh(1,1,1)
+\Hh(0,0)\,(12
-150\,x)
+\Hh(0,1)\,(12
-100\,x)
+\Hh(1,0)\,(12
-50\,x)
\qquad \nn \\[0.5mm] & & \mbox{ \hspn }
+\Hh(1,1)\,(12
-50\,x)
+\,\H(0)\,(38
-634/3\,x
+36\,\z2\,x)
+\,\H(1)\,(38
-317/3\,x
+12\,\z2\,x)
\nn \\[0.5mm] & & \mbox{ \hspn }
+203/3
-8609/54\,x
-12\,\z2
+100\,\z2\,x
+12\,\z3\,x

\:\: .
\eea
The corresponding results for $F_{2,\rm ns}$ are written in a very 
similar manner as
\bea
  c_{2,\rm ns}^{\,(4)}(x) & \!=\! &
  \nfz \mbox{ and } \nfo \mbox{ contributions }
  \nn \\[1mm] & & \mbox{\hspn\hspn\hspn}
  + \cf \* \ca \,\* \nfs \*\, \frct{4}{9} \*\:
    c_{2,\rm ns}^{\,(4)\mathrm{L}}(x)
  + \cf\, (\cf-\frct{1}{2}\,\* \ca) \,\* \nfs \*\, \frct{4}{9} \*\:
    c_{2,\rm ns}^{\,(4)\mathrm{N}}(x)
  + \cf \* \nft \* \,\frct{4}{27} \*\:
    c_{2,\rm ns}^{\,(4)\mathrm{F}}(x) 
\qquad
\eea
with
\bea
\label{c2nsX4L}
\lefteqn{c_{2,\rm ns}^{\,(4)\mathrm{L}}(x) \;=\;  } \nn \\[0.5mm] & & \mbox{ \hspn }
%
%
%
%
%
%
+\Hhhhh(0,0,0,0,0)\,(1951/4
-1951/3\,\xm^{-1}
+1951/4\,x)
+\Hhhhh(0,0,0,0,1)\,(607
-2890/3\,\xm^{-1}
+411\,x
\nn \\[0.5mm] & & \mbox{ \hspn }
+480\,x^2
-720\,x^3)
+\Hhhhh(0,0,0,1,0)\,(1547/3
-2572/3\,\xm^{-1}
+1391/3\,x
+192\,x^2
-288\,x^3)
\nn \\[0.5mm] & & \mbox{ \hspn }
+\Hhhhh(0,0,0,1,1)\,(1637/3
-2752/3\,\xm^{-1}
+1343/3\,x
+240\,x^2
-360\,x^3)
+\Hhhhh(0,0,1,0,0)\,(387
\nn \\[0.5mm] & & \mbox{ \hspn }
-670\,\xm^{-1}
+475\,x
-192\,x^2
+288\,x^3)
+\Hhhhh(0,0,1,0,1)\,(428
-752\,\xm^{-1}
+400\,x
+48\,x^2
\nn \\[0.5mm] & & \mbox{ \hspn }
-72\,x^3)
+\Hhhhh(0,0,1,1,0)\,(390
-676\,\xm^{-1}
+436\,x
-48\,x^2
+72\,x^3)
+\Hhhhh(0,0,1,1,1)\,(404
-704\,\xm^{-1}
\nn \\[0.5mm] & & \mbox{ \hspn }
+404\,x)
+\Hhhhh(0,1,0,0,0)\,(770/3
-1408/3\,\xm^{-1}
+1250/3\,x
-480\,x^2
+720\,x^3)
\nn \\[0.5mm] & & \mbox{ \hspn }
+\Hhhhh(0,1,0,0,1)\,(296
-48\,x^{-2}
-548\,\xm^{-1}
+256\,x
+48\,x^2
-72\,x^3)
+\Hhhhh(0,1,0,1,0)\,(272
-500\,\xm^{-1}
\nn \\[0.5mm] & & \mbox{ \hspn }
+288\,x
-48\,x^2
+72\,x^3)
+\Hhhhh(0,1,0,1,1)\,(300
-556\,\xm^{-1}
+162\,x
+432\,x^2
-360\,x^3)
\nn \\[0.5mm] & & \mbox{ \hspn }
+\Hhhhh(0,1,1,0,0)\,(260
+48\,x^{-2}
-476\,\xm^{-1}
+364\,x
-240\,x^2
+360\,x^3)
+\Hhhhh(0,1,1,0,1)\,(266
\nn \\[0.5mm] & & \mbox{ \hspn }
-488\,\xm^{-1}
+386\,x
-432\,x^2
+360\,x^3)
+\Hhhhh(0,1,1,1,0)\,(240
-436\,\xm^{-1}
+258\,x)
\nn \\[0.5mm] & & \mbox{ \hspn }
+\Hhhhh(0,1,1,1,1)\,(252
-460\,\xm^{-1}
+252\,x)
+\Hhhhh(1,0,0,0,0)\,(671/3
-1342/3\,\xm^{-1}
+671/3\,x)
\nn \\[0.5mm] & & \mbox{ \hspn }
+\Hhhhh(1,0,0,0,1)\,(824/3
-80\,x^{-2}
-1648/3\,\xm^{-1}
+344/3\,x
+480\,x^2
-720\,x^3)
+\Hhhhh(1,0,0,1,0)\,(234
\nn \\[0.5mm] & & \mbox{ \hspn }
-32\,x^{-2}
-468\,\xm^{-1}
+170\,x
+192\,x^2
-288\,x^3)
+\Hhhhh(1,0,0,1,1)\,(270
-40\,x^{-2}
-540\,\xm^{-1}
\nn \\[0.5mm] & & \mbox{ \hspn }
+190\,x
+240\,x^2
-360\,x^3)
+\Hhhhh(1,0,1,0,0)\,(178
+32\,x^{-2}
-356\,\xm^{-1}
+242\,x
-192\,x^2
\nn \\[0.5mm] & & \mbox{ \hspn }
+288\,x^3)
+\Hhhhh(1,0,1,0,1)\,(216
-8\,x^{-2}
-432\,\xm^{-1}
+200\,x
+48\,x^2
-72\,x^3)
+\Hhhhh(1,0,1,1,0)\,(206
\nn \\[0.5mm] & & \mbox{ \hspn }
+8\,x^{-2}
-412\,\xm^{-1}
+222\,x
-48\,x^2
+72\,x^3)
+\Hhhhh(1,0,1,1,1)\,(228
-456\,\xm^{-1}
+228\,x)
\nn \\[0.5mm] & & \mbox{ \hspn }
+\Hhhhh(1,1,0,0,0)\,(530/3
+80\,x^{-2}
-1060/3\,\xm^{-1}
+1010/3\,x
-480\,x^2
+720\,x^3)
\nn \\[0.5mm] & & \mbox{ \hspn }
+\Hhhhh(1,1,0,0,1)\,(196
-8\,x^{-2}
-392\,\xm^{-1}
+180\,x
+48\,x^2
-72\,x^3)
+\Hhhhh(1,1,0,1,0)\,(162
+8\,x^{-2}
\nn \\[0.5mm] & & \mbox{ \hspn }
-324\,\xm^{-1}
+178\,x
-48\,x^2
+72\,x^3)
+\Hhhhh(1,1,0,1,1)\,(188
-376\,\xm^{-1}
+68\,x
+432\,x^2
-360\,x^3)
\nn \\[0.5mm] & & \mbox{ \hspn }
+\Hhhhh(1,1,1,0,0)\,(142
+40\,x^{-2}
-284\,\xm^{-1}
+222\,x
-240\,x^2
+360\,x^3)
+\Hhhhh(1,1,1,0,1)\,(164
\nn \\[0.5mm] & & \mbox{ \hspn }
-328\,\xm^{-1}
+284\,x
-432\,x^2
+360\,x^3)
+\Hhhhh(1,1,1,1,0)\,(140
-280\,\xm^{-1}
+140\,x)
\nn \\[0.5mm] & & \mbox{ \hspn }
+\Hhhhh(1,1,1,1,1)\,(160
-320\,\xm^{-1}
+160\,x)
+\Hhhh(0,0,0,0)\,(2997/2
-20567/9\,\xm^{-1}
+11249/6\,x)
\nn \\[0.5mm] & & \mbox{ \hspn }
+\Hhhh(0,0,0,1)\,(3441/2
+80\,x^{-1}
-24686/9\,\xm^{-1}
+3717/2\,x
+1104\,x^2
-612\,x^3)
\nn \\[0.5mm] & & \mbox{ \hspn }
+\Hhhh(0,0,1,0)\,(23231/18
+32\,x^{-1}
-19634/9\,\xm^{-1}
+30677/18\,x
+288\,x^2)
\nn \\[0.5mm] & & \mbox{ \hspn }
+\Hhhh(0,0,1,1)\,(23231/18
+40\,x^{-1}
-19967/9\,\xm^{-1}
+15163/9\,x
+456\,x^2
-48\,x^3)
\nn \\[0.5mm] & & \mbox{ \hspn }
+\Hhhh(0,1,0,0)\,(7015/9
-32\,x^{-1}
-13373/9\,\xm^{-1}
+11956/9\,x
-672\,x^2
+612\,x^3)
\nn \\[0.5mm] & & \mbox{ \hspn }
+\Hhhh(0,1,0,1)\,(2593/3
+8\,x^{-1}
-4835/3\,\xm^{-1}
+3628/3\,x
+72\,x^2)
+\Hhhh(0,1,1,0)\,(2302/3
\nn \\[0.5mm] & & \mbox{ \hspn }
-8\,x^{-1}
-4283/3\,\xm^{-1}
+7361/6\,x
-168\,x^2
+48\,x^3)
+\Hhhh(0,1,1,1)\,(2261/3
-4270/3\,\xm^{-1}
\nn \\[0.5mm] & & \mbox{ \hspn }
+3473/3\,x)
+\Hhhh(1,0,0,0)\,(3706/9
-80\,x^{-1}
-9995/9\,\xm^{-1}
+8704/9\,x
-720\,x^2)
\nn \\[0.5mm] & & \mbox{ \hspn }
+\Hhhh(1,0,0,1)\,(5740/9
-8\,x^{-2}
-16\,x^{-1}
-12263/9\,\xm^{-1}
+7963/9\,x
+456\,x^2
-612\,x^3)
\nn \\[0.5mm] & & \mbox{ \hspn }
+\Hhhh(1,0,1,0)\,(1546/3
-8\,x^{-1}
-3212/3\,\xm^{-1}
+2506/3\,x
-72\,x^2)
+\Hhhh(1,0,1,1)\,(597
\nn \\[0.5mm] & & \mbox{ \hspn }
-1210\,\xm^{-1}
+1333/2\,x
+456\,x^2
-48\,x^3)
+\Hhhh(1,1,0,0)\,(3946/9
+8\,x^{-2}
-16\,x^{-1}
\nn \\[0.5mm] & & \mbox{ \hspn }
-8495/9\,\xm^{-1}
+8275/9\,x
-744\,x^2
+612\,x^3)
+\Hhhh(1,1,0,1)\,(1412/3
-3022/3\,\xm^{-1}
\nn \\[0.5mm] & & \mbox{ \hspn }
+3128/3\,x
-360\,x^2)
+\Hhhh(1,1,1,0)\,(1321/3
-2558/3\,\xm^{-1}
+4073/6\,x
-96\,x^2
+48\,x^3)
\nn \\[0.5mm] & & \mbox{ \hspn }
+\Hhhh(1,1,1,1)\,(442
-904\,\xm^{-1}
+714\,x)
+\Hhh(0,0,0)\,(22496/9
-152383/36\,\xm^{-1}
+39667/9\,x
\nn \\[0.5mm] & & \mbox{ \hspn }
+720\,x^2
+2890/3\,\z2\,\xm^{-1}
-607\,\z2
-411\,\z2\,x
-480\,\z2\,x^2
+720\,\z2\,x^3)
\nn \\[0.5mm] & & \mbox{ \hspn }
+\Hhh(0,0,1)\,(73799/36
-24\,x^{-1}
-35378/9\,\xm^{-1}
+136907/36\,x
+1740\,x^2
-394\,x^3
\nn \\[0.5mm] & & \mbox{ \hspn }
+752\,\z2\,\xm^{-1}
-428\,\z2
-400\,\z2\,x
-48\,\z2\,x^2
+72\,\z2\,x^3)
+\Hhh(0,1,0)\,(3868/3
-32\,x^{-1}
\nn \\[0.5mm] & & \mbox{ \hspn }
-24305/9\,\xm^{-1}
+53105/18\,x
-120\,x^2
+394\,x^3
+48\,\z2\,x^{-2}
+548\,\z2\,\xm^{-1}
-296\,\z2
\nn \\[0.5mm] & & \mbox{ \hspn }
-256\,\z2\,x
-48\,\z2\,x^2
+72\,\z2\,x^3)
+\Hhh(0,1,1)\,(10021/9
-40\,x^{-1}
-23191/9\,\xm^{-1}
\nn \\[0.5mm] & & \mbox{ \hspn }
+26957/9\,x
+48\,x^2
+488\,\z2\,\xm^{-1}
-266\,\z2
-386\,\z2\,x
+432\,\z2\,x^2
-360\,\z2\,x^3)
\nn \\[0.5mm] & & \mbox{ \hspn }
+\Hhh(1,0,0)\,(11321/18
-24\,x^{-1}
-15200/9\,\xm^{-1}
+12601/6\,x
-540\,x^2
+80\,\z2\,x^{-2}
\nn \\[0.5mm] & & \mbox{ \hspn }
+1648/3\,\z2\,\xm^{-1}
-824/3\,\z2
-344/3\,\z2\,x
-480\,\z2\,x^2
+720\,\z2\,x^3)
+\Hhh(1,0,1)\,(3763/6
\nn \\[0.5mm] & & \mbox{ \hspn }
-8\,x^{-1}
-10459/6\,\xm^{-1}
+10787/6\,x
+912\,x^2
-394\,x^3
+8\,\z2\,x^{-2}
+432\,\z2\,\xm^{-1}
\nn \\[0.5mm] & & \mbox{ \hspn }
-216\,\z2
-200\,\z2\,x
-48\,\z2\,x^2
+72\,\z2\,x^3)
+\Hhh(1,1,0)\,(11303/18
+8\,x^{-1}
\nn \\[0.5mm] & & \mbox{ \hspn }
-27307/18\,\xm^{-1}
+35993/18\,x
-600\,x^2
+394\,x^3
+8\,\z2\,x^{-2}
+392\,\z2\,\xm^{-1}
-196\,\z2
\nn \\[0.5mm] & & \mbox{ \hspn }
-180\,\z2\,x
-48\,\z2\,x^2
+72\,\z2\,x^3)
+\Hhh(1,1,1)\,(9337/18
-25979/18\,\xm^{-1}
+32905/18\,x
\nn \\[0.5mm] & & \mbox{ \hspn }
+328\,\z2\,\xm^{-1}
-164\,\z2
-284\,\z2\,x
+432\,\z2\,x^2
-360\,\z2\,x^3)
+\Hh(0,0)\,(498587/216
\nn \\[0.5mm] & & \mbox{ \hspn }
-842039/162\,\xm^{-1}
+1618205/216\,x
+1260\,x^2
-80\,\z2\,x^{-1}
+24686/9\,\z2\,\xm^{-1}
\nn \\[0.5mm] & & \mbox{ \hspn }
-3441/2\,\z2
-3717/2\,\z2\,x
-1104\,\z2\,x^2
+612\,\z2\,x^3
+1160/3\,\z3\,\xm^{-1}
-769/3\,\z3
\nn \\[0.5mm] & & \mbox{ \hspn }
-1339/3\,\z3\,x
+336\,\z3\,x^2
-504\,\z3\,x^3)
+\Hh(0,1)\,(86396/81
-24\,x^{-1}
\nn \\[0.5mm] & & \mbox{ \hspn }
-1178369/324\,\xm^{-1}
+439538/81\,x
+1222\,x^2
-8\,\z2\,x^{-1}
+4835/3\,\z2\,\xm^{-1}
\nn \\[0.5mm] & & \mbox{ \hspn }
-2593/3\,\z2
-3628/3\,\z2\,x
-72\,\z2\,x^2
-48\,\z3\,x^{-2}
+976/3\,\z3\,\xm^{-1}
-520/3\,\z3
\nn \\[0.5mm] & & \mbox{ \hspn }
+818/3\,\z3\,x
-1824\,\z3\,x^2
+1296\,\z3\,x^3)
+\Hh(1,0)\,(168623/648
+32\,x^{-1}
\nn \\[0.5mm] & & \mbox{ \hspn }
-147071/81\,\xm^{-1}
+2253563/648\,x
+86\,x^2
+8\,\z2\,x^{-2}
+16\,\z2\,x^{-1}
+12263/9\,\z2\,\xm^{-1}
\nn \\[0.5mm] & & \mbox{ \hspn }
-5740/9\,\z2
-7963/9\,\z2\,x
-456\,\z2\,x^2
+612\,\z2\,x^3
-56\,\z3\,x^{-2}
+944/3\,\z3\,\xm^{-1}
\nn \\[0.5mm] & & \mbox{ \hspn }
-472/3\,\z3
-808/3\,\z3\,x
+336\,\z3\,x^2
-504\,\z3\,x^3)
+\Hh(1,1)\,(15961/216
+40\,x^{-1}
\nn \\[0.5mm] & & \mbox{ \hspn }
-46483/27\,\xm^{-1}
+784189/216\,x
+48\,x^2
+3022/3\,\z2\,\xm^{-1}
-1412/3\,\z2
-3128/3\,\z2\,x
\nn \\[0.5mm] & & \mbox{ \hspn }
+360\,\z2\,x^2
-56\,\z3\,x^{-2}
+176\,\z3\,\xm^{-1}
-88\,\z3
+400\,\z3\,x
-1824\,\z3\,x^2
+1296\,\z3\,x^3)
\nn \\[0.5mm] & & \mbox{ \hspn }
+\,\H(0)\,(1578379/2592
-5764837/1296\,\xm^{-1}
+25289039/2592\,x
+1174\,x^2
\nn \\[0.5mm] & & \mbox{ \hspn }
+24\,\z2\,x^{-1}
+35378/9\,\z2\,\xm^{-1}
-73799/36\,\z2
-136907/36\,\z2\,x
-1740\,\z2\,x^2
\nn \\[0.5mm] & & \mbox{ \hspn }
+394\,\z2\,x^3
+56\,\z3\,x^{-1}
+3085/3\,\z3\,\xm^{-1}
-5771/9\,\z3
-24019/18\,\z3\,x
+984\,\z3\,x^2
\nn \\[0.5mm] & & \mbox{ \hspn }
-660\,\z3\,x^3
-2506/3\,\z4\,\xm^{-1}
+5755/12\,\z4
+3463/12\,\z4\,x
+636\,\z4\,x^2
-954\,\z4\,x^3)
\nn \\[0.5mm] & & \mbox{ \hspn }
+\,\H(1)\,(-601625/648
+72\,x^{-1}
-2134163/1296\,\xm^{-1}
+831119/162\,x
+1174\,x^2
\nn \\[0.5mm] & & \mbox{ \hspn }
+8\,\z2\,x^{-1}
+10459/6\,\z2\,\xm^{-1}
-3763/6\,\z2
-10787/6\,\z2\,x
-912\,\z2\,x^2
+394\,\z2\,x^3
\nn \\[0.5mm] & & \mbox{ \hspn }
-8\,\z3\,x^{-2}
+32\,\z3\,x^{-1}
+5879/9\,\z3\,\xm^{-1}
-3823/9\,\z3
+9187/18\,\z3\,x
-816\,\z3\,x^2
\nn \\[0.5mm] & & \mbox{ \hspn }
-660\,\z3\,x^3
-106\,\z4\,x^{-2}
-1400/3\,\z4\,\xm^{-1}
+700/3\,\z4
+64/3\,\z4\,x
+636\,\z4\,x^2
\nn \\[0.5mm] & & \mbox{ \hspn }
-954\,\z4\,x^3)
-12941689/5184
-132728/81\,\xm^{-1}
+4964587/576\,x
+24\,\z2\,x^{-1}
\nn \\[0.5mm] & & \mbox{ \hspn }
+1178369/324\,\z2\,\xm^{-1}
-86396/81\,\z2
-439538/81\,\z2\,x
-1222\,\z2\,x^2
\nn \\[0.5mm] & & \mbox{ \hspn }
+13247/9\,\z3\,\xm^{-1}
-11459/12\,\z3
-120413/36\,\z3\,x
+3900\,\z3\,x^2
-1182\,\z3\,x^3
\nn \\[0.5mm] & & \mbox{ \hspn }
-980/3\,\z3\,\z2\,\xm^{-1}
+209\,\z3\,\z2
+405\,\z3\,\z2\,x
-480\,\z3\,\z2\,x^2
+432\,\z3\,\z2\,x^3
\nn \\[0.5mm] & & \mbox{ \hspn }
+106\,\z4\,x^{-1}
-72041/36\,\z4\,\xm^{-1}
+87439/72\,\z4
+98209/72\,\z4\,x
+1578\,\z4\,x^2
\nn \\[0.5mm] & & \mbox{ \hspn }
-1152\,\z4\,x^3
+1036/3\,\z5\,\xm^{-1}
-206\,\z5
+834\,\z5\,x
-3840\,\z5\,x^2
+2880\,\z5\,x^3
\nn \\[0.5mm] & & \mbox{ \hspn }
+\ddelta\,(-18199451/6912
-5764837/1296\,\z2
-29/648\,\z3
+68705/72\,\z4
\nn \\[0.5mm] & & \mbox{ \hspn }
+4300/9\,\z3\,\z2
+3091/6\,\z5
+35/3\,\zts
+521/2\,\z6)

\:\: ,
\eea
\bea
\label{c2nsX4N}
\lefteqn{c_{2,\rm ns}^{\,(4)\mathrm{N}}(x) \;=\;  } \nn \\[0.5mm] & & \mbox{ \hspn }
%
%
%
%
%
%
+\Hhhhh(-1,-1,-1,-1,0)\,(192
-1024\,\xp^{-1}
-2112\,x)
+\Hhhhh(-1,-1,-1,0,0)\,(-352
+1792\,\xp^{-1}
+3616\,x)
\nn \\[0.5mm] & & \mbox{ \hspn }
+\Hhhhh(-1,-1,0,-1,0)\,(-224
+1152\,\xp^{-1}
+2336\,x)
+\Hhhhh(-1,-1,0,0,0)\,(464
-2080\,\xp^{-1}
-3920\,x)
\nn \\[0.5mm] & & \mbox{ \hspn }
+\Hhhhh(-1,0,-1,-1,0)\,(-224
+1152\,\xp^{-1}
+2336\,x)
+\Hhhhh(-1,0,-1,0,0)\,(432
-1984\,\xp^{-1}
-3792\,x)
\nn \\[0.5mm] & & \mbox{ \hspn }
+\Hhhhh(-1,0,0,-1,0)\,(304
-1344\,\xp^{-1}
-2512\,x)
+\Hhhhh(-1,0,0,0,0)\,(-612
+1800\,\xp^{-1}
+2340\,x)
\nn \\[0.5mm] & & \mbox{ \hspn }
+\Hhhhh(-1,0,0,0,1)\,(-8
+16\,\xp^{-1}
+8\,x)
+\Hhhhh(-1,0,0,1,1)\,(16
-32\,\xp^{-1}
-16\,x)
+\Hhhhh(-1,0,1,1,1)\,(-32
\nn \\[0.5mm] & & \mbox{ \hspn }
+64\,\xp^{-1}
+32\,x)
+\Hhhhh(0,-1,-1,-1,0)\,(-224
+1152\,\xp^{-1}
+2336\,x)
+\Hhhhh(0,-1,-1,0,0)\,(432
\nn \\[0.5mm] & & \mbox{ \hspn }
-1984\,\xp^{-1}
-3792\,x)
+\Hhhhh(0,-1,0,-1,0)\,(272
-1280\,\xp^{-1}
-2480\,x)
+\Hhhhh(0,-1,0,0,0)\,(-1060
\nn \\[0.5mm] & & \mbox{ \hspn }
+1408\,\xm^{-1}
+2152\,\xp^{-1}
+1092\,x)
+\Hhhhh(0,-1,0,0,1)\,(-216
+864\,\xm^{-1}
+16\,\xp^{-1}
-1544\,x)
\nn \\[0.5mm] & & \mbox{ \hspn }
+\Hhhhh(0,-1,0,1,0)\,(-128
+544\,\xm^{-1}
-992\,x)
+\Hhhhh(0,-1,0,1,1)\,(-80
+448\,\xm^{-1}
-32\,\xp^{-1}
-880\,x)
\nn \\[0.5mm] & & \mbox{ \hspn }
+\Hhhhh(0,0,-1,-1,0)\,(296
+16\,\xm^{-1}
-1376\,\xp^{-1}
-2616\,x)
+\Hhhhh(0,0,-1,0,0)\,(-1280
+2040\,\xm^{-1}
\nn \\[0.5mm] & & \mbox{ \hspn }
+2344\,\xp^{-1}
+392\,x)
+\Hhhhh(0,0,-1,0,1)\,(-248
+1056\,\xm^{-1}
+16\,\xp^{-1}
-1960\,x)
\nn \\[0.5mm] & & \mbox{ \hspn }
+\Hhhhh(0,0,0,-1,0)\,(-1012
+1632\,\xm^{-1}
+1736\,\xp^{-1}
+52\,x)
+\Hhhhh(0,0,0,0,0)\,(1951/2
-2102/3\,\xm^{-1}
\nn \\[0.5mm] & & \mbox{ \hspn }
-600\,\xp^{-1}
+751/2\,x)
+\Hhhhh(0,0,0,0,1)\,(884
-3728/3\,\xm^{-1}
-24\,\xp^{-1}
+860\,x)
\nn \\[0.5mm] & & \mbox{ \hspn }
+\Hhhhh(0,0,0,1,0)\,(2458/3
-3872/3\,\xm^{-1}
+2458/3\,x)
+\Hhhhh(0,0,0,1,1)\,(2506/3
-4064/3\,\xm^{-1}
\nn \\[0.5mm] & & \mbox{ \hspn }
+32\,\xp^{-1}
+2602/3\,x)
+\Hhhhh(0,0,1,0,0)\,(698
-1092\,\xm^{-1}
+410\,x)
+\Hhhhh(0,0,1,0,1)\,(656
\nn \\[0.5mm] & & \mbox{ \hspn }
-1104\,\xm^{-1}
+656\,x)
+\Hhhhh(0,0,1,1,0)\,(792
-1376\,\xm^{-1}
+792\,x)
+\Hhhhh(0,0,1,1,1)\,(724
-1208\,\xm^{-1}
\nn \\[0.5mm] & & \mbox{ \hspn }
-32\,\xp^{-1}
+692\,x)
+\Hhhhh(0,1,0,-1,0)\,(-120
+560\,\xm^{-1}
-1080\,x)
+\Hhhhh(0,1,0,0,0)\,(1660/3
\nn \\[0.5mm] & & \mbox{ \hspn }
-2192/3\,\xm^{-1}
-932/3\,x)
+\Hhhhh(0,1,0,0,1)\,(356
-256\,\xm^{-1}
-748\,x)
+\Hhhhh(0,1,0,1,0)\,(616
\nn \\[0.5mm] & & \mbox{ \hspn }
-1256\,\xm^{-1}
+952\,x)
+\Hhhhh(0,1,0,1,1)\,(476
-864\,\xm^{-1}
+476\,x)
+\Hhhhh(0,1,1,0,0)\,(680
-1304\,\xm^{-1}
\nn \\[0.5mm] & & \mbox{ \hspn }
+776\,x)
+\Hhhhh(0,1,1,0,1)\,(504
-840\,\xm^{-1}
+264\,x)
+\Hhhhh(0,1,1,1,0)\,(632
-1256\,\xm^{-1}
+872\,x)
\nn \\[0.5mm] & & \mbox{ \hspn }
+\Hhhhh(0,1,1,1,1)\,(504
-920\,\xm^{-1}
+504\,x)
+\Hhhhh(1,0,-1,-1,0)\,(-16
+32\,\xm^{-1}
-16\,x)
\nn \\[0.5mm] & & \mbox{ \hspn }
+\Hhhhh(1,0,-1,0,0)\,(-216
+1008\,\xm^{-1}
-1944\,x)
+\Hhhhh(1,0,-1,0,1)\,(-96
+576\,\xm^{-1}
-1248\,x)
\nn \\[0.5mm] & & \mbox{ \hspn }
+\Hhhhh(1,0,0,-1,0)\,(-184
+880\,\xm^{-1}
-1720\,x)
+\Hhhhh(1,0,0,0,0)\,(1756/3
-1784/3\,\xm^{-1}
-3428/3\,x)
\nn \\[0.5mm] & & \mbox{ \hspn }
+\Hhhhh(1,0,0,0,1)\,(1024/3
+16/3\,\xm^{-1}
-5168/3\,x)
+\Hhhhh(1,0,0,1,0)\,(432
-640\,\xm^{-1}
-240\,x)
\nn \\[0.5mm] & & \mbox{ \hspn }
+\Hhhhh(1,0,0,1,1)\,(344
-352\,\xm^{-1}
-664\,x)
+\Hhhhh(1,0,1,0,0)\,(532
-1208\,\xm^{-1}
+964\,x)
\nn \\[0.5mm] & & \mbox{ \hspn }
+\Hhhhh(1,0,1,0,1)\,(392
-752\,\xm^{-1}
+296\,x)
+\Hhhhh(1,0,1,1,0)\,(452
-1016\,\xm^{-1}
+788\,x)
\nn \\[0.5mm] & & \mbox{ \hspn }
+\Hhhhh(1,0,1,1,1)\,(368
-736\,\xm^{-1}
+368\,x)
+\Hhhhh(1,1,0,-1,0)\,(-96
+448\,\xm^{-1}
-864\,x)
\nn \\[0.5mm] & & \mbox{ \hspn }
+\Hhhhh(1,1,0,0,0)\,(1540/3
-3080/3\,\xm^{-1}
+1540/3\,x)
+\Hhhhh(1,1,0,0,1)\,(300
-232\,\xm^{-1}
-804\,x)
\nn \\[0.5mm] & & \mbox{ \hspn }
+\Hhhhh(1,1,0,1,0)\,(440
-1008\,\xm^{-1}
+824\,x)
+\Hhhhh(1,1,0,1,1)\,(328
-576\,\xm^{-1}
+88\,x)
\nn \\[0.5mm] & & \mbox{ \hspn }
+\Hhhhh(1,1,1,0,0)\,(460
-1096\,\xm^{-1}
+988\,x)
+\Hhhhh(1,1,1,0,1)\,(328
-592\,\xm^{-1}
+136\,x)
\nn \\[0.5mm] & & \mbox{ \hspn }
+\Hhhhh(1,1,1,1,0)\,(416
-976\,\xm^{-1}
+848\,x)
+\Hhhhh(1,1,1,1,1)\,(320
-640\,\xm^{-1}
+320\,x)
\nn \\[0.5mm] & & \mbox{ \hspn }
+\Hhhh(-1,-1,-1,0)\,(3200/3
+32\,x^{-2}
+3200/3\,\xp^{-1}
+12544/3\,x
+960\,x^2
-288\,x^3)
\nn \\[0.5mm] & & \mbox{ \hspn }
+\Hhhh(-1,-1,0,0)\,(-5056/3
-272/5\,x^{-2}
-5440/3\,\xp^{-1}
-19712/3\,x
-1632\,x^2
+2448/5\,x^3)
\nn \\[0.5mm] & & \mbox{ \hspn }
+\Hhhh(-1,0,-1,0)\,(-3328/3
-176/5\,x^{-2}
-3520/3\,\xp^{-1}
-12992/3\,x
-1056\,x^2
+1584/5\,x^3)
\nn \\[0.5mm] & & \mbox{ \hspn }
+\Hhhh(-1,0,0,0)\,(2584
+576/5\,x^{-2}
+1792\,\xp^{-1}
+4728\,x
-5184/5\,x^3)
+\Hhhh(-1,0,0,1)\,(2608/3
\nn \\[0.5mm] & & \mbox{ \hspn }
+224/5\,x^{-2}
+160/3\,\xp^{-1}
+80/3\,x
-1344\,x^2
-2016/5\,x^3)
+\Hhhh(-1,0,1,0)\,(560
+144/5\,x^{-2}
\nn \\[0.5mm] & & \mbox{ \hspn }
-16\,x
-864\,x^2
-1296/5\,x^3)
+\Hhhh(-1,0,1,1)\,(1504/3
+128/5\,x^{-2}
-320/3\,\xp^{-1}
-352/3\,x
\nn \\[0.5mm] & & \mbox{ \hspn }
-768\,x^2
-1152/5\,x^3)
+\Hhhh(0,-1,-1,0)\,(-3544/3
-192/5\,x^{-2}
-3520/3\,\xp^{-1}
-12968/3\,x
\nn \\[0.5mm] & & \mbox{ \hspn }
-960\,x^2
+288\,x^3)
+\Hhhh(0,-1,0,0)\,(2724
+576/5\,x^{-2}
+1568\,\xm^{-1}
+1840\,\xp^{-1}
+1452\,x
\nn \\[0.5mm] & & \mbox{ \hspn }
+1632\,x^2
-2448/5\,x^3)
+\Hhhh(0,-1,0,1)\,(2800/3
+192/5\,x^{-2}
+640\,\xm^{-1}
+160/3\,\xp^{-1}
\nn \\[0.5mm] & & \mbox{ \hspn }
-6544/3\,x)
+\Hhhh(0,0,-1,0)\,(5164/3
+384/5\,x^{-2}
+4768/3\,\xm^{-1}
+1504\,\xp^{-1}
+532/3\,x
\nn \\[0.5mm] & & \mbox{ \hspn }
+1440\,x^2
-1008/5\,x^3)
+\Hhhh(0,0,0,0)\,(1688
-13852/9\,\xm^{-1}
-1240/3\,\xp^{-1}
+2584/3\,x
\nn \\[0.5mm] & & \mbox{ \hspn }
+5184/5\,x^3)
+\Hhhh(0,0,0,1)\,(2751
-28516/9\,\xm^{-1}
-160/3\,\xp^{-1}
+8981/3\,x
+2064\,x^2
\nn \\[0.5mm] & & \mbox{ \hspn }
+6792/5\,x^3)
+\Hhhh(0,0,1,0)\,(16565/9
-22732/9\,\xm^{-1}
+22841/9\,x
+672\,x^2
+1008/5\,x^3)
\nn \\[0.5mm] & & \mbox{ \hspn }
+\Hhhh(0,0,1,1)\,(18773/9
-25252/9\,\xm^{-1}
+160/3\,\xp^{-1}
+25169/9\,x
+960\,x^2
+4272/5\,x^3)
\nn \\[0.5mm] & & \mbox{ \hspn }
+\Hhhh(0,1,0,0)\,(7988/9
-12250/9\,\xm^{-1}
+7736/9\,x
-432\,x^2
-4344/5\,x^3)
+\Hhhh(0,1,0,1)\,(3130/3
\nn \\[0.5mm] & & \mbox{ \hspn }
-5000/3\,\xm^{-1}
+5194/3\,x
+96\,x^2
+144/5\,x^3)
+\Hhhh(0,1,1,0)\,(1106
-2054\,\xm^{-1}
+2146\,x
\nn \\[0.5mm] & & \mbox{ \hspn }
-288\,x^2
-3264/5\,x^3)
+\Hhhh(0,1,1,1)\,(3190/3
-5372/3\,\xm^{-1}
+5614/3\,x)
\nn \\[0.5mm] & & \mbox{ \hspn }
+\Hhhh(1,0,-1,0)\,(1136/3
+64/5\,x^{-2}
+2048/3\,\xm^{-1}
-6448/3\,x
+384\,x^2
+576/5\,x^3)
\nn \\[0.5mm] & & \mbox{ \hspn }
+\Hhhh(1,0,0,0)\,(10580/9
-12790/9\,\xm^{-1}
-9376/9\,x)
+\Hhhh(1,0,0,1)\,(13904/9
+64/5\,x^{-2}
\nn \\[0.5mm] & & \mbox{ \hspn }
-11770/9\,\xm^{-1}
-8956/9\,x
+720\,x^2
+4776/5\,x^3)
+\Hhhh(1,0,1,0)\,(1988/3
-32/5\,x^{-2}
\nn \\[0.5mm] & & \mbox{ \hspn }
-5668/3\,\xm^{-1}
+6164/3\,x
-192\,x^2
-288/5\,x^3)
+\Hhhh(1,0,1,1)\,(916
-1652\,\xm^{-1}
+1364\,x
\nn \\[0.5mm] & & \mbox{ \hspn }
+192\,x^2
+624\,x^3)
+\Hhhh(1,1,0,0)\,(5348/9
-16/5\,x^{-2}
-15358/9\,\xm^{-1}
+14096/9\,x
\nn \\[0.5mm] & & \mbox{ \hspn }
-432\,x^2
-4344/5\,x^3)
+\Hhhh(1,1,0,1)\,(2468/3
+16/5\,x^{-2}
-4060/3\,\xm^{-1}
+2372/3\,x
\nn \\[0.5mm] & & \mbox{ \hspn }
+96\,x^2
+144/5\,x^3)
+\Hhhh(1,1,1,0)\,(1456/3
-16/5\,x^{-2}
-4652/3\,\xm^{-1}
+5488/3\,x
-288\,x^2
\nn \\[0.5mm] & & \mbox{ \hspn }
-3264/5\,x^3)
+\Hhhh(1,1,1,1)\,(620
-1280\,\xm^{-1}
+1164\,x)
+\Hhh(-1,-1,-1)\,(-512\,\z2\,\xp^{-1}
+96\,\z2
\nn \\[0.5mm] & & \mbox{ \hspn }
-1056\,\z2\,x)
+\Hhh(-1,-1,0)\,(-7648/3
-992/25\,x^{-2}
-128\,x^{-1}
-1984/3\,\xp^{-1}
-5536\,x
\nn \\[0.5mm] & & \mbox{ \hspn }
-1920\,x^2
+11808/25\,x^3
+352\,\z2\,\xp^{-1}
-64\,\z2
+736\,\z2\,x)
+\Hhh(-1,0,-1)\,(576\,\z2\,\xp^{-1}
\nn \\[0.5mm] & & \mbox{ \hspn }
-112\,\z2
+1168\,\z2\,x)
+\Hhh(-1,0,0)\,(289016/45
+2896/25\,x^{-2}
+272/5\,x^{-1}
+9352/9\,\xp^{-1}
\nn \\[0.5mm] & & \mbox{ \hspn }
+321856/45\,x
-2448/5\,x^2
-34704/25\,x^3
-248\,\z2\,\xp^{-1}
+68\,\z2
-404\,\z2\,x)
\nn \\[0.5mm] & & \mbox{ \hspn }
+\Hhh(-1,0,1)\,(16528/9
+992/25\,x^{-2}
-96\,x^{-1}
+304/9\,\xp^{-1}
+2432/9\,x
-2208\,x^2
\nn \\[0.5mm] & & \mbox{ \hspn }
-11808/25\,x^3
-16\,\z2\,\xp^{-1}
+8\,\z2
-8\,\z2\,x)
+\Hhh(0,-1,-1)\,(576\,\z2\,\xp^{-1}
-112\,\z2
\nn \\[0.5mm] & & \mbox{ \hspn }
+1168\,\z2\,x)
+\Hhh(0,-1,0)\,(42214/9
+1904/25\,x^{-2}
+112/5\,x^{-1}
+800\,\xm^{-1}
\nn \\[0.5mm] & & \mbox{ \hspn }
+6088/9\,\xp^{-1}
+63946/45\,x
+1776\,x^2
-11808/25\,x^3
-864\,\z2\,\xm^{-1}
-376\,\z2\,\xp^{-1}
\nn \\[0.5mm] & & \mbox{ \hspn }
+292\,\z2
+844\,\z2\,x)
+\Hhh(0,0,-1)\,(-1048\,\z2\,\xm^{-1}
-704\,\z2\,\xp^{-1}
+396\,\z2
+652\,\z2\,x)
\nn \\[0.5mm] & & \mbox{ \hspn }
+\Hhh(0,0,0)\,(65087/45
-576/5\,x^{-1}
-18371/18\,\xm^{-1}
-5080/9\,\xp^{-1}
-176803/45\,x
\nn \\[0.5mm] & & \mbox{ \hspn }
+5184/5\,x^2
+34704/25\,x^3
+3572/3\,\z2\,\xm^{-1}
+76\,\z2\,\xp^{-1}
-884\,\z2
-808\,\z2\,x)
\nn \\[0.5mm] & & \mbox{ \hspn }
+\Hhh(0,0,1)\,(41267/18
-288/5\,x^{-1}
-23429/9\,\xm^{-1}
-152/9\,\xp^{-1}
+139937/90\,x
+2544\,x^2
\nn \\[0.5mm] & & \mbox{ \hspn }
+36208/25\,x^3
+416\,\z2\,\xm^{-1}
+8\,\z2\,\xp^{-1}
-508\,\z2
+652\,\z2\,x)
+\Hhh(0,1,0)\,(12734/15
\nn \\[0.5mm] & & \mbox{ \hspn }
-112/5\,x^{-1}
-12331/9\,\xm^{-1}
+79654/45\,x
-2856/5\,x^2
-976\,x^3
-104\,\z2\,\xm^{-1}
\nn \\[0.5mm] & & \mbox{ \hspn }
-280\,\z2
+1448\,\z2\,x)
+\Hhh(0,1,1)\,(47296/45
-128/5\,x^{-1}
-13964/9\,\xm^{-1}
+84266/45\,x
\nn \\[0.5mm] & & \mbox{ \hspn }
-1968/5\,x^2
+264\,\z2\,\xm^{-1}
-392\,\z2
+904\,\z2\,x)
+\Hhh(1,0,-1)\,(-560\,\z2\,\xm^{-1}
+88\,\z2
\nn \\[0.5mm] & & \mbox{ \hspn }
+1240\,\z2\,x)
+\Hhh(1,0,0)\,(69457/45
+16/5\,x^{-1}
-17476/9\,\xm^{-1}
+3181/15\,x
+4344/5\,x^2
\nn \\[0.5mm] & & \mbox{ \hspn }
-712/3\,\z2\,\xm^{-1}
-844/3\,\z2
+6356/3\,\z2\,x)
+\Hhh(1,0,1)\,(6267/5
-16/5\,x^{-1}
-1839\,\xm^{-1}
\nn \\[0.5mm] & & \mbox{ \hspn }
+17429/15\,x
+4296/5\,x^2
+976\,x^3
+176\,\z2\,\xm^{-1}
-280\,\z2
+872\,\z2\,x)
\nn \\[0.5mm] & & \mbox{ \hspn }
+\Hhh(1,1,0)\,(32257/45
+16/5\,x^{-1}
-15511/9\,\xm^{-1}
+83923/45\,x
-1176/5\,x^2
-976\,x^3
\nn \\[0.5mm] & & \mbox{ \hspn }
-120\,\z2\,\xm^{-1}
-236\,\z2
+1540\,\z2\,x)
+\Hhh(1,1,1)\,(6205/9
-12119/9\,\xm^{-1}
+11089/9\,x
\nn \\[0.5mm] & & \mbox{ \hspn }
+80\,\z2\,\xm^{-1}
-232\,\z2
+920\,\z2\,x)
+\Hh(-1,-1)\,(16\,\z2\,x^{-2}
+1600/3\,\z2\,\xp^{-1}
\nn \\[0.5mm] & & \mbox{ \hspn }
+1600/3\,\z2
+6272/3\,\z2\,x
+480\,\z2\,x^2
-144\,\z2\,x^3
+512\,\z3\,\xp^{-1}
-96\,\z3
\nn \\[0.5mm] & & \mbox{ \hspn }
+1056\,\z3\,x)
+\Hh(-1,0)\,(4252378/675
+21512/375\,x^{-2}
+1392/25\,x^{-1}
+9712/27\,\xp^{-1}
\nn \\[0.5mm] & & \mbox{ \hspn }
+4279178/675\,x
-15408/25\,x^2
-107496/125\,x^3
-56\,\z2\,x^{-2}
-896/3\,\z2\,\xp^{-1}
\nn \\[0.5mm] & & \mbox{ \hspn }
-3704/3\,\z2
-3352/3\,\z2\,x
+1008\,\z2\,x^2
+504\,\z2\,x^3
-256\,\z3\,\xp^{-1}
+48\,\z3
\nn \\[0.5mm] & & \mbox{ \hspn }
-528\,\z3\,x)
+\Hh(0,-1)\,(-288/5\,\z2\,x^{-2}
-640\,\z2\,\xm^{-1}
-640\,\z2\,\xp^{-1}
-1524\,\z2
\nn \\[0.5mm] & & \mbox{ \hspn }
+20\,\z2\,x
-480\,\z2\,x^2
+144\,\z2\,x^3
-224\,\z3\,\xm^{-1}
-560\,\z3\,\xp^{-1}
+152\,\z3
-728\,\z3\,x)
\nn \\[0.5mm] & & \mbox{ \hspn }
+\Hh(0,0)\,(-350027/5400
-2896/25\,x^{-1}
+103447/324\,\xm^{-1}
-17384/27\,\xp^{-1}
\nn \\[0.5mm] & & \mbox{ \hspn }
-16956149/1800\,x
+26664/25\,x^2
+107496/125\,x^3
+28900/9\,\z2\,\xm^{-1}
\nn \\[0.5mm] & & \mbox{ \hspn }
+32/3\,\z2\,\xp^{-1}
-2751\,\z2
-8449/3\,\z2\,x
-2064\,\z2\,x^2
-1560\,\z2\,x^3
+5260/3\,\z3\,\xm^{-1}
\nn \\[0.5mm] & & \mbox{ \hspn }
+148\,\z3\,\xp^{-1}
-3230/3\,\z3
-1346/3\,\z3\,x)
+\Hh(0,1)\,(1332331/2025
-512/25\,x^{-1}
\nn \\[0.5mm] & & \mbox{ \hspn }
-81845/162\,\xm^{-1}
-2398349/2025\,x
-25672/25\,x^2
-96/5\,\z2\,x^{-2}
+1080\,\z2\,\xm^{-1}
\nn \\[0.5mm] & & \mbox{ \hspn }
-1634\,\z2
+430\,\z2\,x
-576\,\z2\,x^2
-864/5\,\z2\,x^3
+1976/3\,\z3\,\xm^{-1}
-1532/3\,\z3
\nn \\[0.5mm] & & \mbox{ \hspn }
+1348/3\,\z3\,x)
+\Hh(1,0)\,(2535097/1620
+112/5\,x^{-1}
-110893/81\,\xm^{-1}
-46891/324\,x
\nn \\[0.5mm] & & \mbox{ \hspn }
+640\,x^2
-24\,\z2\,x^{-2}
+9562/9\,\z2\,\xm^{-1}
-17192/9\,\z2
+18772/9\,\z2\,x
-1056\,\z2\,x^2
\nn \\[0.5mm] & & \mbox{ \hspn }
-1056\,\z2\,x^3
+2008/3\,\z3\,\xm^{-1}
-1172/3\,\z3
-164/3\,\z3\,x)
+\Hh(1,1)\,(654959/540
\nn \\[0.5mm] & & \mbox{ \hspn }
+128/5\,x^{-1}
-26339/27\,\xm^{-1}
-159709/540\,x
-1968/5\,x^2
-96/5\,\z2\,x^{-2}
\nn \\[0.5mm] & & \mbox{ \hspn }
+820\,\z2\,\xm^{-1}
-1356\,\z2
+1300\,\z2\,x
-576\,\z2\,x^2
-864/5\,\z2\,x^3
+504\,\z3\,\xm^{-1}
\nn \\[0.5mm] & & \mbox{ \hspn }
-332\,\z3
+148\,\z3\,x)
+\,\H(-1)\,(-1488/25\,\z2\,x^{-2}
+32\,\z2\,x^{-1}
-3280/9\,\z2\,\xp^{-1}
\nn \\[0.5mm] & & \mbox{ \hspn }
-28000/9\,\z2
-27344/9\,\z2\,x
+1248\,\z2\,x^2
+17712/25\,\z2\,x^3
-144/5\,\z3\,x^{-2}
\nn \\[0.5mm] & & \mbox{ \hspn }
-480\,\z3\,\xp^{-1}
-784\,\z3
-2032\,\z3\,x
-96\,\z3\,x^2
+1296/5\,\z3\,x^3
-516\,\z4\,\xp^{-1}
+158\,\z4
\nn \\[0.5mm] & & \mbox{ \hspn }
-758\,\z4\,x)
+\,\H(0)\,(-74841839/81000
+7768/375\,x^{-1}
+162721/324\,\xm^{-1}
\nn \\[0.5mm] & & \mbox{ \hspn }
-928/3\,\xp^{-1}
-423561229/81000\,x
+83776/125\,x^2
+80\,\z2\,x^{-1}
+2665\,\z2\,\xm^{-1}
\nn \\[0.5mm] & & \mbox{ \hspn }
-404/9\,\z2\,\xp^{-1}
-41267/18\,\z2
-803/6\,\z2\,x
-2544\,\z2\,x^2
-48016/25\,\z2\,x^3
\nn \\[0.5mm] & & \mbox{ \hspn }
+10856/3\,\z3\,\xm^{-1}
+280\,\z3\,\xp^{-1}
-19552/9\,\z3
-13840/9\,\z3\,x
+96\,\z3\,x^2
\nn \\[0.5mm] & & \mbox{ \hspn }
+5232/5\,\z3\,x^3
-3287/3\,\z4\,\xm^{-1}
+300\,\z4\,\xp^{-1}
+1961/3\,\z4
+2645/3\,\z4\,x)
\nn \\[0.5mm] & & \mbox{ \hspn }
+\,\H(1)\,(6570689/4050
+1472/25\,x^{-1}
-239633/648\,\xm^{-1}
-16747063/8100\,x
\nn \\[0.5mm] & & \mbox{ \hspn }
-15832/25\,x^2
-496/25\,\z2\,x^{-2}
+336/5\,\z2\,x^{-1}
+4525/3\,\z2\,\xm^{-1}
-37921/15\,\z2
\nn \\[0.5mm] & & \mbox{ \hspn }
+24091/15\,\z2\,x
-9096/5\,\z2\,x^2
-30304/25\,\z2\,x^3
-88/5\,\z3\,x^{-2}
+11626/9\,\z3\,\xm^{-1}
\nn \\[0.5mm] & & \mbox{ \hspn }
-10424/9\,\z3
+1180/9\,\z3\,x
+6528/5\,\z3\,x^3
-484/3\,\z4\,\xm^{-1}
+812/3\,\z4
\nn \\[0.5mm] & & \mbox{ \hspn }
-2608/3\,\z4\,x)
-79641581/162000
+35192/375\,x^{-1}
-161929/432\,\xm^{-1}
\nn \\[0.5mm] & & \mbox{ \hspn }
-113391919/162000\,x
+162936/125\,x^2
+1904/25\,\z2\,x^{-1}
+52709/162\,\z2\,\xm^{-1}
\nn \\[0.5mm] & & \mbox{ \hspn }
+4856/27\,\z2\,\xp^{-1}
-1332331/2025\,\z2
+15235883/2025\,\z2\,x
+25672/25\,\z2\,x^2
\nn \\[0.5mm] & & \mbox{ \hspn }
-107496/125\,\z2\,x^3
+232/5\,\z3\,x^{-1}
+31993/9\,\z3\,\xm^{-1}
+356\,\z3\,\xp^{-1}
-42073/30\,\z3
\nn \\[0.5mm] & & \mbox{ \hspn }
-4933/90\,\z3\,x
-24/5\,\z3\,x^2
+8736/5\,\z3\,x^3
-3628/3\,\z3\,\z2\,\xm^{-1}
-52\,\z3\,\z2\,\xp^{-1}
\nn \\[0.5mm] & & \mbox{ \hspn }
+842\,\z3\,\z2
-10\,\z3\,\z2\,x
-47453/18\,\z4\,\xm^{-1}
+3488/3\,\z4\,\xp^{-1}
+69301/36\,\z4
\nn \\[0.5mm] & & \mbox{ \hspn }
+144733/36\,\z4\,x
+1812\,\z4\,x^2
+4062/5\,\z4\,x^3
+44/3\,\z5\,\xm^{-1}
+508\,\z5\,\xp^{-1}
\nn \\[0.5mm] & & \mbox{ \hspn }
-448\,\z5
+1660\,\z5\,x
+\ddelta\,(-61555/128
+212833/324\,\z2
-707833/162\,\z3
\nn \\[0.5mm] & & \mbox{ \hspn }
+123449/36\,\z4
+7322/9\,\z3\,\z2
+1394/3\,\z5
+976/3\,\zts
+1124/3\,\z6)

\eea
and
\bea
\label{c2nsX4F}
\lefteqn{c_{2,\rm ns}^{\,(4)F}(x) \;=\;  } \nn \\[0.5mm] & & \mbox{\hspn}
%
%
%
%
%
%
+\Hhhh(0,0,0,0)\,(-119
+238\,\xm^{-1}
-119\,x)
+\Hhhh(0,0,0,1)\,(-96
+192\,\xm^{-1}
-96\,x)
+\Hhhh(0,0,1,0)\,(-72
\nn \\[0.5mm] & & \mbox{\hspn}
+144\,\xm^{-1}
-72\,x)
+\Hhhh(0,0,1,1)\,(-72
+144\,\xm^{-1}
-72\,x)
+\Hhhh(0,1,0,0)\,(-48
+96\,\xm^{-1}
-48\,x)
\nn \\[0.5mm] & & \mbox{\hspn}
+\Hhhh(0,1,0,1)\,(-48
+96\,\xm^{-1}
-48\,x)
+\Hhhh(0,1,1,0)\,(-48
+96\,\xm^{-1}
-48\,x)
+\Hhhh(0,1,1,1)\,(-48
\nn \\[0.5mm] & & \mbox{\hspn}
+96\,\xm^{-1}
-48\,x)
+\Hhhh(1,0,0,0)\,(-24
+48\,\xm^{-1}
-24\,x)
+\Hhhh(1,0,0,1)\,(-24
+48\,\xm^{-1}
-24\,x)
\nn \\[0.5mm] & & \mbox{\hspn}
+\Hhhh(1,0,1,0)\,(-24
+48\,\xm^{-1}
-24\,x)
+\Hhhh(1,0,1,1)\,(-24
+48\,\xm^{-1}
-24\,x)
+\Hhhh(1,1,0,0)\,(-24
\nn \\[0.5mm] & & \mbox{\hspn}
+48\,\xm^{-1}
-24\,x)
+\Hhhh(1,1,0,1)\,(-24
+48\,\xm^{-1}
-24\,x)
+\Hhhh(1,1,1,0)\,(-24
+48\,\xm^{-1}
-24\,x)
\nn \\[0.5mm] & & \mbox{\hspn}
+\Hhhh(1,1,1,1)\,(-24
+48\,\xm^{-1}
-24\,x)
+\Hhh(0,0,0)\,(-985/3
+1706/3\,\xm^{-1}
-1621/3\,x)
\nn \\[0.5mm] & & \mbox{\hspn}
+\Hhh(0,0,1)\,(-240
+420\,\xm^{-1}
-408\,x)
+\Hhh(0,1,0)\,(-152
+268\,\xm^{-1}
-272\,x)
+\Hhh(0,1,1)\,(-152
\nn \\[0.5mm] & & \mbox{\hspn}
+268\,\xm^{-1}
-272\,x)
+\Hhh(1,0,0)\,(-64
+116\,\xm^{-1}
-136\,x)
+\Hhh(1,0,1)\,(-64
+116\,\xm^{-1}
-136\,x)
\nn \\[0.5mm] & & \mbox{\hspn}
+\Hhh(1,1,0)\,(-64
+116\,\xm^{-1}
-136\,x)
+\Hhh(1,1,1)\,(-64
+116\,\xm^{-1}
-136\,x)
+\Hh(0,0)\,(-1130/3
\nn \\[0.5mm] & & \mbox{\hspn}
+2096/3\,\xm^{-1}
-1116\,x
-192\,\z2\,\xm^{-1}
+96\,\z2
+96\,\z2\,x)
+\Hh(0,1)\,(-632/3
\nn \\[0.5mm] & & \mbox{\hspn}
+1282/3\,\xm^{-1}
-2240/3\,x
-96\,\z2\,\xm^{-1}
+48\,\z2
+48\,\z2\,x)
+\Hh(1,0)\,(-148/3
\nn \\[0.5mm] & & \mbox{\hspn}
+470/3\,\xm^{-1}
-1120/3\,x
-48\,\z2\,\xm^{-1}
+24\,\z2
+24\,\z2\,x)
+\Hh(1,1)\,(-148/3
+470/3\,\xm^{-1}
\nn \\[0.5mm] & & \mbox{\hspn}
-1120/3\,x
-48\,\z2\,\xm^{-1}
+24\,\z2
+24\,\z2\,x)
+\,\H(0)\,(-4474/27
+14321/27\,\xm^{-1}
\nn \\[0.5mm] & & \mbox{\hspn}
-37000/27\,x
-420\,\z2\,\xm^{-1}
+240\,\z2
+408\,\z2\,x
-44\,\z3\,\xm^{-1}
+22\,\z3
+22\,\z3\,x)
\nn \\[0.5mm] & & \mbox{\hspn}
+\,\H(1)\,(1300/27
+4429/27\,\xm^{-1}
-18518/27\,x
-116\,\z2\,\xm^{-1}
+64\,\z2
+136\,\z2\,x)
\nn \\[0.5mm] & & \mbox{\hspn}
+14939/81
+25279/162\,\xm^{-1}
-79606/81\,x
-1282/3\,\z2\,\xm^{-1}
+632/3\,\z2
\nn \\[0.5mm] & & \mbox{\hspn}
+2240/3\,\z2\,x
-436/3\,\z3\,\xm^{-1}
+266/3\,\z3
+386/3\,\z3\,x
+102\,\z4\,\xm^{-1}
-51\,\z4\,(1+x)
\nn \\[0.5mm] & & \mbox{\hspn}
+\ddelta\,(281971/864
+14321/27\,\z2
-1/6\,\z3
+4\,\z3\,\z2
+1027/6\,\z4
+2\,\z5)

\:\: .
\eea

The structure of these expressions reflects the $N$-space results 
presented above. The large-$\nc$ coefficients (\ref{cLnsX4L}), 
(\ref{cLnsX4F}), (\ref{c2nsX4L}) and (\ref{c2nsX4F}) include only HPLs 
with non-negative indices corresponding to the non-alternating harmonic 
sums in eqs.~(\ref{cLns4L}), (\ref{cLns4F}), (\ref{c2ns4L}) and 
(\ref{c2ns4F}).
At the highest weight, $w=5$, the $\nfs$ coefficients 
(\ref{cLnsX4L}) and (\ref{c2nsX4L}) do not only appear with factors $x$ 
for ${\cal C}_L$ and $\xm^{-1} = 1/(1-x)$, $1$ and $x$ for ${\cal C}_2$, 
but also include additional fixed combinations of $x^a$ with 
$a = -2,\,-1,\,2,\,3$.
These do not occur with $w=5$ HPLs in the $\cfs \nfs$ coefficients 
(\ref{cLnsX4N}) and (\ref{c2nsX4N}) corresponding to eqs.~(\ref{cLns4N}) 
and (\ref{c2ns4N}), which however include HPLs with negative indices 
and terms with $\xp^{-1} = 1/(1+x)$.

Up to terms that vanish in the corresponding limit,
the large-$x$ and small-$x$ behaviour of the coefficient functions is 
given by plus-distributions, 
\beq
  \Dplus(n) \;=\; \left[ \frac{\Lnt(n)}{1-x} \right]_+
  \:\: , 
\eeq
the $\delta$-function $\delta(1\!-\!x)$ and powers of the logarithms
\beq
  L_1 \; = \; \ln(1\!-\!x)
  \:\: , \;\;
  L_0 \; = \; \ln x
\:\: .
\eeq
The $\nfs$ and $\nft$ coefficients of $\Dplus(n)$ and $\delta(1\!-\!x)$ for 
${\cal C}_2$ are
\bea
c_{2,\rm ns}^{\,(4)}(x)\Big|_{\Dplus(5)} &\! = \!& \mbox{}
%
%
%
%
%
%
 \frac{64}{27}\,\*\cfs\,\nfs\
\label{c2ns4D5}
 , \\[1mm]
c_{2,\rm ns}^{\,(4)}(x)\Big|_{\Dplus(4)} &\! = \!& \mbox{}
%
%
%
%
%
%
-\frac{44}{9}\,\*\cf\*\ca\,\nfs\
-\frac{640}{27}\,\*\cfs\,\nfs\
+\frac{8}{27}\,\*\cf\,\nft\
\label{c2ns4D4}
 , \\[2mm]
c_{2,\rm ns}^{\,(4)}(x)\Big|_{\Dplus(3)} &\! = \!& \mbox{}
%
%
%
%
%
%
 \cf\*\ca\,\nfs\,\Big(\frac{1540}{27}
-\frac{32}{9}\,\z2\Big)
+\cfs\,\nfs\,\Big(\frac{24238}{243}
-\frac{928}{27}\,\z2\Big)
\nn \\ & & \mbox{\hspn}
-\frac{232}{81}\,\*\cf\,\nft\
\label{c2ns4D3}
 ,  \\[1.5mm]
c_{2,\rm ns}^{\,(4)}(x)\Big|_{\Dplus(2)} &\! = \!& \mbox{}
%
%
%
%
%
%
 \cf\*\ca\,\nfs\,\Big(-\frac{7403}{27}
+\frac{688}{9}\,\z2
+16\,\z3\Big)
\nn \\[0.5mm] & & \mbox{\hspn}
+\cfs\,\nfs\,\Big(-\frac{52678}{243}
+\frac{6104}{27}\,\z2
+\frac{304}{9}\,\z3\Big)
+\cf\,\nft\,\Big(\frac{940}{81}
-\frac{32}{9}\,\z2\Big)
\label{c2ns4D2}
\nn \\[2mm]
c_{2,\rm ns}^{\,(4)}(x)\Big|_{\Dplus(1)} &\! = \!& \mbox{}
%
%
%
%
%
%
 \cf\*\ca\,\nfs\,\Big(\frac{315755}{486}
-\frac{9848}{27}\,\z2
+\frac{64}{5}\,\zss
-\frac{688}{9}\,\z3\Big)
\nn \\[0.5mm] & & \mbox{\hspn}
+\cfs\,\nfs\,\Big(\frac{239633}{1458}
-\frac{50140}{81}\,\z2
+\frac{1312}{27}\,\zss
-\frac{19304}{81}\,\z3\Big)
\nn \\[0.5mm] & &  \mbox{\hspn}
+\cf\,\nft\,\Big(-\frac{17716}{729}
+\frac{464}{27}\,\z2\Big)
\label{c2ns4D1}
 \; , \\[2mm]
c_{2,\rm ns}^{\,(4)}(x)\Big|_{\Dplus(0)} &\! = \!& \mbox{}
%
%
%
%
%
%
 \cf\*\ca\,\nfs\,\Big(-\frac{3761509}{5832}
+\frac{131878}{243}\,\z2
-\frac{616}{9}\,\zss
+\frac{6092}{81}\,\z3
-\frac{400}{9}\,\z3\,\z2
\nn \\[0.5mm] & & \mbox{\hspn}
+\frac{1192}{9}\,\z5\Big)
+\cfs\,\nfs\,\Big(-\frac{161929}{972}
+\frac{385300}{729}\,\z2
-\frac{19904}{135}\,\zss
+\frac{3812}{9}\,\z3
\nn \\[0.5mm] & & \mbox{\hspn}
-\frac{1376}{27}\,\z3\,\z2
-\frac{64}{9}\,\z5\Big)
+\cf\,\nft\,\Big(\frac{50558}{2187}
-\frac{1880}{81}\,\z2
+\frac{16}{9}\,\zss
+\frac{80}{81}\,\z3\Big)
\label{c2ns4D0}
 , \qquad \\[2mm]
c_{2,\rm ns}^{\,(4)}(x)\Big|_{\delta} &\! = \!& \mbox{}
%
%
%
%
%
%
 \cf\,\ca\,\nfs\,\Big(-\frac{8268733}{7776}
-\frac{2063501}{972}\,\z2
-\frac{18248}{135}\,\zss
\nn \\[0.5mm] & & \mbox{}
+\frac{17477}{18}\,\z3
+\frac{284}{9}\,\z3\,\z2
-\frac{604}{9}\,\zts
+\frac{3394}{27}\,\z5
+\frac{878}{27}\,\z6\Big)
\nn \\[0.5mm] & & \mbox{\hspn}
+\cfs\,\nfs\,\Big(-\frac{61555}{288}
+\frac{212833}{729}\,\z2
+\frac{246898}{405}\,\zss
-\frac{1415666}{729}\,\z3
\nn \\[0.5mm] & & 
+\frac{29288}{81}\,\z3\,\z2
+\frac{3904}{27}\,\zts
+\frac{5576}{27}\,\z5
+\frac{4496}{27}\,\z6\Big)
\nn \\ & & \mbox{\hspn}
+\cf\*\nft\,\Big(\frac{281971}{5832}
+\frac{57284}{729}\,\z2
+\frac{4108}{405}\,\zss
-\frac{2}{81}\,\z3
+\frac{16}{27}\,\z3\,\z2
+\frac{8}{27}\,\z5\Big)
\; .
\label{c2ns4de}

\eea
The coefficients of $\Dplus(n)$ have been obtained from the soft-gluon
exponentiation in eqs.~(5.6) - (5.9) of ref.~\cite{Moch:2005ba} for 
$n = 2,\,\ldots,\,5$, and in eq.~(A.4) of ref.~\cite{Das:2019btv} for 
$n=1$ and $n=0$; our results in eqs.~(\ref{c2ns4D5}) - (\ref{c2ns4D0}) 
agree with these predictions.
On the other hand, the coefficient of $\delta(1\!-\!x)$ in 
eq.~(\ref{c2ns4de}) is a new result of the present article.

The corresponding terms with powers of $\LntO$, which are subleading for 
${\cal C}_2$ but leading for ${\cal C}_L$ in the threshold limit, 
are given by
\bea
c_{2,\rm ns}^{\,(4)}(x)\Big|_{L_1^5} &\! = \!& \mbox{}
%
%
%
%
%
%
-\frac{64}{27}\,\*\cfs\,\nfs\
\label{c2ns4L5}
 , \\[1mm]
c_{2,\rm ns}^{\,(4)}(x)\Big|_{L_1^4} &\! = \!& \mbox{}
%
%
%
%
%
%
 \frac{44}{9}\,\*\cf\,\ca\,\nfs\
+\frac{1352}{27}\,\*\cfs\,\nfs\
-\frac{8}{27}\,\*\cf\,\nft\
\label{c2ns4L4}
 , \\[2mm]
c_{2,\rm ns}^{\,(4)}(x)\Big|_{L_1^3} &\! = \!& \mbox{}
%
%
%
%
%
%
 \cf\,\ca\,\nfs\,\Big(-\frac{3652}{27}
+\frac{448}{27}\,\z2\Big)
+\cfs\,\nfs\,\Big(-\frac{70132}{243}
+\frac{224}{27}\,\z2\Big)
\nn \qquad \\[0.5mm] & & \mbox{\hspn}
+\frac{592}{81}\,\*\cf\,\nft\
\label{c2ns4L3}
 \; , \\[2mm]
c_{2,\rm ns}^{\,(4)}(x)\Big|_{L_1^2} &\! = \!& \mbox{}
%
%
%
%
%
%
 \cf\,\ca\,\nfs\,\Big(\frac{32249}{27}
-209\,\z2
-112\,\z3\Big)
+\cfs\,\nfs\,\Big(\frac{122221}{243}
\nn \\[0.5mm] & & 
-\frac{6800}{27}\,\z2
+\frac{848}{9}\,\z3\Big)
+\cf\,\nft\,\Big(-\frac{4432}{81}
+\frac{32}{9}\,\z2\Big)
\label{c2ns4L2}
 \; ,  \\[2mm]
c_{2,\rm ns}^{\,(4)}(x)\Big|_{L_1^1} &\! = \!& \mbox{}
%
%
%
%
%
%
 \cf\,\ca\,\nfs\,\Big(-\frac{1120828}{243}
+\frac{30136}{27}\,\z2
-\frac{512}{45}\,\zss
+\frac{6734}{9}\,\z3\Big)
\nn \\[0.5mm] & & \mbox{\hspn}
+\cfs\,\nfs\,\Big(\frac{360863}{729}
+\frac{98632}{81}\,\z2
+\frac{800}{27}\,\zss
-\frac{35056}{81}\,\z3\Big)
\nn \\[0.5mm] & & \mbox{\hspn}
+\cf\,\nft\,\Big(\frac{136624}{729}
-\frac{1184}{27}\,\z2\Big)
\label{c2ns4L1}
 \; , \\[2mm]
c_{2,\rm ns}^{\,(4)}(x)\Big|_{L_1^0} &\! = \!& \mbox{}
%
%
%
%
%
%
 \cf\,\ca\,\nfs\,\Big(\frac{36059503}{5832}
-\frac{544594}{243}\,\z2
+\frac{572}{9}\,\zss
-\frac{37004}{81}\,\z3
\nn \\[0.5mm] & & \mbox{}
+\frac{2704}{9}\,\z3\,\z2
-\frac{1912}{9}\,\z5\Big)
+\cfs\,\nfs\,\Big(-\frac{706090}{729}
-\frac{1170910}{729}\,\z2
\nn \\[0.5mm] & & \mbox{}
+\frac{6496}{45}\,\zss
-\frac{19336}{81}\,\z3
-\frac{6688}{27}\,\z3\,\z2
+\frac{1504}{9}\,\z5\Big)
\nn \\[0.5mm] & & \mbox{\hspn}
+\cf\,\nft\,\Big(-\frac{583016}{2187}
+\frac{8864}{81}\,\z2
-\frac{16}{9}\,\zss
-\frac{128}{81}\,\z3\Big)
\label{c2ns4L0}

\eea
and
\bea
c_{L,\rm ns}^{\,(4)}(x)\Big|_{L_1^4} &\! = \!& \mbox{}
%
%
%
%
%
%
 \frac{16}{3}\,\*\cfs\,\nfs\
\label{cLns4L4}
 ,  \\[1mm]
c_{L,\rm ns}^{\,(4)}(x)\Big|_{L_1^3} &\! = \!& \mbox{}
%
%
%
%
%
%
 \cf\*\ca\,\nfs\,\Big(-\frac{880}{27}
+\frac{352}{27}\,\z2\Big)
+\cfs\,\nfs\,\Big(-\frac{184}{27}
-\frac{704}{27}\,\z2\Big)
\nn \\[-0.5mm] & & \mbox{\hspn}
+\frac{32}{27}\,\*\cf\,\nft\
 \; , \\[1mm]
\label{cLns4L3}
c_{L,\rm ns}^{\,(4)}(x)\Big|_{L_1^2} &\! = \!& \mbox{}
%
%
%
%
%
%
 \cf\*\ca\,\nfs\,\Big(\frac{3200}{9}
-64\,\z2
-\frac{320}{3}\,\z3\Big)
\nn \\[0.5mm] & & \mbox{\hspn}
+\cfs\,\nfs\,\Big(-\frac{15172}{81}
+\frac{736}{9}\,\z2
+128\,\z3\Big)
-\frac{304}{27}\,\*\cf\,\nft\
\label{cLns4L2}
 \; , \\[2mm]
c_{L,\rm ns}^{\,(4)}(x)\Big|_{L_1^1} &\! = \!& \mbox{}
%
%
%
%
%
%
 \cf\*\ca\,\nfs\,\Big(-\frac{35846}{27}
+\frac{6592}{27}\,\z2
+\frac{1216}{45}\,\zss
+\frac{2944}{9}\,\z3\Big)
\nn \\[0.5mm] & & \mbox{\hspn}
+\cfs\,\nfs\,\Big(\frac{494242}{729}
-\frac{2032}{27}\,\z2
+\frac{704}{9}\,\zss
-\frac{12512}{27}\,\z3\Big)
\nn \\[0.5mm] & & \mbox{\hspn}
+\cf\,\nft\,\Big(\frac{3248}{81}
-\frac{64}{9}\,\z2\Big)
\label{cLns4L1}
 \; ,   \\[2mm]
c_{L,\rm ns}^{\,(4)}(x)\Big|_{L_1^0} &\! = \!& \mbox{}
%
%
%
%
%
%
 \cf\,\ca\,\nfs\,\Big(\frac{275278}{243}
-\frac{12320}{27}\,\z2
+\frac{704}{15}\,\zss
+\frac{304}{3}\,\z3
+256\,\z3\,\z2
\nn \\[0.5mm] & & \mbox{}
-\frac{560}{3}\,\z5\Big)
+\cfs\,\nfs\,\Big(-\frac{15803}{243}
-\frac{1480}{27}\,\z2
-\frac{17728}{135}\,\zss
+\frac{7984}{81}\,\z3
\qquad \nn \\[0.5mm] & & \mbox{}
-\frac{896}{3}\,\z3\,\z2
+160\,\z5\Big)
+\cf\,\nft\,\Big(-\frac{39640}{729}
+\frac{608}{27}\,\z2\Big)
\label{cLns4L0}
 \; .

\eea
The coefficients (\ref{c2ns4L5}) and (\ref{c2ns4L4}) have been 
predicted in ref.~\cite{Moch:2009hr} up to one number that was 
determined as $\xi_{\rm DIS_4}^{} = 100/3$ a little later
\cite{Grunberg:2009vs,Almasy:2010wn}; eq.~(\ref{c2ns4L4}) 
provides the first check of this result by a four-loop diagram 
calculation.
Also the coefficient (\ref{cLns4L4}) has been predicted before
\cite{Moch:2009mu,Moch:2009hr}; all other results in 
eqs.~(\ref{c2ns4L5}) -- (\ref{cLns4L0}) are new.

The small-$x$ limit of the non-singlet splitting functions and
coefficient function is given by the terms with 
$x^{\:\!0} \ln^{\:\!\ell} x = x^{\:\!0} L_0^{\:\!\ell}$.
Their $\nfs$ and $\nft$ contributions read
\bea
c_{2,\rm ns}^{\,(4)}(x)\Big|_{L_0^5} &\! =\!& \mbox{}
%
%
%
%
%
%
-\frac{1951}{1620}\,\*\cfs\,\nfs\
\label{c2ns4Lx5}
 , \\[1mm]
c_{2,\rm ns}^{\,(4)}(x)\Big|_{L_0^4} &\!=\!& \mbox{}
%
%
%
%
%
%
-\frac{1309}{108}\,\*\cf\,\ca\,\nfs\
-\frac{1190}{243}\,\*\cfs\,\nfs\
+\frac{119}{162}\,\*\cf\,\nft\
\label{c2ns4Lx4}
 , \\[1mm]
c_{2,\rm ns}^{\,(4)}(x)\Big|_{L_0^3} &\!=\!& \mbox{}
%
%
%
%
%
%
 \cf\,\ca\,\nfs\,\Big(-\frac{10051}{81}
+\frac{110}{9}\,\z2\Big)
\nn \\[0.5mm] & & \mbox{\hspn}
+\cfs\,\nfs\,\Big(-\frac{3533}{243}
+\frac{2296}{81}\,\z2\Big)
+\frac{1442}{243}\,\*\cf\,\nft\
\label{c2ns4Lx3}
 \; , \\[2mm]
c_{2,\rm ns}^{\,(4)}(x)\Big|_{L_0^2} &\!=\!& \mbox{}
%
%
%
%
%
%
 \cf\,\ca\,\nfs\,\Big(-\frac{386531}{648}
+\frac{1654}{9}\,\z2
-\frac{188}{3}\,\z3\Big)
\nn \\[0.5mm] & & \mbox{\hspn}
+\cfs\,\nfs\,\Big(-\frac{271195}{2916}
+\frac{8474}{81}\,\z2
+\frac{4948}{27}\,\z3\Big)
\nn \\[0.5mm] & & \mbox{\hspn}
+\cf\,\nft\,\Big(\frac{644}{27}
-\frac{64}{9}\,\z2\Big)
\label{c2ns4Lx2}
 \; , \\[2mm]
c_{2,\rm ns}^{\,(4)}(x)\Big|_{L_0^1} &\!=\!& \mbox{}
%
%
%
%
%
%
 \cf\,\ca\,\nfs\,\Big(-\frac{2989295}{1944}
+\frac{20702}{27}\,\z2
-224\,\z3
-\frac{1139}{9}\,\z4\Big)
\nn \\[0.5mm] & &  \mbox{\hspn}
+\cfs\,\nfs\,\Big(-\frac{460601}{1458}
+\frac{458}{3}\,\z2
+\frac{62144}{81}\,\z3
-\frac{568}{9}\,\z4\Big)
\nn \\[0.5mm] & & \mbox{\hspn}
+\cf\,\nft\,\Big(\frac{39388}{729}
-\frac{80}{3}\,\z2
-\frac{88}{27}\,\z3\Big)
\label{c2ns4Lx1}
 \; , \\[2mm]
c_{2,\rm ns}^{\,(4)}(x)\Big|_{L_0^0} &\!=\!& \mbox{}
%
%
%
%
%
%
 \cf\,\ca\,\nfs\,\Big(-\frac{19112737}{11664}
+\frac{30782}{27}\,\z2
-328\,\z3
+\frac{368}{9}\,\z3\,\z2
\nn \\[0.5mm] & & \mbox{}
-\frac{12785}{27}\,\z4
+\frac{136}{3}\,\z5\Big)
+\cfs\,\nfs\,\Big(-\frac{269059}{729}
-\frac{3638}{243}\,\z2
\nn \\[0.5mm] & & \mbox{}
+\frac{90502}{81}\,\z3
-\frac{5032}{27}\,\z3\,\z2
+\frac{5417}{27}\,\z4
+\frac{896}{27}\,\z5\Big)
\nn \\[0.5mm] & & \mbox{\hspn}
+\cf\,\nft\,\Big(\frac{110314}{2187}
-\frac{2600}{81}\,\z2
-\frac{680}{81}\,\z3
+\frac{68}{9}\,\z4\Big)

\eea
and
\bea
c_{L,\rm ns}^{\,(4)}(x)\Big|_{L_0^3}  &\!=\!& \mbox{}
%
%
%
%
%
%
-\frac{920}{81}\,\*\cfs\,\nfs\
\label{cLns4Lx3}
 , \\[1mm]
c_{L,\rm ns}^{\,(4)}(x)\Big|_{L_0^2}  &\!=\!& \mbox{}
%
%
%
%
%
%
-\frac{176}{3}\,\*\cf\,\ca\,\nfs\
-\frac{5140}{81}\,\*\cfs\,\nfs\
+\frac{32}{9}\,\*\cf\,\nft\
\label{cLns4Lx2}
 , \\[1mm]
c_{L,\rm ns}^{\,(4)}(x)\Big|_{L_0^1}  &\!=\!& \mbox{}
%
%
%
%
%
%
 \cf\,\ca\,\nfs\,\Big(-\frac{4064}{9}
+\frac{160}{3}\,\z2\Big)
\nn \\[0.5mm] & & \mbox{\hspn}
+\cfs\,\nfs\,\Big(-\frac{2456}{27}
+\frac{80}{3}\,\z2\Big)
+\frac{608}{27}\,\*\cf\,\nft\
\label{cLns4Lx1}
 \; , \\[2mm]
c_{L,\rm ns}^{\,(4)}(x)\Big|_{L_0^0}  &\!=\!& \mbox{}
%
%
%
%
%
%
 \cf\,\ca\,\nfs\,\Big(-\frac{26422}{27}
+\frac{752}{3}\,\z2
-64\,\z3\Big)
\nn \\[0.5mm] & & \mbox{\hspn}
+\cfs\,\nfs\,\Big(-\frac{2062}{729}
+\frac{1040}{27}\,\z2
+\frac{3632}{27}\,\z3\Big)
\nn \\[0.5mm] & & \mbox{\hspn}
+\cf\,\nft\,\Big(\frac{3248}{81}
-\frac{64}{9}\,\z2\Big)
\label{cLns4Lx0}

\; .
\eea
Eqs.~(\ref{c2ns4Lx5}) and (\ref{cLns4Lx3}) agree with the resummation 
predictions (4.5) and (4.6) of ref.~\cite{Davies:2022ofz} after using 
that $\ln^{\:\!\ell\!} x$ transforms to $(-1)^\ell\, \ell\, !\, 
N^{\,-\ell-1}$.  All other small-$x$ coefficients are~new.

\begin{figure}[bth]
\vspace*{2mm}
\centerline{\epsfig{file=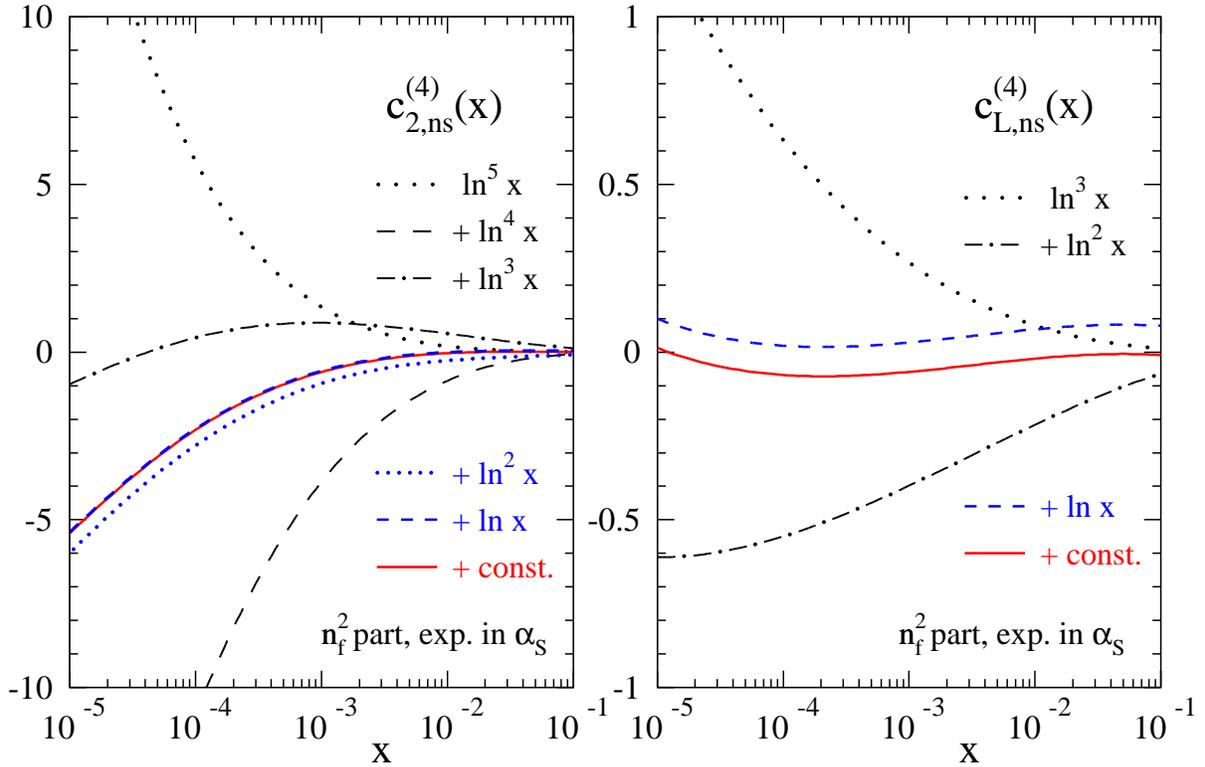,width=16.0cm,angle=0}\quad}
\vspace{-3mm}
\caption{ \label{Fig3} \small
 Successive small-$x$ approximations for the $\nfs$ contributions to 
 the fourth-order coefficient functions for $F_{2,\rm ns}$ (left panel)
 and $F_{L,\rm ns}$ (right panel). The coefficients in eqs.~(\ref{c2xto0}) 
 and (\ref{cLxto0}) have been converted to an expansion in $\als$. }
\vspace{-2mm}
\end{figure}

In the case of QCD, eqs.~(\ref{c2ns4Lx5}) -- (\ref{cLns4Lx0}) lead
to the numerical small-$x$ expansions
\bea
\label{c2xto0}
  c_{2,\rm ns}^{\,(4)}(x) \Big|_{\nfs} &\!=\!& \mbox{}
    - 2.1410151 \,\ln^{\:\! 5} \! x
    - 57.187471 \,\ln^{\:\! 4} \! x
    - 358.88192 \,\ln^{\:\! 3} \! x
  \nn \\[-2.5mm] & & \mbox{}
    - 945.87933 \,\ln^{\:\! 2} \! x
    - 1327.8897 \,\ln x 
    - 688.88341
  + {\cal O} \left( x \right) 
\eea
and
\bea
\label{cLxto0}
  c_{L,\rm ns}^{\,(4)}(x) \Big|_{\nfs} &\!=\!& \mbox{}
    - 20.192044 \,\ln^{\:\! 3} \! x
    - 347.47874 \,\ln^{\:\! 2} \! x
    - 1539.0328 \,\ln x 
  \quad \nn \\[-2.5mm] & & \mbox{}
    - 2177.6994 
  + {\cal O} \left( x \right)
\eea
which yield the successive approximations shown in fig.~\ref{Fig3}.
The pattern seen in this figure is the same seen for the complete
third-order non-singlet coefficient functions in the right parts
of fig.~2 and fig.~7 in ref.~\cite{Vermaseren:2005qc} and for the
four- and five-loop resummation predictions in figs.~1 -- 3 of 
ref.~\cite{Davies:2022ofz}: the leading, next-to-leading and
next-to-next-to-leading logarithmic small-$x$ approximations wildly
oscillate and thus, in general, cannot be used to obtain any reliable 
prediction for even the rough shape of high-order coefficient 
functions at any physically relevant small values of $x$.

%
\section{Summary and outlook}
%

As a step towards the determination of the fourth-order QCD coefficient
functions $c_{a}^{\,(4)}(x)$ in inclusive deep-inelastic scattering 
(DIS), we have computed the double-fermionic ($\nfs$) non-singlet 
contributions to the structure functions $F_2$ and $F_L$. Our results
are applicable to electromagnetic DIS (where we have ignored the very 
small \cite{Larin:1996wd,Vermaseren:2005qc} contributions in which the 
photon couples to different quark lines) and to charged-current 
$(W^+ \!+\! W^-)$ exchange.
The analytic dependence of these coefficient functions on Mellin-$N$,
 and hence Bjorken-$x$, has been reconstructed from a very large number 
of even values of $N$, up to $N \simeq 1200$. 

Our calculations verify the corresponding contributions to the 
next-to-next-to-leading order splitting function $P_{\rm ns}^{\,+(3)}(x)$
obtained in ref.~\cite{Davies:2016jie}, and the much simpler (and much 
smaller) $\cf \nft$ leading large-$\nf$ contributions to $c_{L}^{\,(4)}(x)$
and $c_{\:\!2}^{\,(4)}(x)$ \cite{Gracey:1995aj,Mankiewicz:1997gz}.
The coefficients of all large-$x$ plus-distributions in ${\cal C}_2$
at this order, 
$[ (1\!-\!x)^{-1} \ln^{\:\!\ell} (1\!-\!x)]_+$ with $0 \leq \ell \leq 5$,
have been predicted by the soft-gluon exponentiation 
\cite{Moch:2005ba, Das:2019btv}; we agree with these predictions.
Our results for both structure functions make contact, just, with the
resummations of $\ln (1\!-\!x)$ and $\ln x$ double logarithms in refs.~%
\cite{Moch:2009hr,Grunberg:2009vs,Almasy:2010wn,Moch:2009mu,Davies:2022ofz},
as the leading terms of the $\nfs$ parts contribute to the overall
next-to-next-to-leading logarithms; also here we find full agreement.
Finally we have predicted the analytic form of the $\z4 \:\! \nft$ 
contributions to the Mellin-$N$ moments of the five-loop splitting function 
$P_{\rm ns}^{\,+(4)}(x)$, and verified the corresponding $\z4 \:\! \nff$
parts of ref.~\cite{Gracey:1994nn}.

To compute the large number of moments required for the reconstruction of the 
all-$N$ coefficient functions, we have developed a new approach based on the 
method of integration by parts, differential equations and the optical theorem.
By employing the knowledge that the Mellin moments of the forward scattering 
amplitude correspond to the coefficients of the Taylor series around 
$\omega=1/x=0$, we derive a system of recurrence relations which holds for any 
linear system of differential equations, whose matrix contains no higher-order 
poles in~$\omega$. 
While such higher poles are generically present, we find a simple algorithm
which allows to change the basis of master integrals into a form where only 
simple poles appear in the differential matrix. The thereby obtained 
recurrence relations for the master integrals allow to obtain the Mellin 
moments at, in principle, arbitrary high $N$ from the knowledge of the 
boundary conditions, which we compute using the \textsc{Forcer} program 
\cite{Ruijl:2017cxj}.

In practice, we find that the algorithm starts to become computationally 
demanding for the problem under consideration at values of $N$ around about 
1000. 
One element which allowed us to speed up the calculation considerably was to 
convert the basis of master integrals into one where the $\ep$-dependence of 
the reduction coefficients factorizes from the dependence on $x$.
An important feature of the method is the simplicity with which the 
recurrences can be solved - a procedure which we have implemented in 
\textsc{Form} \cite{Vermaseren:2000nd,Kuipers:2012rf,Ruijl:2017dtg}.
  
In spirit the method is not dissimilar from the `method of arbitrary high 
moments' \cite{Blumlein:2017dxp}, which was used recently to recompute the 
3-loop DIS coefficient functions \cite{Blumlein:2022gpp}. 
Their method also derives recurrence relations from differential equations, 
but differs from our method in several ways. 
It relies on more advanced combinatorial algorithms to construct and 
recursively solve the differential equations for each master integral (while 
our method is based on solving a recurrence relation for the entire system). 
Furthermore it requires that the differential equations are first-order 
factorizable, a property not needed in our approach. 
While it is possible that their approach leads to a faster algorithm at 
very large $N$, it is difficult at this stage to comment on timings without 
explicit performance benchmarks.

We plan to present the $\nfs$ fourth-order contributions to the structure
function $F_3$ in $(W^+ \!+\! W^-)$ charged-current DIS in a forthcoming
publication. The computation of its `standard' $\cf \ca\:\! \nfs$ and 
$\cfs\:\!\nfs$ parts is not much more difficult than that presented in 
this article, but some additional contributions with the group invariant 
$d^{\:\!abc}d_{abc\:\!}$, for the third-order results see 
refs.~\cite{Retey:2000nq,Moch:2008fj,Blumlein:2022gpp}, prove to be more
challenging. 
We have also started  to work on a first set of $\nfo$ contributions, the
$\cft\, \nf$ terms, where we hope to be able to provide the first exact 
results of $\nfo$ parts of $P_{\rm ns}^{\,+(3)}(x)$ beyond the large-$\nc$ 
limit of ref.~\cite{Moch:2017uml}.

\vspace*{4mm}
\noindent
{\sc Form} files with our results can be obtained from the preprint server 
https://arXiv.org by downloading the source of this article.
They are also available from the authors upon request.
These files include also the analytic $N$-dependence of the third-order
quantities $a_{2,\rm ns}^{(n)}(N)$ and $a_{L,\rm ns}^{(n)}(N)$ in
eq.~(\ref{CexpD}) which we have not included in section 4 above.

%
\subsection*{Acknowledgements}
\vspace*{-1mm}
We would like to thank Jos Vermaseren, Johann Usovitsch and Sven Moch 
for useful discussions.
This work has been supported by the Vidi grant 680-47-551 of the 
Dutch Research Council (NWO), the UK Research and Innovation (UKRI) 
FLF Mr/S03479x/1 and the Consolidated Grant ST/T000988/1 of the UK
Science and Technology Facilities Council (STFC). 
A.B.S. is grateful to Nikhef for the additional support provided 
to him during the final stages of this project.

%
%
%
%

{\small
\setlength{\baselineskip}{0.5cm}

}
\end{document}